\documentclass{aa}

\usepackage{geometry}
\usepackage{todonotes}

\usepackage{graphicx}
\usepackage{txfonts}
\usepackage[colorlinks]{hyperref}
\hypersetup{allcolors={blue}}
\usepackage{placeins}

\addto\extrasenglish{%
}

\newcommand{\supernova}[1] {SN~#1}
\newcommand{\grbsn}[1] {GRB-SN~#1}
\newcommand{\xrfsn}[1] {XRF-SN~#1}
\newcommand{\kms}{\ensuremath{\text{km}~\text{s}^{-1}}}
\newcommand{\ergs}{\ensuremath{\text{erg}~\text{s}^{-1}}}
\newcommand{\erg}{\ensuremath{\text{erg}}}
\newcommand{\IcBL}[1]{Type Ic-BL~#1}
\newcommand{\Ib}[1]{Type Ib~#1}

\newcommand{\IIb}[1]{Type IIb~#1}

\newcommand{\IIn}[1]{Type IIn~#1}
\newcommand{\Mni}{\ensuremath{M_\text{Ni}}}

\begin{document}

   \title{\supernova{2019odp} -- A massive oxygen-rich Type Ib supernova}

   \author{
     T. Schweyer\inst{\ref{suokc}} \and
     J. Sollerman\inst{\ref{suokc}} \and
     A. Jerkstrand\inst{\ref{suokc}} \and
     M. Ergon\inst{\ref{suokc}} \and
     T.-W. Chen\inst{\ref{suokc},\ref{tum},\ref{mpa}} \and
     C. M. B. Omand\inst{\ref{suokc}} \and
     S. Schulze\inst{\ref{suph}} \and
     M. W. Coughlin\inst{\ref{umin}} \and
     I. Andreoni\inst{\ref{umdastro}} \and
     C. Fremling\inst{\ref{caltech}} \and
     A. Rau\inst{\ref{mpe}} \and
     Y. Sharma\inst{\ref{caltech}} \and
     N. L. Strotjohann\inst{\ref{weiz}} \and
     L. Yan\inst{\ref{caltech}} \and
     M. J. Graham\inst{\ref{caltech}} \and
     M. M. Kasliwal\inst{\ref{caltech}} \and
     R. R. Laher\inst{\ref{ipac}} \and
     J. Purdum\inst{\ref{caltechobs}} \and
     P. Rosnet\inst{\ref{uca}} \and
     B. Rusholme\inst{\ref{ipac}} \and
     R. Smith\inst{\ref{caltechobs}}
   }

\institute{
  Department of Astronomy, The Oskar Klein Centre, Stockholm University, AlbaNova, 10691 Stockholm, Sweden,\label{suokc} \email{tassilo.schweyer@astro.su.se}
  \and
  Technische Universit{\"a}t M{\"u}nchen, TUM School of Natural Sciences, Physik-Department, James-Franck-Stra{\ss}e 1, 85748 Garching, Germany\label{tum}
  \and
  Max-Planck-Institut f{\"u}r Astrophysik, Karl-Schwarzschild Stra{\ss}e 1, 85748 Garching, Germany\label{mpa}
  \and
  Department of Physics, The Oskar Klein Center, Stockholm University, AlbaNova, 10691 Stockholm, Sweden\label{suph}
  \and
  School of Physics and Astronomy, University of Minnesota, Minneapolis, Minnesota 55455, USA\label{umin}
  \and
  Department of Astronomy, University of Maryland, College Park, MD 20742, USA\label{umdastro}
  \and
  Division of Physics, Mathematics and Astronomy, California Institute of Technology, Pasadena, CA 91125, USA\label{caltech}
  \and
  Max-Planck-Institut f{\"u}r extraterrestrische Physik, Gie{\ss}enbachstra{\ss}e 1, D-85748 Garching, Germany\label{mpe}
  \and
  Department of Particle Physics and Astrophysics, Weizmann Institute of Science, 234 Herzl St, 76100 Rehovot, Israel\label{weiz}
  \and
  IPAC, California Institute of Technology, 1200 E. California Blvd, Pasadena, CA 91125, USA\label{ipac}
  \and
  Caltech Optical Observatories, California Institute of Technology, Pasadena, CA  91125, USA\label{caltechobs}
  \and
  Universit\'e Clermont Auvergne, CNRS/IN2P3, LPC, F-63000 Clermont-Ferrand, France\label{uca}
}

   \date{}

  \abstract
   {}
   {
     Stripped envelope (SE) supernovae are explosions of stars that have somehow lost most of their outer envelopes. We present the discovery and analyse the observations of the \Ib{supernova~2019odp} (a.k.a. ZTF19abqwtfu) covering epochs within days of the explosion to late nebular phases at 360\,d post-explosion.     
   }
   {
     Our observations include an extensive set of photometric observations and low- to medium-resolution spectroscopic observations, both covering the complete observable time range.
     We analysed the data using analytic models for the recombination cooling emission of the early excess emission and the diffusion of the peak light curve.
     We expanded on existing methods to derive oxygen mass estimates from nebular phase spectroscopy, and briefly discuss progenitor models based on this analysis.
   }
   {
     Our spectroscopic observations confirm the presence of He in the supernova ejecta and we thus (re)classify SN 2019odp as a \Ib{supernova}.
     From the pseudo-bolometric light curve, we estimate a high ejecta mass of $M_\text{ej} \sim 4 - 7~M_\odot$. %
     The high ejecta mass, large nebular $[\ion{O}{I}]/[\ion{Ca}{II}]$ line flux ratio ($1.2 - 1.9$), and an oxygen mass above $\gtrapprox  0.5\, M_\odot$ point towards a progenitor with a pre-explosion mass higher than $18\,M_\odot$. Whereas a majority of analysed SE supernovae in the literature seem to have low ejecta masses, indicating stripping in a binary star system, SN 2019odp instead has parameters that are consistent with an origin in a single massive star. 
     The compact nature of the progenitor ($\lesssim 10\,R_\odot$) suggests that a Wolf-Rayet star is the progenitor.
   }
   {}


   \maketitle

\section{Introduction}
\label{sec:intro}

Supernovae are luminous transients marking the end of the life cycles of certain stars.
The traditional supernova classification scheme \citep{1997ARA&A..35..309F} is based on the presence or absence of spectral features close to peak brightness.
If no hydrogen lines are present, they are classified as Type I supernovae.
The presence of helium distinguishes helium-rich Type Ib supernovae from helium-poor Type Ic supernovae.
Based on the high expansion velocities,  %
broad-lined Type Ic (Type Ic-BL) supernovae can be distinguished.
Type Ib and Ic supernovae are often considered together, since their estimated explosion parameters %
overlap \citep{2016MNRAS.457..328L}, but the actual connection to progenitors and explosion mechanisms are still under debate \citep{2019NatAs...3..717M}.
There also exist transitional transients that change their type over time, motivating additional classification schemes \citep{2017MNRAS.469.2672P,2019ApJ...880L..22W}.

For Type Ibc supernovae, the progenitor stars have to lose %
their outer envelope, stripping away most of the hydrogen or helium.
The exact mechanisms are still under debate, but %
one possible progenitor channel is single massive stars that eject their outer atmosphere in strong stellar winds \citep{2008A&ARv..16..209P,1993ApJ...411..823W}.
One key issue here is that only the most massive stars ($M_\text{ZAMS} \gtrsim 40\,M_\odot$) are able to strip their hydrogen envelope completely by this mechanism.
Alternatively, the evolution in a binary system could transfer the outer envelope to the companion star \citep[e.g.,][]{2015MNRAS.451.2123T}.
Direct detections of progenitors for Type Ib supernovae are still rare, with iPTF13bvn \citep{2016A&A...593A..68F} and \supernova{2019yvr} \citep{2021MNRAS.504.2073K} being examples, but
\cite{2012A&A...544L..11Y} argue that the existing data on progenitor luminosities are not %
constraining for the binary versus single massive star question. %
A monotonically increasing tracer for the zero-age main sequence (ZAMS) mass is the oxygen mass \citep{2021AA...656A..58L}, which may be estimated using nebular phase spectra \citep[e.g.,][]{2014MNRAS.439.3694J}.

Observations shortly after first light can also yield valuable information about shock cooling, recombination effects, and nickel mixing, which can be used to constrain aspects of the outer structure of the progenitor.
After a short ($\sim$ hours) shock-breakout flash in the ultra-violet (UV) and X-ray, such as that seen in \supernova{2008D} \citep{2009ApJ...702..226M, 2008ApJ...683L.135C}, follows the `shock cooling envelope' (SCE) emission, which can be seen as an early excess or plateau before the main peak for stripped envelope supernovae, as was discussed in connection with the \Ib{supernovae} \supernova{2008D} and \supernova{1999ex} \citep{2002AJ....124.2100S}.
For Type IIb supernovae, this has been seen in a number of objects, with \supernova{2011dh} \citep{2011ApJ...742L..18A,2015A&A...580A.142E}, \supernova{2016gkg} \citep{2018Natur.554..497B}, and \supernova{2017jgh} \citep{2021MNRAS.507.3125A} being well-studied examples.
However, despite more and more transients being discovered at ever earlier times thanks to large-area high-cadence survey programmes \citep{2021ApJ...912...46B}, not all supernovae show these cooling features early on, and iPTF13bvn for example showed no signs of any early excess.

In this paper, we present and discuss \supernova{2019odp} and attempt to infer some clues about the progenitor based on observations from very early to very late times.
We reclassify \supernova{2019odp} as a \Ib{supernova} instead of a \IcBL{supernova}.
From the light curve, we deduce a large ejecta mass and a compact progenitor.
The nebular spectra allow us to put strict limits on the oxygen mass.

The paper is structured as follows.
In Sect. \ref{sec:discovery}, we outline the initial discovery.
In Sect. \ref{sec:observations}, we present the photometric and spectroscopic observations of the supernova, and we discuss the evolution of observables in Sect. \ref{sec:evo}.
In Sect. \ref{sec:model}, we apply semi-analytical models to estimate physical parameters, such as the ejecta mass, progenitor radius, and oxygen mass.
Finally, in Sect. \ref{sec:summary} we discuss these properties in the context of different progenitor scenarios and summarise our findings. 

We use the following unless otherwise specified:
the supernova phase is in observer-frame days relative to the $g$-band peak,
all quantities have been corrected for the estimated line-of-sight Milky Way extinction (see Sect. \ref{sec:extinction}),
all magnitudes are given in the AB magnitude system, and error bars denote 1$\sigma$ uncertainties.

\section{Observations and data reduction}
\label{sec:observations}

\subsection{Discovery and initial classification}
\label{sec:discovery}

\begin{figure}
  \centering
  \includegraphics[width=\linewidth]{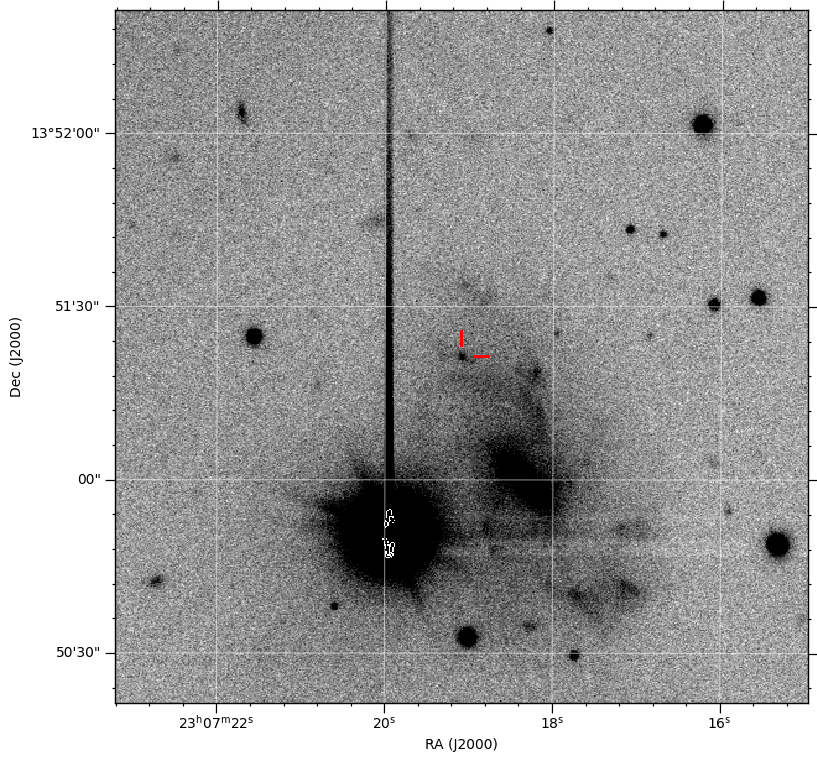}
  \caption{Stacked $r^\prime$-band GROND image using images taken between $+29$d and $+79$d showing the field of the supernova. The supernova is marked with the two red markers. The transient is located approximately $25"$ from the core of the galaxy, corresponding to a projected separation of around 8 kpc. The bottom left of the image shows a saturated star with heavy blooming.}
  \label{fig:obs:finder}
\end{figure}

The transient SN 2019odp (ZTF19abqwtfu) was discovered as part of the %
\textit{Zwicky} Transient Facility survey \citep[ZTF;][]{2019PASP..131a8002B,2019PASP..131g8001G} and was first reported to the Transient Name Server  (TNS\footnote{\url{https://www.wis-tns.org/}}) by \cite{2019TNSTR1585....1N}.
The discovery was made on August 21 2019 (MJD = 58716.38) in the $r$ band with a magnitude of $18.7$.
The previous epoch on August 18 2019 (MJD = 58713.44) shows a 2$\sigma$ flux excess in the $i$ band with a magnitude of $20.9$.
The last non-detection was on August 17 2019 (MJD = 58712.48) in the $g$ band.
We define the explosion epoch, $t_\text{expl}$, to be at MJD $58714.5 \pm 2$ -- the centre point between the last non-detection and the first significant ($> 3\sigma$) detection.

We estimated the $g$-band peak epoch, $t_g^\text{peak}$, using the interpolated $g$-band light curve (see Sect. \ref{sec:obs:interpolated}) to be at $\text{MJD}=58734 \pm 1$ days.
We specify the phase, $\Delta t^\text{peak}_g$, relative to this $g$-band peak epoch in the rest of the paper.

The transient is located at a right ascension of 23:07:19.090 (h:m:s) and declination of +13:51:21.42 
(\degr:\arcmin:\arcsec; J2000.0)
in the spiral galaxy UGC 12373 (see Fig. \ref{fig:obs:finder}).
The transient is located approximately $25\arcsec$ from the core of the galaxy, corresponding to a projected separation of around 8 kpc.
Adopting the \ion{H}{I} based redshift of $z=0.01435$ from \cite{1990ApJS...72..245S}, we used the derived Hubble flow distance of $D=64 \pm 5$ Mpc and distance modulus of $\mu = 34.0 \pm 0.2$ mag from the NASA/IPAC Extragalactic Database (NED).\footnote{This adopts the following cosmology parameters: $H_0 = 67.8$ $\text{km}~\text{s}^{-1}~\text{Mpc}^{-1}$, $\Omega_\text{matter} = 0.308$, and $\Omega_\text{vacuum}=0.692$. We used the value for the peculiar velocities that include the Virgo, great attractor, and Shapley supercluster velocity fields from NED. \citep{2000ApJ...529..786M} 
}

On August 23 2019, (MJD = 58718.2) \supernova{2019odp} was classified as a \IcBL{supernova} by \cite{2019TNSCR1595....1B} as part of the ePESSTO+ survey \citep{2015A&A...579A..40S}. However, based on further observations, we reclassify it as a \Ib{supernova} in Sect. \ref{sec:classification}.

\subsection{Photometry}
\label{sec:obs:phot}

\begin{figure*}
  \includegraphics[width=\linewidth]{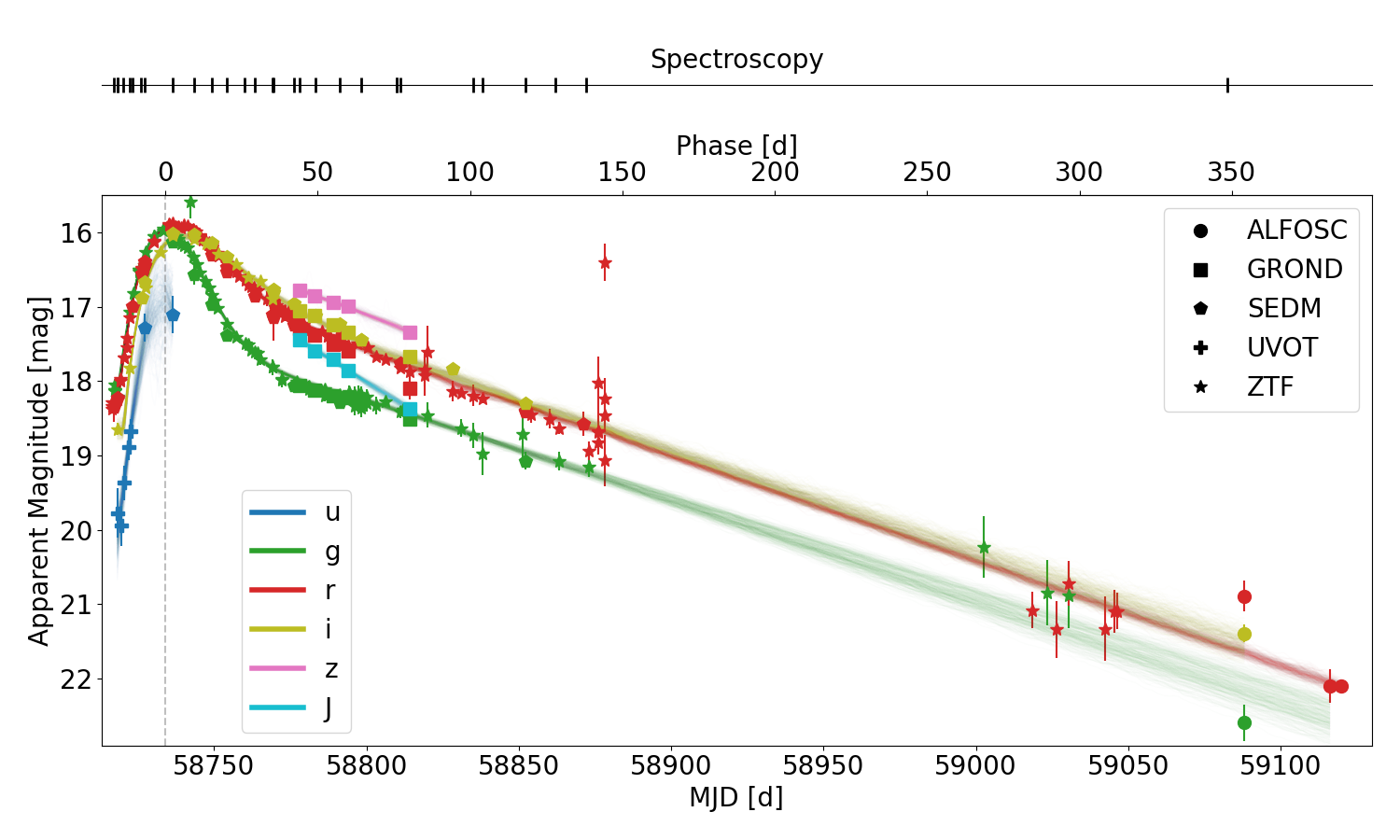}
  \caption{Photometric evolution of \supernova{2019odp} in the $ugri z^\prime J$ bands using the combined photometry dataset. The colour denotes the band and the different markers denote the source instrument. The faintly overlaid light curves are light curve realisations drawn from the conditioned Gaussian process interpolation kernel (see Sect. \ref{sec:obs:interpolated} for a detailed description). The light curves have not been corrected for extinction. The times of spectroscopic observations are marked in the top bar by vertical lines.}
  \label{fig:lc:combined}
\end{figure*}

\begin{figure}
  \includegraphics[width=\linewidth]{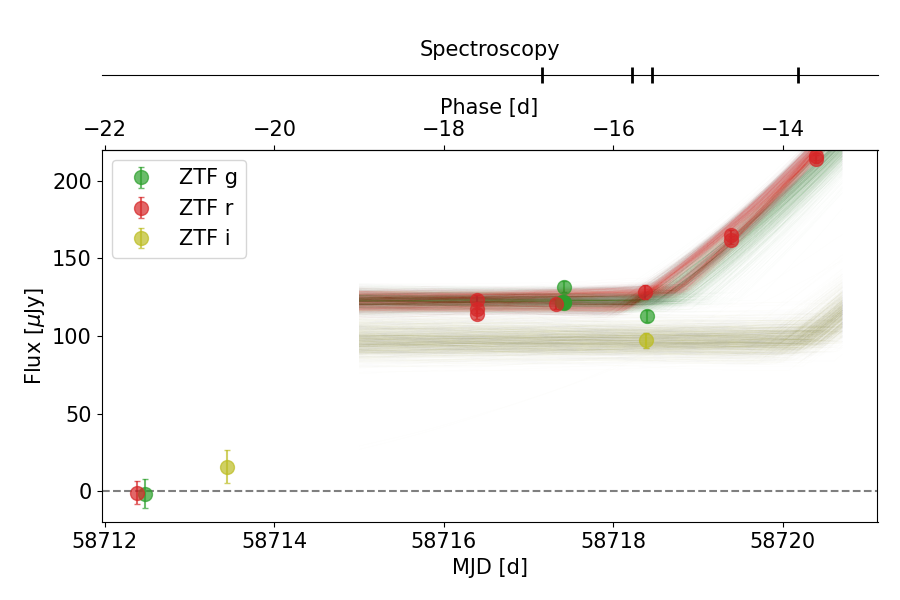}
  \caption{Photometric evolution of \supernova{2019odp} around the discovery epoch. The light curve is given in flux units to show the pre-discovery upper limits on the same scale. The different colours denote the different bands. The zero flux level is denoted by a dashed line. The faintly overlaid light curves are light curve realisations drawn from the conditioned Gaussian process interpolation kernel (see Sect. \ref{sec:obs:interpolated} for a detailed description). For illustration purposes, they are drawn beyond their validity time range (turning into a extrapolation). The light curves have not been corrected for extinction. The times of spectroscopic observations are marked in the top bar by vertical lines.}
  \label{fig:lc:early}
\end{figure}

Follow-up photometry in the $g$, $r$, and $i$ bands was obtained using the ZTF camera \citep{2020PASP..132c8001D} mounted on the Palomar 48-inch telescope (P48) as part of the ZTF survey \citep{2019PASP..131f8003B, 2019PASP..131a8002B}.
The obtained data were processed using the ZTF pipeline \citep{2019PASP..131a8003M}, which detrends the images and does PSF-matching image-subtraction against stacked template images and automatic photometric calibration against field stars using the Pan-STARRS 1 \citep[PS1;][]{2016arXiv161205560C} survey catalogue.
We used \texttt{ztflc}\footnote{\url{https://github.com/MickaelRigault/ztflc} by M. Rigault.} to perform forced photometry for all epochs. 
Based on this photometry, we see no outbursts before the main explosion; however, we notice a small plateau before the main peak (see Fig. \ref{fig:lc:early}).

In addition, we obtained manually triggered observations in the $u$, $g$, $r$, and $i$ bands using the Spectral Energy Distribution Machine \citep[SEDM;][]{2018PASP..130c5003B} Rainbow Camera mounted on the Palomar 60-inch telescope (P60).
The obtained data were automatically processed using the SEDM-RC pipeline \citep{2016A&A...593A..68F}.

We also obtained some post-peak follow-up photometry in the $g^\prime r^\prime i^\prime z^\prime J H K_s$ bands using the Gamma-ray Burst Optical/Near-infrared Detector \citep[GROND;][]{2008PASP..120..405G} mounted on the MPG 2.2m telescope located at the ESO La Silla observatory.
The data were reduced using a \textit{pyraf/IRAF}-based\footnote{\cite{2012ascl.soft07011S,1999ascl.soft11002N}.} pipeline \citep{2008ApJ...685..376K}.
For the near-infrared (NIR)  bands, aperture photometry was performed.
The $g^\prime r^\prime i^\prime z^\prime$ bands were calibrated against the Sloan Digital Sky Survey Data Release 15 catalogue \citep[SDSS DR15;][]{2019ApJS..240...23A} and the NIR $J H K_S$ bands were calibrated against the Two Micron All Sky Survey catalogue \citep[2MASS;][]{2006AJ....131.1163S}.
The GROND NIR Vega magnitudes were converted to AB magnitudes using \cite{2007AJ....133..734B}.

Late-time optical photometry in the $g$, $r$, and $i$ bands were obtained using the Alhambra Faint Object Spectrograph and Camera instrument (ALFOSC) mounted on the Nordic Optical Telescope (NOT).
The observations were reduced using the \textit{PyNOT}\footnote{\url{https://github.com/jkrogager/PyNOT} by Jens-Kristian Krogager.} pipeline.
We then performed image subtraction using \textit{hotpants} \citep{2015ascl.soft04004B} against matching PS1 images as a template.
Resampling of the template image to the same pixel scale as the science images was performed using \textit{SWarp} \citep{2002ASPC..281..228B}.
Aperture photometry was then performed on the difference image using \textit{photutils} \citep{larry_bradley_2021_5525286}.

To extend the wavelength coverage to the UV, we utilised the 30 cm UV Optical Telescope \citep[UVOT;][]{2005SSRv..120...95R} on board the \textit{Neil Gehrels Swift} Observatory \citep{2004ApJ...611.1005G}.
We retrieved science-ready data from the \textit{Swift} archive\footnote{\url{https://www.swift.ac.uk/swift_portal}}.
We first co-added all sky exposures for a given epoch and filter using \textit{uvotimsum} in HEAsoft\footnote{\url{https://heasarc.gsfc.nasa.gov/docs/software/heasoft/}} version 6.26.1.
Afterwards, we measured the brightness of \supernova{2019odp} with the \textit{Swift} tool {\tt uvotsource}.
The source aperture had a radius of $3''$, while the background region had a significantly larger radius.
The photometry was calibrated with the latest calibration files from September 2020 and converted to the AB system using \cite{2011AIPC.1358..373B}.

The combined light curve is shown in Fig. \ref{fig:lc:combined} and a list of photometric measurements is provided in \autoref{tab:obslog:phot}.
For the first 70 days, the light curve has an average cadence of 2 days in the $g$ band (no gap larger than 4 days), 1 day in the $r$ band (no gap larger than 5 days), and 3 days in the $i$ band (no gap larger than 7 days).

\subsection{Spectroscopy}
\label{sec:obs:spec}

\begin{figure}[p]
  \resizebox{\hsize}{!}{\includegraphics[width=0.8\linewidth]{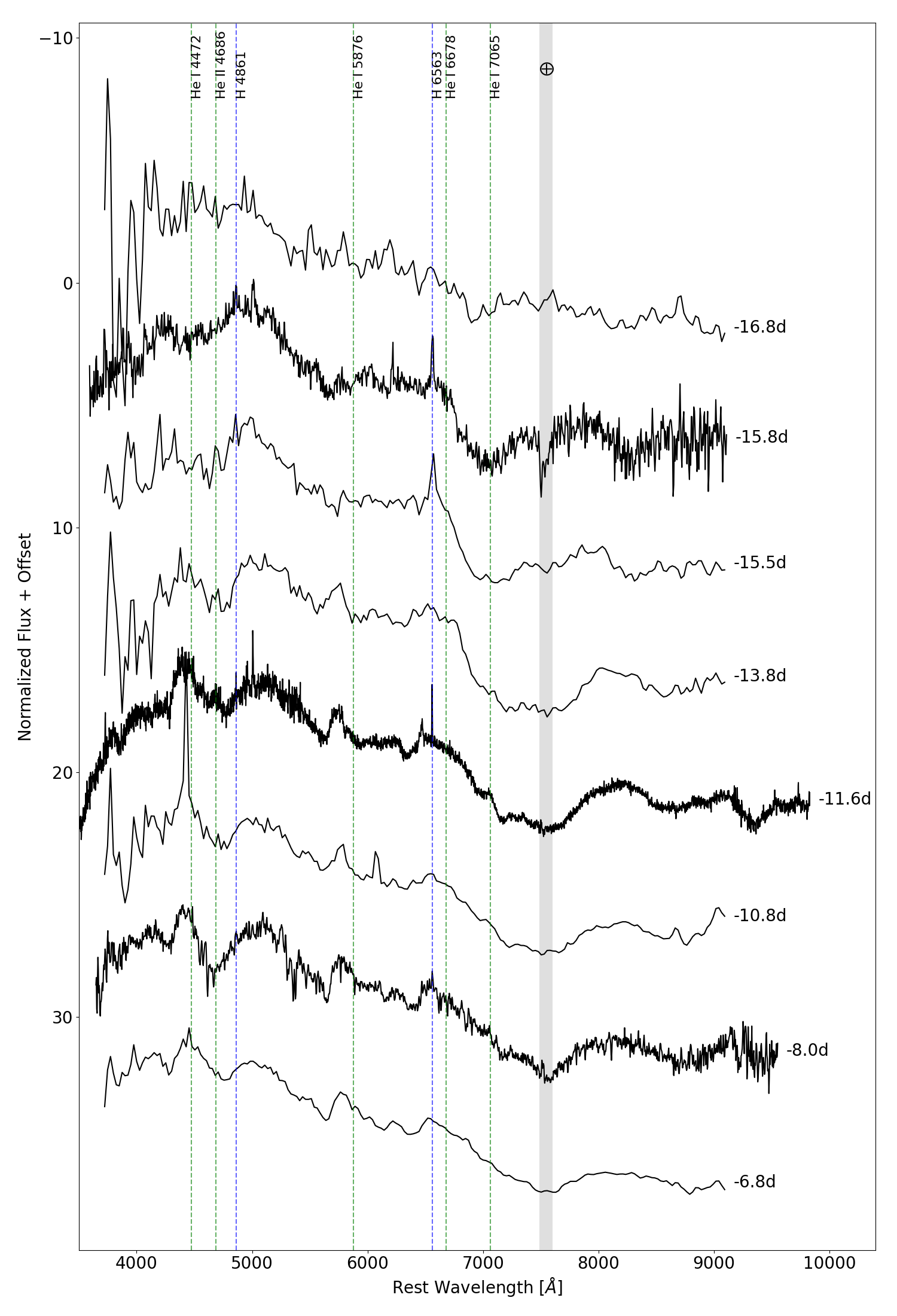}}
  \caption{
    Spectral sequence of obtained spectra from discovery to before peak.
    The observation phase of each spectrum is denoted on the right of each spectrum.
    Rest wavelengths of strong \ion{He}{I} features have been marked in green, and the positions of Balmer lines are denoted in blue.
    A Telluric absorption feature has been denoted with a shaded region and a $\oplus$ symbol.
  }
  \label{fig:specs:seq:breakout}
\end{figure}
\begin{figure}[p]
  \resizebox{\hsize}{!}{\includegraphics[width=0.8\linewidth]{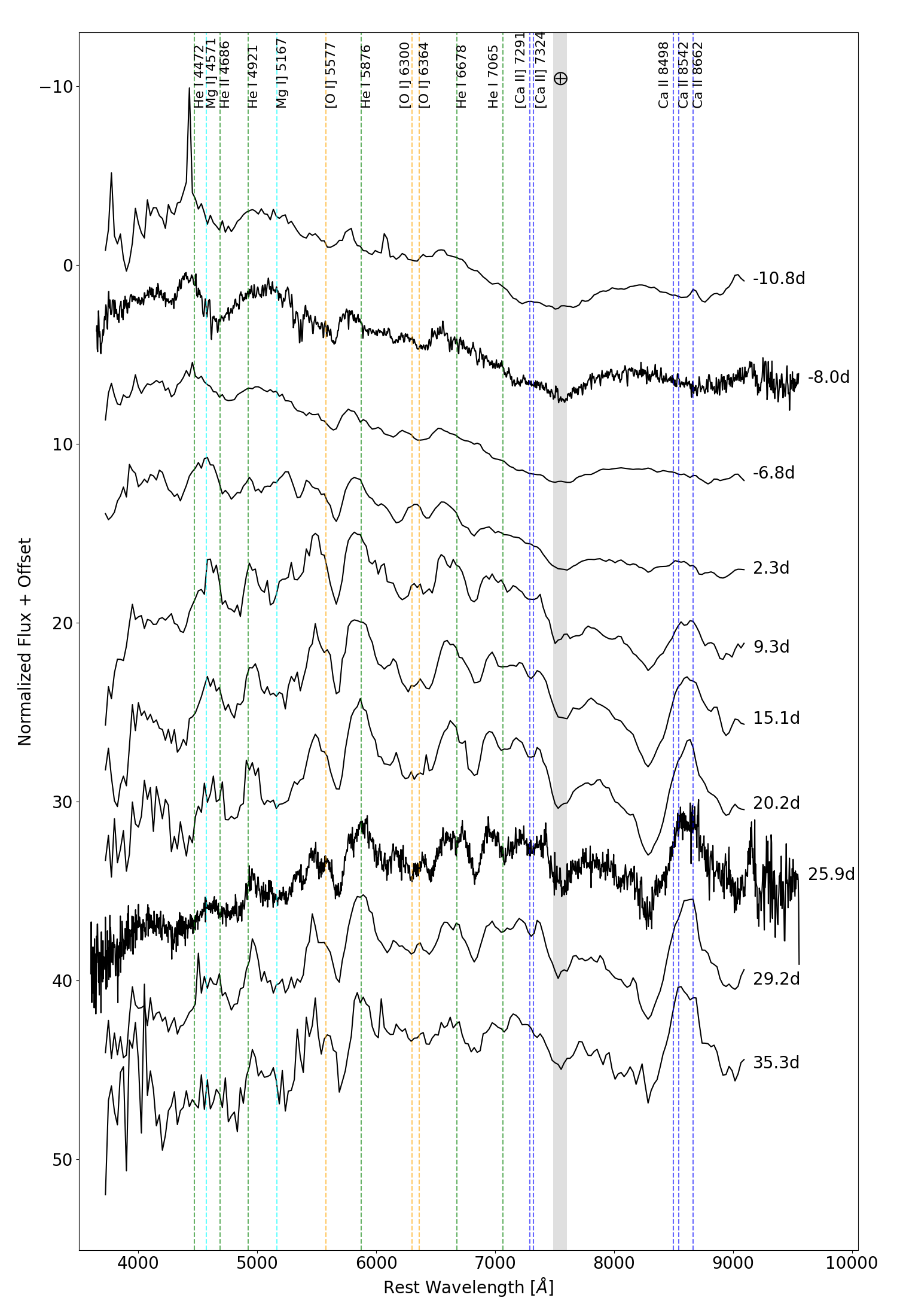}}
  \caption{
    Spectral sequence of obtained spectra around the photospheric phase.
    The notation is the same as in Fig. \ref{fig:specs:seq:breakout}, but here we have also marked the wavelengths of several intermediate-mass elements. 
  }
  \label{fig:specs:seq:photospheric}
\end{figure}

\begin{figure}[p]
  \resizebox{\hsize}{!}{\includegraphics[width=0.8\linewidth]{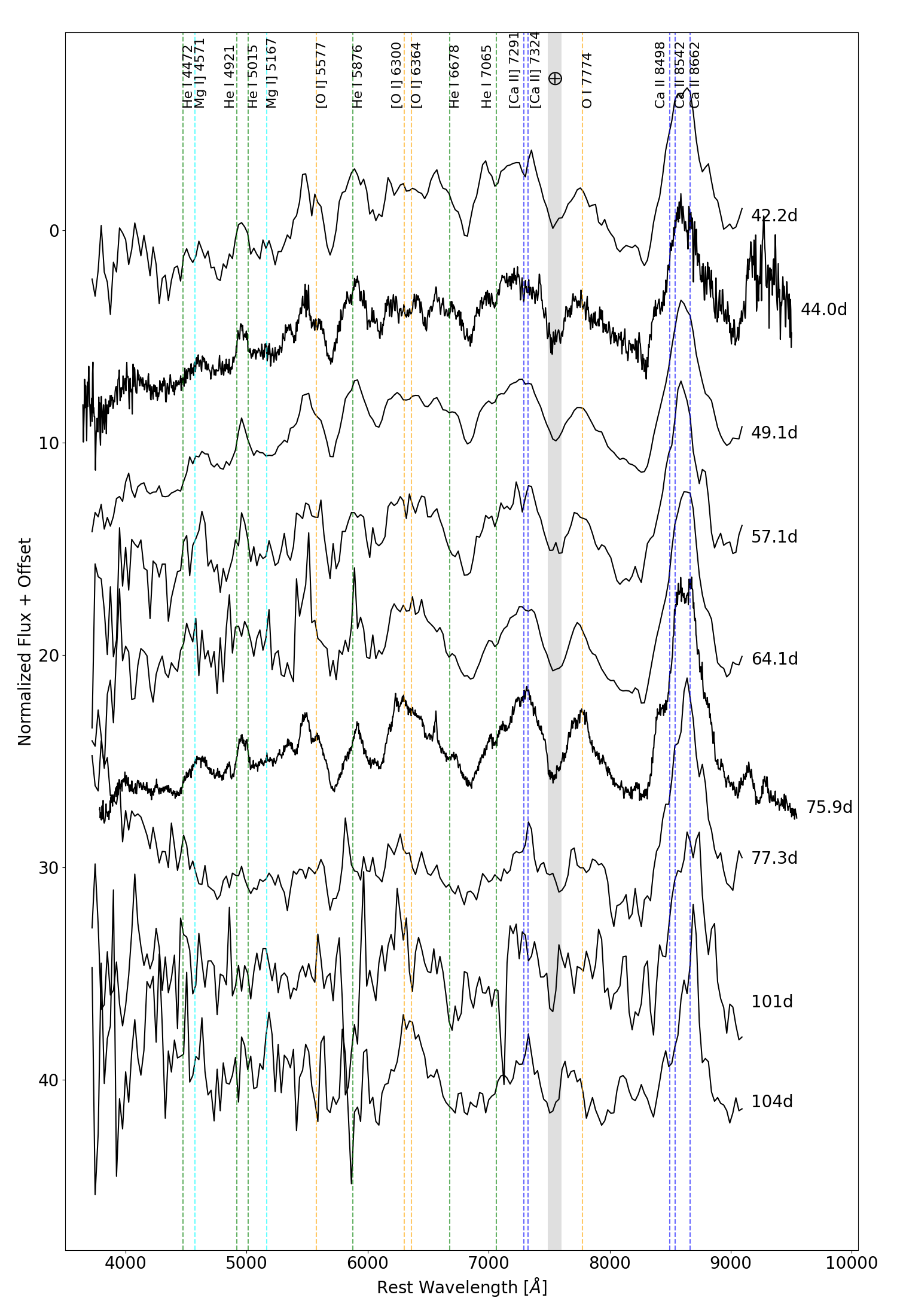}}
  \caption{
    Spectral sequence of obtained spectra in the pre-nebular phase.
    The notation is the same as in Fig. \ref{fig:specs:seq:breakout}, but here we have also marked the wavelengths of several intermediate-mass elements. 
  }
  \label{fig:specs:seq:prenebular}
\end{figure}

\begin{figure}[p]
  \resizebox{\hsize}{!}{\includegraphics[width=0.8\linewidth]{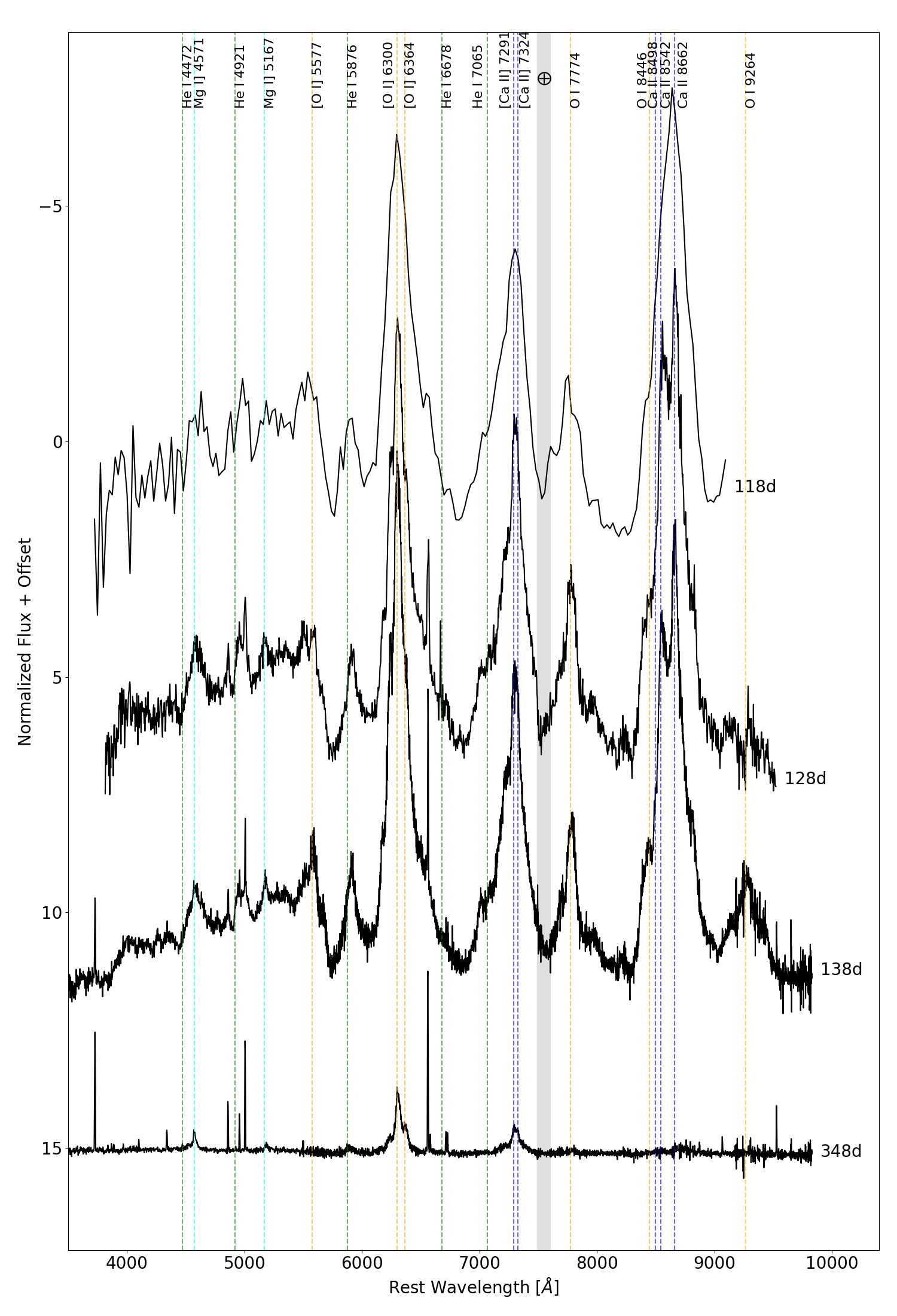}}
  \caption{Spectral sequence of \supernova{2019odp} in the nebular phase. We have marked several of the $\ion{O}{I}$ lines as well as some $\ion{Ca}{II}$ lines.
  }
  \label{fig:specs:seq:nebular}
\end{figure}

The first spectrum was obtained on August 21 2019 -- less than a day after the discovery -- using the SEDM  Integral Field Unit \citep[SEDM IFU;][]{2018PASP..130c5003B}.
The observations were reduced using the \textit{pysedm} package \citep{2019AA...627A.115R}.

We obtained further follow-up spectroscopy using the SEDM and the NOT ALFOSC spectrograph.
In addition, we obtained one pre-peak spectrum using the Double Spectrograph \citep[DBSP;][]{1982PASP...94..586O} mounted on the Palomar 200-inch telescope (P200).
In total, we obtained 8 spectra before the peak, and 30 spectra in total.
This also includes the public NTT classification spectrum from \cite{2019TNSCR1595....1B} under the ePESSTO programme.

We also obtained two high signal-to-noise (S/N) late-time spectra using the Low Resolution Imaging Spectrograph \citep[LRIS;][]{1995PASP..107..375O} mounted on the \textit{Keck 1} telescope.
The LRIS observations were reduced using the fully automated pipeline by \cite{2019PASP..131h4503P}.

All spectra were absolute flux-calibrated using synthetic $r$-band photometry, derived using the \textit{speclite}\footnote{\url{https://github.com/desihub/speclite}} package, against the interpolated light curve dataset (see Sect. \ref{sec:obs:interpolated}).

We show the spectral sequence split into the early phase (Fig. \ref{fig:specs:seq:breakout}), photospheric phase (Fig. \ref{fig:specs:seq:photospheric}), pre-nebular phase (Fig. \ref{fig:specs:seq:prenebular}), and nebular phase (Fig. \ref{fig:specs:seq:nebular}).
The full log of spectroscopic observations can be found in \autoref{tab:obslog:spec}.
The observation epochs are also indicated in the upper part of the light curve figure (Fig. \ref{fig:lc:combined}).
The final reduced and flux-calibrated spectra are available on \texttt{WISeREP}\footnote{\url{https://wiserep.weizmann.ac.il/}} \citep{2012PASP..124..668Y}.

\subsection{Light curve interpolation and parameter estimation}
\label{sec:obs:interpolated}

For \supernova{2019odp} and the comparison sample (see Sect. \ref{sec:comparisonsample}), we performed light curve dataset combination, interpolation, and fitting using the same framework.
For each transient, we performed the following procedure.
First, we pre-processed the individual instrument light curves by transforming all photometry to the AB system and correcting for the extinction with instrument-specific coefficients (see Sect. \ref{sec:extinction} for \supernova{2019odp} and Sect. \ref{sec:comparisonsample} for the values used for the comparison sample).
We used Gaussian process interpolation (see \citealt{gortler2019a} for a review) to produce per-band light curves combining the different photometric datasets from the different instruments.
Simultaneously, we estimated empirical light curve observables, such as the late-time decline rate and peak time, by fitting empirical model functions to the light curves.
This was done by using them as the mean function in the Gaussian process.
These photometric model functions are described in \autoref{appendix:photmodel}.

We used the \textit{dynesty} dynamic nested sampler \citep{2020MNRAS.493.3132S,2004AIPC..735..395S,10.1214/06-BA127,2019S&C....29..891H} to estimate the posterior distribution of the model parameters as well as the amplitude and length-scale parameters of the Gaussian process Matérn-3/2 kernel.
When more than one photometric instrument (with a nominally similar photometric filter system) and overlapping observations were available, we also included an offset parameter in the parameter estimation.
The offsets are stated relative to the photometric instrument with the best coverage (for instance, ZTF in the case of \supernova{2019odp}).
Additional photometric observables, such as the peak magnitude, $M_\text{max}$, the light curve rise parameter, $\Delta m_{-10}$, or the light curve decline parameter, $\Delta m_{15}$, were estimated by Monte Carlo sampling the derived Gaussian process function.

\FloatBarrier


\section{Analysis}
\label{sec:evo}

\begin{table*}[h]
  \caption{Adopted parameters for the comparison objects.}
  \centering
      {
        \renewcommand{\arraystretch}{1.2}
        \begin{tabular}{|c|c|c|c|c|c|c|c|}
          \hline
          Supernova & Type & $t_\text{expl}$ & $t_\text{peak}$ & Distance & $E(B-V)_\text{MW}$ & $E(B-V)_\text{Host}$ & Phot. Velocity $v_\text{ph}$ \\
          & & (d) & (d) & (Mpc) & (mag) & (mag) & (1000\,\kms) \\
          \hline
          \hline
          \supernova{1998bw} & Ic-BL & $50928.909$ (8) & $50945 \pm 3$ & $40.84 \pm 2.86$ (0) & $0.047 - 0.06$ (1) & 0 (5) & $19.5_{-1}^{+1.7}$ (3) \\
          \supernova{2002ap} & Ic-BL & $52300 \pm 0.5$ (9) & $52313 \pm 3$ & $10.69 \pm 0.75$ (0) & $0.0585 - 0.0661$ (1) & $0.01 - 0.02$ (6) & $13_{-1}^{+2}$ (3) \\
          \supernova{2008D} & Ib-pec & $54474.564$ (7) & $54493 \pm 3$ & $33.69 \pm 2.36$ (0) & $0.0193 \pm 0.0002$ (1) & $0.4 - 0.8$ (4) & $9.5_{-1}^{+2.1}$ (3) \\
          iPTF13bvn         & Ib     & $56458.7 \pm 0.1$ (2) & $56477 \pm 3$ & $26.8 \pm 2.6$ (2) & $0.0421 - 0.0448$ (1) & $0.04 - 0.15$ (2) & $8 \pm 1$ (3) \\
          \hline
        \end{tabular}
      }
  
      \tablefoot{The photospheric velocity is the estimated velocity at the light curve peak. When a range is specified, a uniform prior is used; in the other case, a (asymmetric) Gaussian is used as a prior. The source of each parameter has been denoted in parentheses.}
      \tablebib{(0) NED; (1) IRSA DUST Service using \cite{2011ApJ...737..103S} map; (2) \cite{2016A&A...593A..68F}; (3) \cite{2016MNRAS.457..328L}; (4) \cite{2008Natur.453..469S}; (5) \cite{2011AJ....141..163C}; (6) \cite{2002PASJ...54..899T}; (7) \cite{2009ApJ...702..226M}; (8) \cite{1998IAUC.6884....1S}; (9) \cite{2002ApJ...572L..61M}.
  }
  \label{tab:ana:comparison:params}
\end{table*}

\subsection{Extinction}
\label{sec:extinction}

Based on the dust maps by \cite{2011ApJ...737..103S}, we can estimate the Milky Way extinction, $E(B-V)$, at the position of the transient to be in the range from 0.14 to 0.20 mag.
For analysis requiring extinction-corrected values, we propagated the uncertainty using Monte Carlo methods.
We assume no host extinction based on the lack of any visible sodium absorption features.
In addition, we have compared the colours against the intrinsic colour templates at +10 d from \cite{2018A&A...609A.135S} and notice that our supernova has bluer $g-r$ and $g-i$ colours than any supernova class in that study. 
Comparing our light curve against the light curves of the sample supernovae  that the Stritzinger colour templates are based on, we notice that \supernova{2019odp} is %
bluer at virtually any time.
This further strengthens our assumption of no host galaxy extinction for this event. 

We used the computed $A_x/E(B-V)$ values for the different filters from \cite{2011ApJ...737..103S} for $R_V = 3.1$ for all photometric extinction corrections.
We used the \textit{extinction} \citep{barbary_kyle_2016_804967} Python implementation of the \cite{1989ApJ...345..245C} extinction law to extinction-correct all spectra.

\subsection{Comparison datasets}
\label{sec:comparisonsample}

We have compared the properties of \supernova{2019odp} against a selected sample of well-observed objects from the literature that are prototypes for the different supernova classes: \supernova{1998bw} (Ic-BL) and \supernova{iPTF13bvn} (Ib).
We also included \supernova{2002ap} (Ic-BL), since it has the same peak brightness and might be a more suitable comparison object for \IcBL supernovae than the much brighter \supernova{1998bw}.
We also included \supernova{2008D} (Ib), since it was a very close match spectroscopically and showed a somewhat similar unusual behaviour right after discovery.
The adopted supernova parameters for the comparison objects are presented in \autoref{tab:ana:comparison:params}.

For \supernova{1998bw}, we used the $UBVRI$ light curves compiled by \cite{2011AJ....141..163C}.
For iPTF13bvn, we used the $gri$ light curve by \cite{2016A&A...593A..68F} with additional $U$-band UVOT photometry points from \cite{2014Ap&SS.354...89B}.
For \supernova{2008D}, we used the $BV g^\prime r^\prime$ light curves from \cite{2014ApJS..213...19B} and the UVOT $U$-band light curve from \cite{2014Ap&SS.354...89B}.
For \supernova{2002ap}, we used the $UBVRI$ light curves compiled by \cite{2003PASP..115.1220F}.
Where necessary, we converted the magnitudes from Vega to AB magnitudes and we used the same interpolation procedure as is described in Sect. \ref{sec:obs:interpolated}.

\subsection{Photometric evolution}
\label{sec:evo:phot}

\begin{figure*}
  \includegraphics[width=\linewidth]{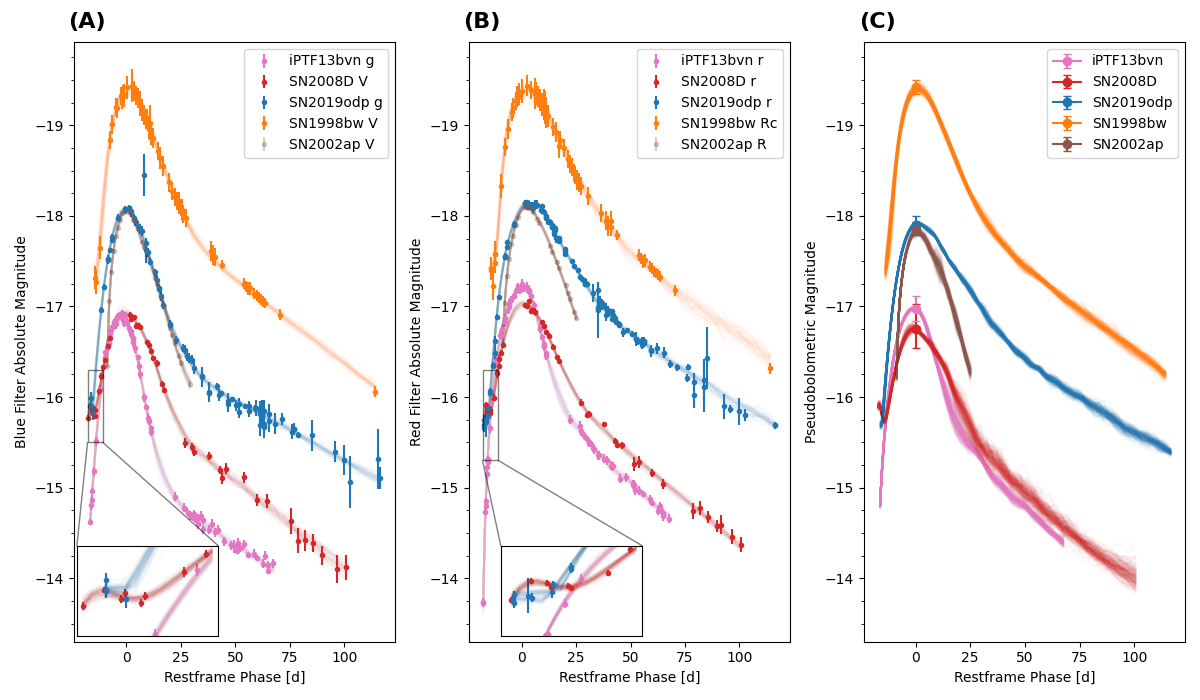}
  \caption{
    Comparison of the evolution of the absolute light curves in $g/V$ (Panel A, left) and $r/R$ (B, middle) bands between \supernova{2019odp} and selected comparison objects.
    In addition, the pseudo-bolometric light curve derived using the Lyman method is shown (C, right).
    In the pseudo-bolometric light curve (C), the error bars around the peak represent the combined error from the distance uncertainty, extinction uncertainty, and the scatter from the Lyman relation.
    The interpolated light curves for each supernova have been placed over the observed datapoints in the same colour. These
    were used to derive the pseudo-bolometric light curves in the right panel. The bottom left corners of panels A and B zoom in on the very early evolution.
  }
  \label{fig:ana:evo:lc}
\end{figure*}

In this section, we compare the photometric evolution of \supernova{2019odp} with our previously introduced comparison supernovae.
A light curve comparison is shown in Fig. \ref{fig:ana:evo:lc}.
We estimated the light curve parameters using the method described in Sect. \ref{sec:obs:interpolated} that uses the empirical photometric models described in \autoref{appendix:photmodel}.
The resulting parameter estimates for the light curve observables, which we discuss below, are summarised in \autoref{tab:ana:observables}.

Zooming in on the time period around the discovery (Fig. \ref{fig:lc:early}) shows that the initial light curve evolution for \supernova{2019odp} is more consistent with a plateau than with an exponential rise.
This behaviour is seen in both the $g$ and $r$ bands ($i$-band data has too low cadence).
Comparing the absolute $r$- and $g$-band light curves between \supernova{2019odp} and \supernova{2008D} (see inset in Fig. \ref{fig:ana:evo:lc}), which is another supernova that showed signs of an early plateau or a shock cooling peak, the bump appears at roughly the same relative phase and is marginally fainter in \supernova{2019odp} than in \supernova{2008D} (when taking distance uncertainty into account, the difference is not statistically significant).
This in stark contrast to iPTF13bvn, which shows no early feature down to a limit several magnitudes deeper.
We estimate the absolute plateau magnitude for \supernova{2019odp} to be $-16.27_{-0.36}^{+0.34}\,$mag in the $g$ band and $-16.16_{-0.33}^{+0.32}\,$mag in the $r$ band.
For \supernova{2019odp}, this is about $1.9\,$mag fainter than the peak magnitude, while for \supernova{2008D} the difference is only about $1\,$mag, and for iPTF13bvn the difference is more than $2\,$mag in the $g$ band and more than $3\,$mag in $r$ band.

Following the plateau, \supernova{2019odp} rises to the main peak in $14-15$ days in the $g$ band, $17-21$ days in the $r$ band, and $17-21$ days in the $i$ band (measured as time between the first data point 
$\geq 3 \sigma$
above the plateau level and peak).
The light curve rise timescale, $\tau_\text{rise}$ (see \autoref{appendix:photmodel} for the definition), is also much longer in all bands for \supernova{2019odp} than in any of the comparison objects, except perhaps \supernova{2008D}.
The light curve for \supernova{2019odp} peaks at an absolute magnitude of $\sim -18$\,mag in the $gri$ bands (see \autoref{tab:ana:observables} for the exact values for each band).
This is towards the brighter end of the luminosity distribution previously established for \Ib{supernovae} \citep{2015A&A...574A..60T}, but
within the bulk of the distribution for \IcBL{supernovae} \citep{2019A&A...621A..71T}.
In the $r$ band, the supernova shows a rather flat peak at an apparent magnitude of $15.9$ that is at least 7 days in duration.
This is not clearly seen %
in the $g$ or the $i$ bands (limited by the sampling period at peak).
The main peak is several days wider in all bands compared to the light curves of iPTF13bvn and \supernova{2008D} and more closely resembles the width of \supernova{1998bw}. The light curve width can be an indicator of a higher ejecta mass \citep[e.g.,][]{2023A&A...678A..87K}, as is discussed in more detail in Sect. \ref{sec:mod:phot:results}.
This difference is most pronounced in the $r$ and $i$ bands.
The peak width, $\sigma$, of the fitted Gaussian is given in \autoref{tab:ana:observables}.

The late-time linear-slope decline parameter, $\beta$ (see \autoref{appendix:photmodel} for the definition), in all bands is significantly slower for \supernova{2019odp} than for the comparison objects.
The measured linear slopes are provided in \autoref{tab:ana:observables}.
This can also 
be an indicator of a large ejecta mass. %

{
  \renewcommand{\arraystretch}{1.2}
  \begin{table*}
  \caption{Basic light curve observables extracted for SN 2019odp and the comparison objects.}
  \centering
  \begin{tabular}{|c|c|c|c|c|c|c|c|}
    \hline
    Supernova & Band & $\Delta m_{-10}$ & $\Delta m_{15}$ & Linear Slope & $M_\text{max}$ & Peak Width & Rise Timescale \\
    & & (mag) & (mag) & (mmag/d) & (mag) & (d) & (d) \\
    \hline
    \hline

    \supernova{2019odp} & g & $0.81_{ -0.01 }^{ +0.01 }$ & $0.85_{ -0.01 }^{ +0.01 }$ & $13.5_{ -0.3 }^{ +0.3 }$ & $-18.08_{ -0.10 }^{ +0.08 }$  & $16.1_{ -0.8 }^{ +0.6 }$ & $13.1_{ -2.6 }^{ +1.9 }$ \\
    
    \supernova{1998bw} & V & $0.84_{ -0.06 }^{ +0.08 }$ & $0.72_{ -0.03 }^{ +0.03 }$ & $18.6_{ -0.1 }^{ +0.1 }$ & $-19.37_{ -0.08 }^{ +0.08 }$  & $16.6_{ -0.5 }^{ +0.4 }$ & $7.1_{ -1.9 }^{ +1.7 }$ \\
    
    \supernova{2002ap} & V & < 3.5 & $0.90_{ -0.03 }^{ +0.03 }$ & $19.2_{ -0.2 }^{ +0.2 }$ & $-18.05_{ -0.06 }^{ +0.07 }$  & $15.5_{ -0.8 }^{ +0.6 }$ & $1.7_{ -0.2 }^{ +0.3 }$ \\
    
    \supernova{2008D} & V & $0.53_{ -0.02 }^{ +0.02 }$ & $0.67_{ -0.02 }^{ +0.02 }$ & $19.1_{ -0.7 }^{ +0.6 }$ & $-16.93_{ -0.13 }^{ +0.17 }$  & $14.1_{ -0.7 }^{ +0.7 }$ & < 16.4 \\
    
    iPTF13bvn & g & $1.15_{ -0.03 }^{ +0.03 }$ & $1.34_{ -0.02 }^{ +0.02 }$ & $19.7_{ -0.2 }^{ +0.2 }$ & $-16.89_{ -0.11 }^{ +0.11 }$  & $11.0_{ -0.2 }^{ +0.2 }$ & $4.7_{ -0.7 }^{ +1.0 }$ \\
    
    \hline

    \supernova{2019odp} & r & $0.42_{ -0.02 }^{ +0.02 }$ & $0.44_{ -0.01 }^{ +0.01 }$ & $14.1_{ -0.1 }^{ +0.1 }$ & $-18.13_{ -0.09 }^{ +0.10 }$  & $21.2_{ -0.6 }^{ +0.4 }$ & $10.0_{ -0.6 }^{ +0.7 }$ \\
    
    \supernova{1998bw} & Rc & $0.72_{ -0.07 }^{ +0.07 }$ & $0.49_{ -0.03 }^{ +0.03 }$ & $16.3_{ -0.1 }^{ +0.1 }$ & $-19.37_{ -0.07 }^{ +0.08 }$  & $22.6_{ -1.0 }^{ +1.0 }$ & $5.9_{ -1.3 }^{ +1.4 }$ \\
    
    \supernova{2002ap} & R & $0.96_{ -0.06 }^{ +0.05 }$ & $0.70_{ -0.01 }^{ +0.01 }$ & $15.7_{ -0.1 }^{ +0.2 }$ & $-18.10_{ -0.07 }^{ +0.08 }$  & $16.9_{ -0.8 }^{ +0.7 }$ & $2.8_{ -0.4 }^{ +1.4 }$ \\
    
    \supernova{2008D} & r & $0.50_{ -0.02 }^{ +0.02 }$ & $0.56_{ -0.01 }^{ +0.01 }$ & $19.8_{ -0.4 }^{ +0.4 }$ & $-17.02_{ -0.14 }^{ +0.15 }$  & $15.4_{ -0.8 }^{ +0.7 }$ & $9.3_{ -1.5 }^{ +1.9 }$ \\
    
    iPTF13bvn & r & $0.69_{ -0.03 }^{ +0.03 }$ & $0.98_{ -0.03 }^{ +0.03 }$ & $18.7_{ -0.2 }^{ +0.2 }$ & $-17.24_{ -0.11 }^{ +0.12 }$  & $11.9_{ -0.6 }^{ +0.7 }$ & $6.5_{ -1.9 }^{ +1.9 }$ \\
    
    \hline

    \supernova{2019odp} & i & $0.37_{ -0.02 }^{ +0.02 }$ & $0.33_{ -0.02 }^{ +0.02 }$ & $13.8_{ -0.3 }^{ +0.2 }$ & $-18.00_{ -0.09 }^{ +0.08 }$  & $26.6_{ -1.2 }^{ +1.1 }$ & $7.7_{ -0.3 }^{ +0.4 }$ \\
    
    \supernova{1998bw} & Ic & $0.41_{ -0.06 }^{ +0.07 }$ & $0.42_{ -0.05 }^{ +0.05 }$ & $16.9_{ -0.1 }^{ +0.2 }$ & $-19.34_{ -0.08 }^{ +0.08 }$  & $21.2_{ -1.7 }^{ +2.0 }$ & $6.1_{ -1.6 }^{ +1.8 }$ \\
    
    \supernova{2002ap} & I & $0.58_{ -0.05 }^{ +0.05 }$ & $0.45_{ -0.02 }^{ +0.02 }$ & $18.5_{ -0.1 }^{ +0.2 }$ & $-17.96_{ -0.07 }^{ +0.07 }$  & $19.2_{ -1.1 }^{ +1.0 }$ & $3.4_{ -0.6 }^{ +1.3 }$ \\
    
    \supernova{2008D} & i & $0.43_{ -0.02 }^{ +0.02 }$ & $0.41_{ -0.01 }^{ +0.01 }$ & $20.6_{ -0.4 }^{ +0.4 }$ & $-17.00_{ -0.13 }^{ +0.11 }$  & $17.6_{ -1.1 }^{ +0.9 }$ & $7.5_{ -0.5 }^{ +0.8 }$ \\
    
    iPTF13bvn & i & $0.58_{ -0.03 }^{ +0.03 }$ & $0.79_{ -0.03 }^{ +0.03 }$ & $16.7_{ -0.4 }^{ +0.4 }$ & $-17.10_{ -0.10 }^{ +0.12 }$  & $13.0_{ -0.9 }^{ +1.0 }$ & $3.6_{ -0.4 }^{ +0.4 }$ \\
    
    \hline
    
  \end{tabular}
  \tablefoot{$\Delta m_{-10}$ denotes the magnitude difference from the peak to 10 days before the peak. $\Delta m_{15}$ denotes the magnitude difference from the peak to 15 days past the peak. The linear slope is the late-time decline slope. $M_\text{max}$ denotes the peak absolute magnitude including extinction and distance uncertainties. The peak width denotes the width of the Gaussian around the peak. The rise timescale denotes the timescale factor from the Contardo model (see \autoref{appendix:photmodel}).}
  \label{tab:ana:observables}
\end{table*}

}

\subsection{Colour evolution}
\label{sec:evo:color}

\begin{figure}
  \includegraphics[width=\linewidth]{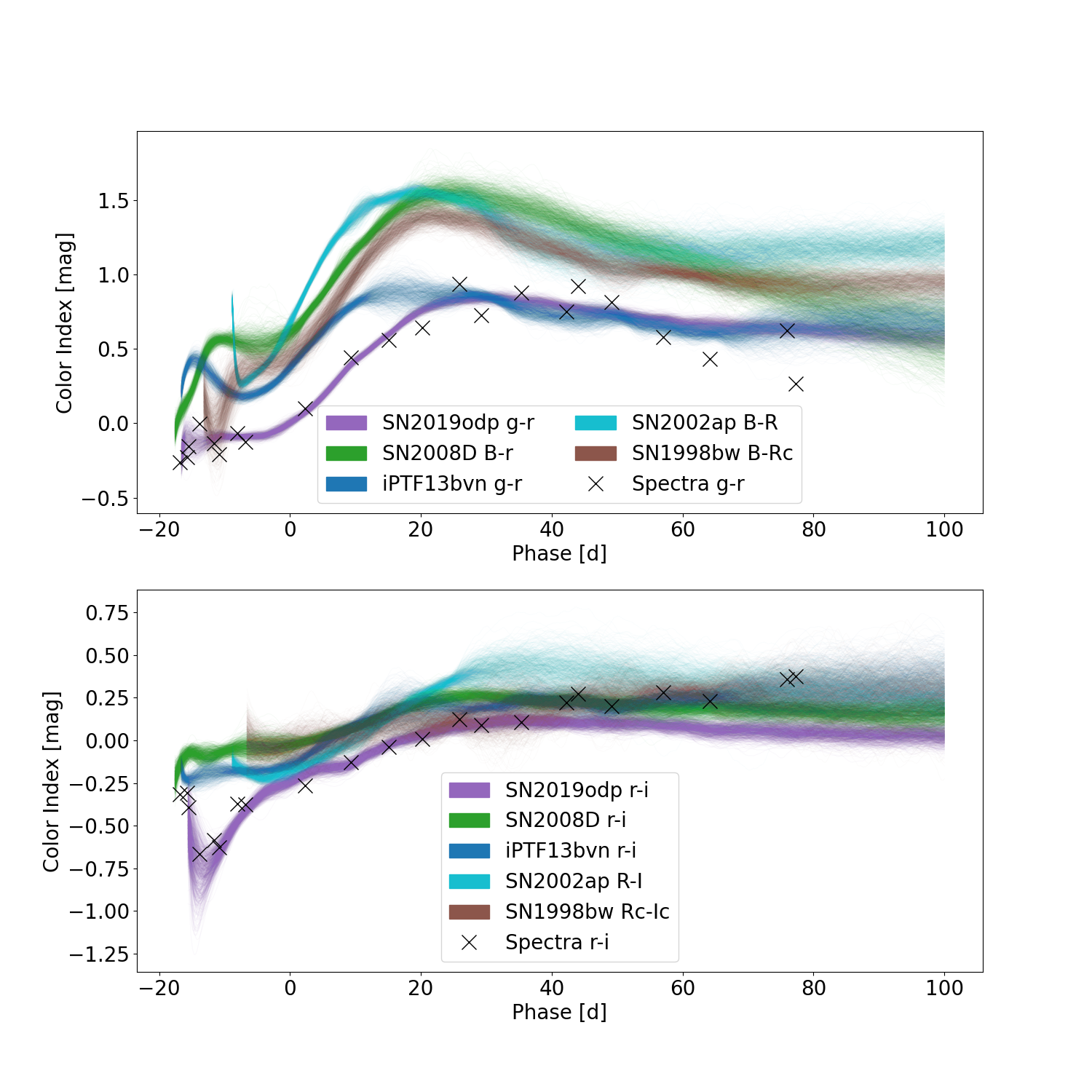}
  \caption{Colour evolution of \supernova{2019odp} and the comparison objects. The top panel shows the $g-r$ colour (or corresponding Johnson colour). The lower panel shows the $r-i$ colour (or corresponding Johnson colour). For each transient, we sampled 1000 realisations of the supernova light curve from the interpolation kernel. For \supernova{2019odp}, we also included synthetic photometry based on the observed spectra as a consistency check (using the same filter curve as for the photometric dataset). They are denoted with black crosses.}
  \label{fig:lc:color:evo}
\end{figure}

During the very early $\gtrsim$ 2-day
plateau-phase, both the $g-r$ colours (shown in the upper panel of Fig. \ref{fig:lc:color:evo}) and the $r-i$ colours (shown in Fig. \ref{fig:lc:color:evo}, lower panel) are bluer than for any of the comparison objects.
The $g-r$ colour stays constant during this phase, showing no bump, unlike all of the comparison objects.
Afterwards, the $g-r$ colour of \supernova{2019odp} evolves to a redder colour, but this evolution happens later and is slower than in the comparison objects.
It gets redder until it meets the colour light curve of iPTF13bvn at around 25 days post-peak and afterwards follows the same evolution.
The $r-i$ colour shows an early time valley around 10 days before peak, which is unseen or much weaker in the comparison objects before becoming redder again and joining the %
evolution of the other supernovae (which mostly stay constant after 40 days post-peak).

The initial $U-r$ colour evolution of \supernova{2019odp} matches that of iPTF13bvn.
However, at around $-5$ days for iPTF13bvn, the $U-r$ colour quickly starts to get redder, while \supernova{2019odp} only starts getting redder after the peak.
Due to the limited $U$-band observations, the exact inflection point is not known.
One interesting observation is that while \supernova{2019odp} has bluer colours in all other colour indices, this is not true in the $U-r$ colour, where \supernova{2008D} starts almost 1 magnitude bluer and monotonically gets redder.

\subsection{Blackbody evolution}
\label{sec:evo:bb}

\begin{figure}
  \includegraphics[width=\linewidth]{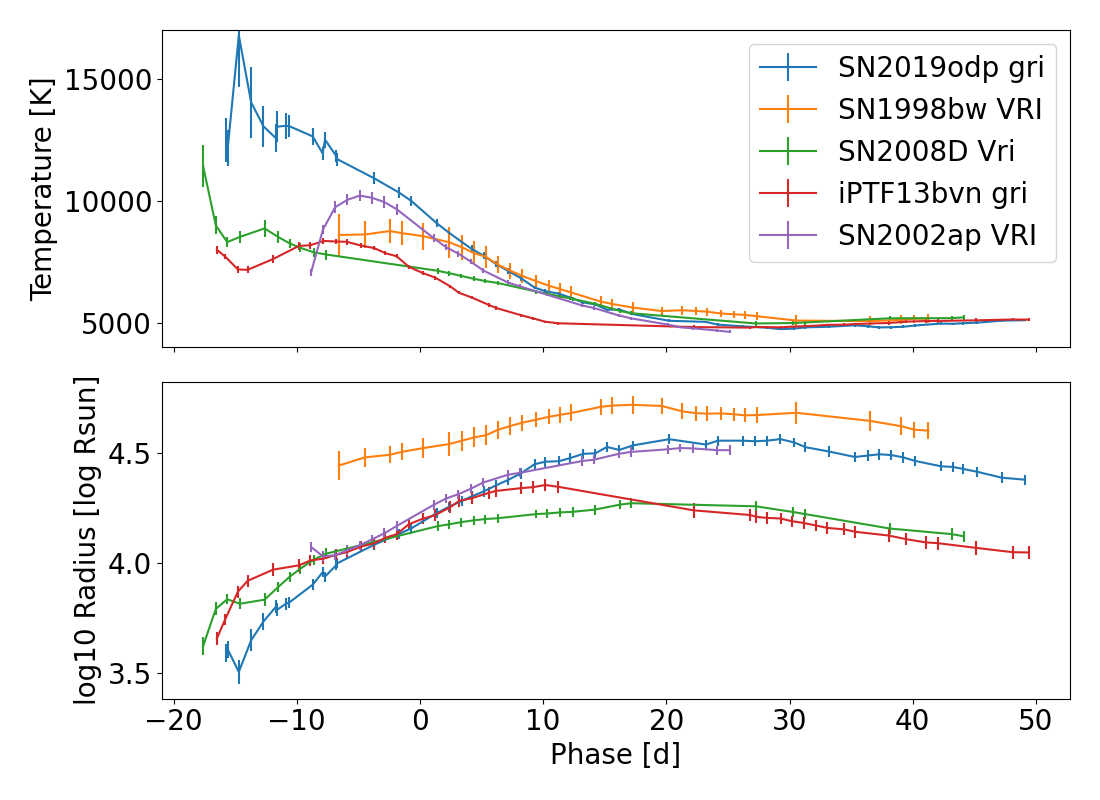}
  \caption{
    Time evolution of the inferred blackbody parameters from the photometry for \supernova{2019odp} and selected comparison objects.
    We chose the closest matching filter sets for all transients and performed our own fitting for all objects.
    The error bars contain all uncertainties including the distance and extinction uncertainties.
    The upper panel shows the temperature time evolution and the lower panel the photospheric radius time evolution.
  }
  \label{fig:evo:bb:comparison}
\end{figure}

We can estimate blackbody parameters using the interpolated photometry datasets for all supernovae in the comparison sample (Sect. \ref{sec:obs:interpolated}).
We used the $gri$ bands when they were available and the closest matches in wavelength if they were not.
The detailed description %
of the method can be found in \autoref{appendix:bbfit}.
Whereas 
the emission from these supernovae is not really a blackbody (see 
the spectral sequences in Fig. \ref{fig:specs:seq:breakout} and Fig. \ref{fig:specs:seq:photospheric}), the derived
blackbody parameters are suitable mainly for a rough qualitative comparison between the objects.
The time evolution of the photospheric temperature and radius is shown in Fig. \ref{fig:evo:bb:comparison}.

Initially, \supernova{2019odp} is both hotter and the initial photospheric radius is smaller than for any of the comparison objects.
This is most likely an underestimate of the true photospheric temperature (and at the same time a slight overestimate of the radius), since the blue part of the spectral energy distribution (SED) seems to be suppressed (only considering the $ri$ bands yields a closer match to the full $grizJH$ photometry modelling for the time-period post-peak when it is available).
Roughly 20 days after the peak, the temperature evolution for SN 2019odp joins that of most other supernovae in the comparison sample.
\supernova{2019odp} shows a larger peak photospheric radius compared to all comparison transients except \supernova{1998bw}.

\subsection{Pseudo-bolometric light curve}
\label{sec:evo:qbol}

Using the photometric datasets (Sect. \ref{sec:obs:interpolated}) for \supernova{2019odp} and the comparison sample, we computed pseudo-bolometric light curves with the \cite{2014MNRAS.437.3848L} method using the $g$ and $r$ bands (or closely corresponding bands for some of the comparison transients).
Based on a sample of well-observed stripped envelope supernovae, \cite{2014MNRAS.437.3848L} has calculated bolometric correction factors for multiple pairs of photometric bands.
Using the provided polynomial fits to the sample, we can apply the derived corrections to our supernova:
\begin{equation}
  M_\text{pbol,g-r} + D_m = m_g + BC_{g-r} = m_g + \sum_{i=0}^2 c_i (m_g-m_r)^i,
\end{equation}
where $M$ is the absolute magnitude, $m$ is the extinction-corrected apparent magnitude, $D_m$ is the distance modulus in magnitudes,
$BC$ is the bolometric correction, and $c_i$ are the polynomial fit parameters from \cite{2014MNRAS.437.3848L} for the $g-r$ band pair.
Pseudo-bolometric refers to the neglect of any UV contribution, which we deem to be more uncertain.
This means that the derived pseudo-bolometric luminosities constitute only a lower limit.

The derived pseudo-bolometric light curves are shown in the right panel of Fig. \ref{fig:ana:evo:lc}.
We first estimated the peak epoch of the pseudo-bolometric light curve by sampling the interpolated light curve (excluding any uncertainties that affect the light curve globally) in a 10-day window around the $g$-band peak epoch and then selecting the time of the brightest point for each sampled light curve.
We estimated the peak magnitude by sampling the interpolated light curve at the peak epoch (including global uncertainties), which yields
\begin{equation}
  M_\text{pbol,peak} = -17.9 \pm 0.2 \,\text{mag}.
\end{equation}

We converted the absolute pseudo-bolometric magnitudes to luminosities using
the bolometric luminosity of the Sun,
$L_{bol,\odot} = 3.828 \times 10^{33}$ erg s$^{-1}$, and $M_{bol,\odot} = 4.74$ mag, to get
\begin{equation}
  L_\text{pbol,peak} = 4.4_{-0.7}^{+1} \times 10^{42} \text{erg}\,\text{s}^{-1}.
\end{equation}

This is well in line with peak pseudo-bolometric luminosities seen previously for \Ib{supernovae} \citep{2016MNRAS.457..328L} or \IcBL{supernovae} \citep{2019A&A...621A..71T}.
As one would expect based on the photometric comparison (see Sect. \ref{sec:evo:phot}), the pseudo-bolometric main peak is also wider than for most of the comparison objects (see right panel in Fig. \ref{fig:ana:evo:lc}).
The total integrated radiated luminosity (radiated energy) is
\begin{equation}
  E_\text{radiated,pbol} = 2.1_{-0.4}^{+0.4} \times 10^{49} \text{erg}.
\end{equation}

We also measured the duration of the early plateau to be in the range of 2 to 5 days.
Using direct integration of the photometric measurements in the ZTF $gri$ bands, we estimated the plateau pseudo-bolometric luminosity to be
\begin{equation}
  L_\text{plateau,pbol} = 2.1_{-0.4}^{+0.4} \times 10^{41} \text{erg}\,\text{s}^{-1},
  \label{eqn:evo:phot:lums:plat}
\end{equation}
which corresponds to $M_\text{plateau,pbol} \simeq -14.6$\,mag.

\newpage

\subsection{Spectroscopic evolution}
\label{sec:evo:spec}

\begin{figure*}[t!]
  \includegraphics[width=\linewidth]{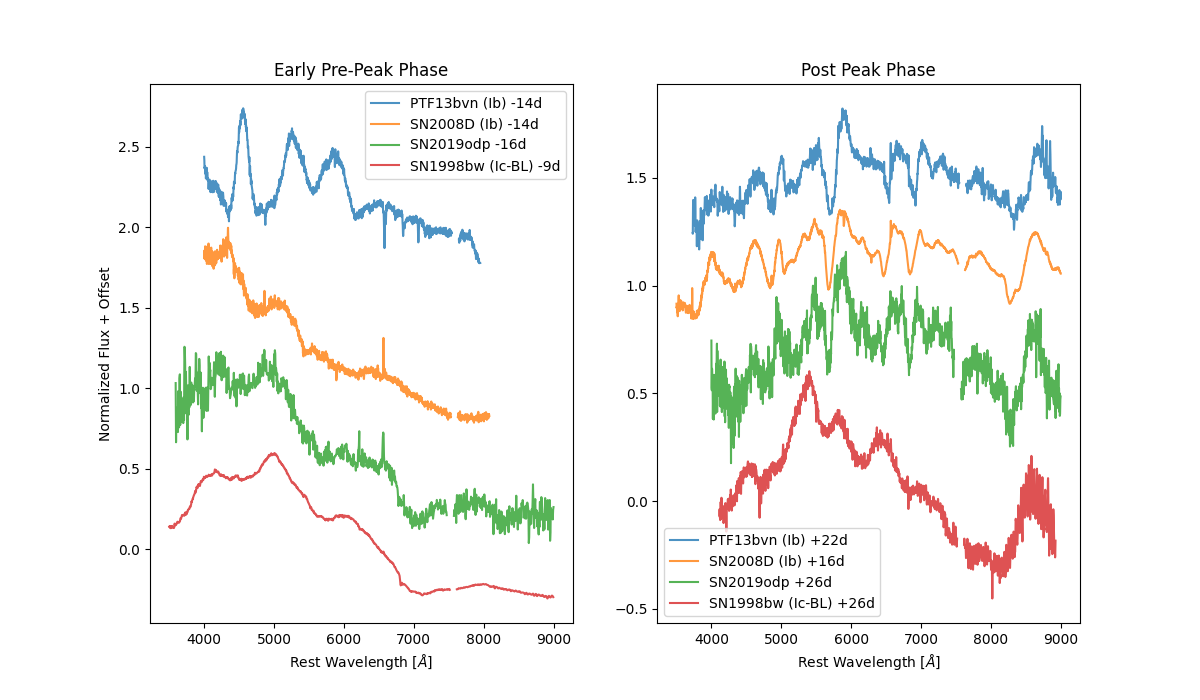}
  \caption{
    Comparison of  spectra of \supernova{2019odp} before peak (left panel) and after peak (right panel) against selected comparison objects.
    The phases with respect to the respective $g/V$-band peak epochs are denoted in the legend.
    Telluric absorptions have been marked and masked, and the spectra have been corrected for host extinction.
    At the early phase, the spectrum of \supernova{2019odp} (the NTT classification spectrum) looks very much like the \IcBL{\supernova{1998bw}} \citep[shown in red,][]{2001ApJ...555..900P}, and not at all like the typical Type Ib iPTF13bvn \citep[shown in blue,][]{2016A&A...593A..68F}.
    This is the reason for the initial classification on TNS as a \IcBL{supernova}.
    However it looks quite similar to the Type Ib \supernova{2008D} \citep[shown in orange,][]{2019MNRAS.482.1545S,2014AJ....147...99M}, which has earlier observations than iPTF13bvn.
    At later phases (right panel), \supernova{2019odp} is instead similar to typical Type Ib supernovae such as iPTF13bvn, showing narrower lines and clear helium features, as opposed to SN 1998bw. In fact, the most similar object might be the Type Ib \supernova{2008D}.
    Some of the spectra suffer from host contamination.
  }
  \label{fig:evo:spec:classification}
\end{figure*}

We present the early spectral evolution between discovery and peak in Fig. \ref{fig:specs:seq:breakout}.
The first spectrum, taken 16.8 days before peak, is mostly blue and featureless with some suppression of the blue side of the spectrum.
The second spectrum, taken one day later, is dominated by a staircase-shaped continuum that has flat regions in the ranges of $5570-6670~\AA$ and $7050-9000~\AA$ with some narrow emission lines that we identify as host galaxy contamination.
While the first spectrum of the sequence might be explained by a single (partially absorbed) blackbody, the later spectra clearly show signs of %
emission or absorption lines.

The first identifiable %
line features are seen in the spectrum at a phase of $-12$ days.
We can identify two P-Cygni features that we associate with $\ion{He}{I}~\lambda\lambda5876,6678$ and possibly 
 $\ion{He}{I}~\lambda4471$ 
at roughly 14000 $\kms$ and with blueshifted emission peaks. %
We investigate the velocity evolution of the helium feature in more detail in Sect. \ref{sec:ana:evo:velocity}.

Post-peak, the lines gain in strength, the helium P-Cygni features become much more pronounced, and additional lines such as the various forbidden calcium lines become visible.
The spectral evolution in the photospheric phase is shown in Fig. \ref{fig:specs:seq:photospheric}.
A feature that is possibly associated with $\ion{Mg}{I}]~\lambda4571$ becomes more obvious.
At around $5000~\AA$, a feature becomes visible, which could be [$\ion{O}{III}]~\lambda\lambda4959,5007$.
This line was first identified by \cite{2016ApJ...831..144L} in super-luminous supernovae (SLSNe) and is seen in SLSN model spectra by \cite{2017ApJ...835...13J}.
We can identify similar features in the comparison between \Ib{supernovae} SN 2008D and iPTF13bvn, which suggests that this may not be an uncommon feature for \Ib{supernovae}.
However, both helium and iron have lines quite close in wavelength to those features as well.

We show the pre-nebular phase covering from +42d to +104d post-peak in Fig. \ref{fig:specs:seq:prenebular}.
The calcium NIR triplet $\ion{Ca}{II}~\lambda\lambda8498,8542,8662$ becomes the strongest emission feature in the spectrum, with some oxygen recombination features, such as $\ion{O}{I}~\lambda7774$, also being quite pronounced.
The helium features, $\ion{He}{I}~\lambda\lambda6678,7065$, become less noticeable due to overlap with the broad oxygen doublet $[\ion{O}{I}]~\lambda\lambda6300,6364$ and the calcium doublet $[\ion{Ca}{II}]~\lambda\lambda7291,7324$.

We show the nebular phase spectra from +101d to +348d post-peak in Fig. \ref{fig:specs:seq:nebular}.
The most conspicuous emission features are the $[\ion{O}{I}]~\lambda\lambda 6300,6364$ complex and the [$\ion{Ca}{II}]~\lambda\lambda7292,7324$ and $\ion{Ca}{II}~\lambda\lambda8493,8542,8662$ lines.
In the spectra, there are also oxygen recombination lines visible at $\ion{O}{I}~\lambda\lambda7772,7774,7775$, and $\ion{O}{I}~\lambda8446$ and another feature at $9264~\AA$.
The oxygen lines show substantial structure, which complicates the measurements.
The $\ion{Mg}{I}]~\lambda4571$ and $\ion{Mg}{I}]~\lambda5167$ lines become more pronounced.
There is a quite pronounced $\ion{Na}{I}\,D$ line visible, which slowly replaces the $\ion{He}{I}~\lambda5876$ feature.

The late-time Keck spectrum (+348d) is dominated by the $[\ion{O}{I}]~\lambda\lambda6300,6364$ complex, by  [$\ion{Ca}{II}]~\lambda\lambda7292,7324$, and a little bit of emission of $\ion{Mg}{I}]~\lambda4571$. %
All other previously mentioned emission lines have vanished at this epoch.

One can use the oxygen-to-calcium abundance ratio as a diagnostic of the core mass \citep{1989ApJ...343..323F}, since while oxygen monotonically increases with increasing core mass, calcium does not \citep{2003ApJ...592..404L}.
The $[\ion{O}{I}]$ to $[\ion{Ca}{II}]$ line-flux ratio is often used as a proxy for the abundance ratio \citep{2019NatAs...3..434F}.
Using integration, we estimated the ratio for \supernova{2019odp} to be in the range of 1.2--1.9, increasing towards the last spectroscopic epoch.
This means that this transient belongs to the class of calcium-poor (or oxygen-rich) supernovae \citep{2022MNRAS.514.5686P} and is likely to have a more massive progenitor.
However, as \cite{1989ApJ...343..323F} have already pointed out, it is only a very crude estimator at best, since a given abundance ratio does not translate into a unique line-flux ratio, but is strongly influenced by mixing and also evolves over time.

\subsubsection{Classification}
\label{sec:classification}

\begin{figure}
  \includegraphics[width=\linewidth]{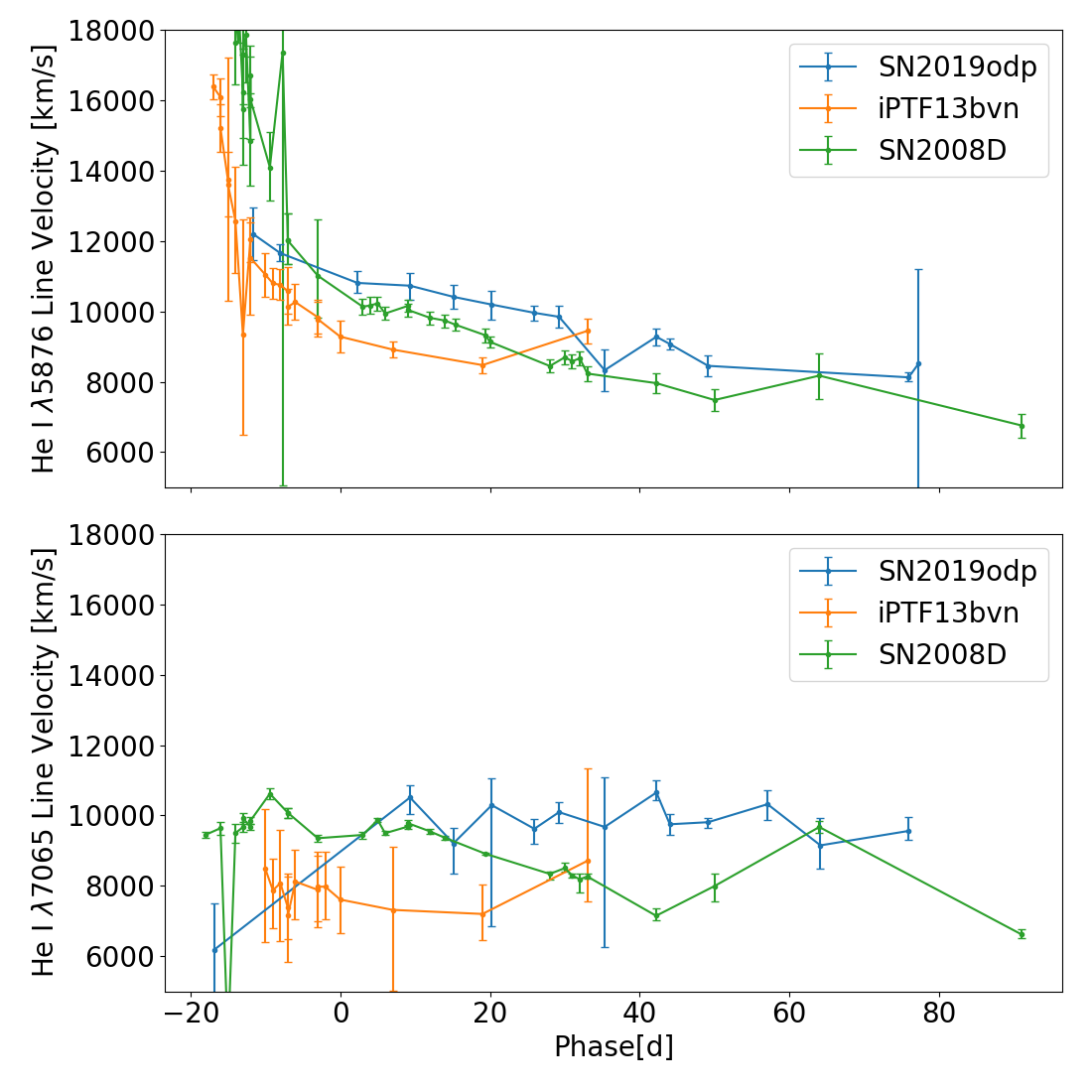}
  \caption{Velocity evolution of the \ion{He}{I} lines in \supernova{2019odp} and the comparison \Ib{supernovae}.}
  \label{fig:spec:he:velocity}
\end{figure}

As was mentioned in Sect. \ref{sec:discovery}, \supernova{2019odp} was initially classified as a \IcBL{supernova} by \cite{2019TNSCR1595....1B}.
When comparing the early spectra obtained shortly after discovery against \Ib (iPTF13bvn) and \IcBL (\supernova{1998bw}) supernovae (Fig. \ref{fig:evo:spec:classification}, left panel), it indeed seems quite suggestive to classify this supernova as a \IcBL due to the overall structure of the continuum and the absence of obvious identifiable lines.
Comparing against typical supernova databases used for classifications such as \textit{SNID} \citep{2007ApJ...666.1024B} supports this conclusion.
However, later spectra (Fig. \ref{fig:evo:spec:classification}, right panel) showed the presence of lines with much smaller line widths than are typically associated with \IcBL{supernovae} as well as the presence of helium lines (see Figs.\ \ref{fig:specs:seq:photospheric} and \ref{fig:specs:seq:prenebular}), which clearly show  that this is a \Ib{supernova}.

Using our own classification database and code,\footnote{We implemented the spectral flattening or smoothing algorithm that \textit{SNID} uses in Python and implemented a simpler comparison metric given the known redshift of the supernova.}
we searched for other supernovae that were a close spectral match for the early pre-peak spectra and found that the \Ib{\supernova{2008D}} is a quite good match (Fig. \ref{fig:evo:spec:classification}, left panel) as well.
This suggests that this lack of features at early phases of the supernova is not completely unseen for \Ib{supernovae} and additional care should be taken when classifying transients at very early phases.

It is not obvious if the lack of helium lines and the broad spectral shape at early phases is due to very high velocities or due to other effects.
\cite{2016ApJ...832..108M} argue that the lack of observed helium lines in \IcBL{supernovae} cannot solely be explained by the helium lines being smeared out by the high velocities in the ejecta.
Further transitional objects, such as \supernova{2016coi} \citep{2017ApJ...837....1Y,2018MNRAS.478.4162P}, have been discovered that show helium lines initially, but evolve to more closely resemble \IcBL{supernovae} later on.
This is the inverse to what is seen in \supernova{2008D} and in \supernova{2019odp}, where initially there were no visible helium lines, but they appeared later on.
\supernova{2017ens} \citep{2018ApJ...867L..31C} is another supernova that shows characteristics of a \IcBL{supernova} early on, but then changes its class to a \IIn{supernova}.
On the other hand, there are normal \Ib{supernovae}, such as iPTF13bvn, that have been discovered at similarly early phases 
and that show no hints of broad lines.
\cite{2022ApJ...930...31B} argue for a split in the \Ib{class}, which might explain these different behaviours.

\subsubsection{Helium velocity evolution}
\label{sec:ana:evo:velocity}

We estimated the helium line velocities as the absorption minima of the P-Cygni features of the individual helium lines.
We used the nested sampler \textit{dynesty} to fit the following P-Cygni model function, $M_\lambda$, to the individual helium features:
\begin{equation}
  M_\lambda = C_\lambda + E~\mathcal{G}_\lambda(\Lambda+\Delta\Lambda_E,\sigma_E)-A~\mathcal{G}_\lambda(\Lambda+\Delta\Lambda_A, \sigma_A),
\end{equation}
where $C_\lambda$ is the linear continuum function, $\mathcal{G}_\lambda$ is the (normalised) Gaussian function, $E$ is the amplitude of the emission feature, $A$ is the amplitude of the absorption feature, $\Lambda$ is the rest-wavelength of the feature, $\Delta_{A,E}$ are the velocity-offsets of the absorption and emission peak, respectively, and, $\sigma_{A,E}$ are the widths of the absorption or emission Gaussians.
The measured velocity evolutions for \supernova{2019odp} and the comparison \Ib{supernovae} are shown in Fig. \ref{fig:spec:he:velocity}.

We estimated the \ion{He}{I} $\lambda\,5876$ line velocity at the peak by fitting the following exponential model to the individual velocity estimates using a GP kernel as a likelihood function:
\begin{equation}
  v_\text{abs}(t) = v_0\,\exp\left(t-t_0\right)^\alpha,
\end{equation}
where $v_0$, $t_0$, and $\alpha$ are free parameters (within reasonably chosen priors).
We only included significant measurements in the time range from $-15.5$ d to 30 d. %
Our estimate for the absorption velocity at the peak is thus
\begin{equation}
  v_\text{abs}(\ion{He}{I}~\lambda5876) = 10977 \pm 400~\kms
.\end{equation}

\newpage
\FloatBarrier

\section{Modelling}
\label{sec:model}

\subsection{Photospheric phase}
\label{sec:model:main}

In Sect.~\ref{sec:mod:phot:results}, we model the diffusion part of the light curve, while in Sect.~\ref{sec:model:plateau} we focus on the earliest part of the light curve. Section~\ref{sec:oxygen} includes  detailed modelling of the nebular spectra.

\subsubsection{Arnett modelling results}
\label{sec:mod:phot:results}

{
  \renewcommand{\arraystretch}{1.2}
  \begin{table*}[ht]
    \caption{Supernova parameter estimates for different pseudo-bolometric light curve models.}
    \begin{tabular}{|c|c|c|c|c|c|c|c|}
      \hline
      SN Name & Model/Reference & Phase fitted & $\tau_m$ & \Mni & $M_\text{ej}$ & $E_\text{K}$ & $R(t=0)$ \\
      & & (d) & (d) & $(M_\odot)$ & $(M_\odot)$ & $(10^{51}$~erg)  & $(R_\odot)$ \\
      \hline
      \hline
      SN2019odp & A82-c & -11 -- 25 & $18.7 \pm 2$ & $0.246 \pm 0.04$ & $5.4 \pm 1.5$ & $6.4 \pm 1.5$ & \\
      SN2019odp & A82-e & -11 -- 25 & $18.9 \pm 2$ & $0.253 \pm 0.04$ & $5.7 \pm 1.2$ & & $9_{-9}^{+9}$ \\
      \hline
      iPTF13bvn & \cite{2016MNRAS.457..328L} & -9 -- 10 & $12.5-13.4$ & $0.06_{-0.01}^{+0.02}$ & $1.7_{-0.4}^{+0.5}$ & $0.7_{-0.2}^{+0.3}$ & \\
      iPTF13bvn & A82-c & -13 -- 30 & $11.8 \pm 1.2$ & $0.0676 \pm 0.02$ & $1.6 \pm 0.4$ & $1.1 \pm 0.3$ & \\
      iPTF13bvn & A82-e & -15 -- 25 & $12.3 \pm 1.1$ & $0.0746 \pm 0.02$ & $1.8 \pm 0.3$ & & $8_{-8}^{+7}$ \\
      \hline
    \end{tabular}
    \tablefoot{Model estimates were computed using the algorithm described in Sect. \ref{sec:mod:phot:fitting}. We also include literature comparison values for iPTF13bvn in this table, from  \cite{2016MNRAS.457..328L}. For iPTF13bvn, hydrodynamical modelling also provided an ejecta mass of $\approx2~M_\odot$ \citep{2016A&A...593A..68F,2014AJ....148...68B}.
    The models are: \textbf{A82-c} (c for compact; Sect. \ref{sec:model:main:arnett}) for small $R_0$ and \textbf{A82-e} (e for extended; Sect. \ref{sec:model:main:arnett:large}) for large $R_0$.}
    \label{tab:mod:phot:results}
  \end{table*}
}

Using the \citet[][hereafter A82]{1982ApJ...253..785A} models, which we %
summarise in Sect. \ref{sec:model:main:arnett}, we %
estimated the supernova parameters using the light curve diffusion peak.
As the input, we used the pseudo-bolometric light curves derived in Sect. \ref{sec:evo:qbol} and the Bayesian parameter estimation algorithm described in Sect. \ref{sec:mod:phot:fitting}.
We considered both the canonical A82-c model (c for compact), which assumes a small initial radius ($R_0 \ll 1000~R_\odot$) described in Sect. \ref{sec:model:main:arnett}, and the A82-e model (e for extended) for a large initial radius, which is described in Sect. \ref{sec:model:main:arnett:large}.
As a cross-check of the overall method, we also performed the same procedure for the comparison object iPTF13bvn.

We summarise the parameter estimation results in \autoref{tab:mod:phot:results} and also include parameter estimates taken from the literature for iPTF13bvn.
The high nickel mass ($M_\text{Ni} \sim 0.25~M_\odot$), ejecta mass ($M_\text{ej} \sim 5~M_\odot$), and kinetic energy ($E_K \sim 6\times 10^{51}$~erg) of \supernova{2019odp} point towards a very energetic explosion of a massive star.
Compared to other samples of Type Ib supernovae, such as \cite{2018A&A...609A.136T} or \cite{2016MNRAS.457..328L}, \supernova{2019odp} is near the edge of the distribution, but there are objects in these samples that have similar ejecta masses and nickel masses.
The inferred kinetic energy ($E_K \sim 6\times 10^{51}$~erg) is significantly higher than for most other \Ib{supernovae}, with a few exceptions (with \supernova{2008D} being in a similar range), 
and is more comparable to values seen in \IcBL{supernovae}.
While the A82-e model does not place a strict constraint on the initial radius, $R_0$, it does indicate a more compact progenitor.
We investigate the progenitor radius further in the context of shock cooling models in Sect. \ref{sec:model:plateau}.

For comparison purposes, we also calculated the ejecta mass and kinetic energy under the (often used) assumption of $v_\text{sc} = v_\text{ph}(t_\text{peak})$ \citep{2016MNRAS.457..328L,2015MNRAS.450.1295W} instead of the method based on \cite{2016MNRAS.458.1618D}, which is described in more detail in Sect. \ref{sec:model:main:arnett}.
This yields a lower ejecta mass of $\sim 4.2\,M_\odot$ and a kinetic energy of $\sim 3.8 \times 10^{51}$\,erg.
We note that this energy also exceeds the values commonly derived in theoretical works for neutrino-driven explosions, which do not exceed an explosion energy of $\sim 2\times10^{51}$\,erg \citep{2020ApJ...890...51E}.

\subsection{Early plateau}
\label{sec:model:plateau}

As was noted in Sect. \ref{sec:evo:phot}, the first few observations 
show an excess or plateau %
rather than a smoothly rising light curve.
Other supernovae that have shown an early excess or light curve `bump' include the \Ib{\xrfsn{2008D}}, 
the \IcBL{\grbsn{2006aj}}, and the \IIb{\supernova{2016gkg}} \citep{2017ApJ...836L..12T,2017ApJ...837L...2A}.
Common features among these three are blue colours, strong UV emission, high colour temperatures ($T_\text{color} \gtrsim 12000$~K) that decrease right from discovery, and the early excess being shaped like an additional (usually smaller) early peak in the light curve.
While \supernova{2019odp} does show a very blue colour (Sect. \ref{sec:evo:color}; Fig. \ref{fig:lc:color:evo}) and high temperatures early on (Sect. \ref{sec:evo:bb}; Fig. \ref{fig:evo:bb:comparison}), it shows neither strong UV emission initially (the $U-r$ colour in fact gets bluer instead of redder) nor a similar initial light curve (no early peak, but an actual plateau; Sect. \ref{sec:evo:phot}).

One commonly invoked scenario to explain this early excess is shock cooling emission \citep{2008ApJ...683L.135C,2017ApJ...837L...2A}.
Several (semi-)analytical shock cooling emission models have been developed to explore the UV/optical light curve \citep{2010ApJ...725..904N,2011ApJ...728...63R,2021ApJ...909..209P}.
As the shock cools, at some point the temperature will drop below the recombination temperature of the shocked material \citep{2011MNRAS.414.2985D,2011ApJ...728...63R} and then
the luminosity will start to plateau until the radioactive energy input becomes significant \citep{2013ApJ...769...67P}.
Since we do not see any excess UV emission in \supernova{2019odp} and the luminosity plateaued right after discovery, we conclude that in this scenario we have missed the shock cooling emission itself and are only seeing emission from the recombination phase.
Based on the numerically recalibrated shock model by \cite{2010ApJ...725..904N}, \cite{2013ApJ...769...67P} use the plateau luminosity and explosion parameters to estimate the progenitor radius, $R_0$.
We solved Eq. 5 from \cite{2013ApJ...769...67P} for $R_0$:
\begin{equation}
  R_0 \approx \left( \frac{L_\text{plateau}}{7\times10^{40}~\text{erg}~\text{s}^{-1}} \frac{ \left( \frac{\kappa}{0.2 \text{cm}^2 \text{g}^{-1}} \right)^{0.69} \left( \frac{M_\text{ej}}{1~M_\odot} \right)^{0.67} }{ \left( \frac{E_K}{10^{51} \text{erg}} \right)^{0.85}} \right)^{1.28} R_\odot
  \label{eqn:model:plateau:piro:r0}
.\end{equation}
\cite{2013ApJ...769...67P} predicted a plateau following the shock cooling emission peak based on a diffusion wave travelling inwards in the expanding supernova ejecta.
The predicted start of the plateau after the explosion for a $< 10~R_\odot$ progenitor and the estimated explosion parameters in Sect. \ref{sec:mod:phot:results} is between 20 hours and 4 days -- both of which fit easily in the observation gap before first light.
Using the estimated plateau luminosity (Sect. \ref{sec:evo:qbol}; Eq. \ref{eqn:evo:phot:lums:plat}), we used Eq. \ref{eqn:model:plateau:piro:r0} to estimate the progenitor radius to be $0.6 - 1.4~R_\odot$.
This is comparable to the estimate of $R_0 \approx 2~R_\odot$ for \supernova{2008D} by \cite{2011ApJ...728...63R}.
This value is also compatible with the upper limit of $18~R_\odot$ (\autoref{tab:mod:phot:results}) derived using the large-radius formulation of the Arnett equation (Sect. \ref{sec:model:main:arnett:large}).

Another possible scenario for this early emission is imperfect 
mixing of the nickel throughout the ejecta \citep{2019ApJ...872..174Y}.
Depending on the mixing fraction, they argue that different light curve plateaus or `pre-bumps' are possible.
A lack of nickel in the outer part of the ejecta would also be consistent with the helium absorption lines only becoming visible later on (Sect. \ref{sec:evo:spec}) due to the lack of non-thermal excitation of the helium by the nickel decays \citep{1991ApJ...383..308L}.
\cite{2015A&A...574A..60T} explore plateau duration, luminosity, and the magnitude difference with the peak ($\Delta M \equiv M_\text{plat} - M_\text{peak}$) using a small grid of hydrodynamical models.
While the inferred ejecta mass and energy for \supernova{2019odp} (Sect. \ref{sec:mod:phot:results}) fall %
outside the model grid, the observed plateau duration of $2-5$\,d, plateau luminosity, $\log L_\text{plateau} = 41.3$ $[\log(\text{erg}\,\text{s}^{-1})]$, and magnitude difference, $\Delta M \approx 3.3$\,mag (Sect. \ref{sec:evo:qbol}), fall well into the range of values seen in their model grid.
Extrapolating the trend seen for the ejecta mass and nickel mixing beyond the model grid suggests that very low nickel mixing scenarios are not an obvious candidate, since both low nickel mixing and high ejecta mass increase the duration of the plateau.

\subsection{Constraining the oxygen mass from nebular spectra}
\label{sec:oxygen}

\begin{table}[ht]
  \caption{Oxygen mass estimates for stripped envelope supernovae from the literature.}
  \begin{tabular}{|l|c|}
    \hline
    Transient & Oxygen Mass \\
    & $(M_\odot)$ \\
    \hline
    \multicolumn{2}{|c|}{Type Ib} \\
    \hline
    \supernova{2022hgk} (0) & $2.6 - 7.4~$  \\
    \supernova{2017iro} (1) & $0.35~$  \\
    iPTF13bvn (2) & $0.33~$  \\
    \supernova{2009jf} (3) & $1.34~$  \\
    \supernova{2007Y} (4) & $0.2~$  \\
    \supernova{1996N} (5) & $0.11 - 0.21$  \\
    \supernova{1990I} (6) & $0.7-1.35$  \\
    \hline
    \hline
    \multicolumn{2}{|c|}{Type IIb} \\
    \hline
    \supernova{2020acat} (7) & $1.0-3.1$  \\
    Sample (8) & $0.005 - 0.1$ \\ %
    \hline
    \hline
    \multicolumn{2}{|c|}{Type Ic} \\
    \hline
    \supernova{2011bm} (9) & $5-10$  \\
    \hline
    \hline
    \multicolumn{2}{|c|}{Type Ic-BL} \\
    \hline
    \supernova{1998bw} (10) & $1.4-3$  \\
    \hline
  \end{tabular}
  
  \tablebib{(0) \cite{2023arXiv230604698D}; (1) \cite{2022ApJ...927...61K}; (2) \cite{2015A&A...579A..95K}; (3) \cite{2011MNRAS.413.2583S}; (4) \cite{2009ApJ...696..713S}; (5) \cite{1998A&A...337..207S}; (6) \cite{2004A&A...426..963E}; (7) \cite{2022MNRAS.513.5540M}; (8) \cite{2023ApJ...959...12D}; (9) \cite{2012ApJ...749L..28V}; (10) \cite{2001ApJ...559.1047M}.} 
  \label{tab:mod:oxygen:litcompare}
\end{table}

Nebular spectra provide important diagnostics as they allow one to peer into the optically thin ejecta \citep{1989ApJ...343..323F}. %
Some observable parameters of supernova progenitors, such as the oxygen mass, provide information on the initial stellar mass, which in turn can constrain the possible progenitor channels \citep{1996ApJ...456..811H,1986ApJ...310L..35U,1992ApJ...387..309L}.
One technique to estimate the ZAMS mass of the progenitor is to use the oxygen mass as a proxy, as it has good sensitivity to the pre-explosion mass in stellar evolution models \citep{2003ApJ...592..404L,2010ApJ...724..341H,2005A&A...433.1013H}.
A method of estimating an upper limit on the oxygen mass is described in \cite{2014MNRAS.439.3694J} based on measured nebular spectroscopy.
For this, we integrated the flux in a $100~\AA$ window around the $[\ion{O}{I}]~\lambda5577$ line and the $[\ion{O}{I}]~\lambda\lambda6300,6364$ lines,\footnote{With the range starting $50~\AA$ blueward of $6300~\AA$ and ending $50~\AA$ redward of $6364~\AA$.} while subtracting a background baseline.
Based on the line ratio, we estimate a temperature in the range from $3888\,\text{K}$ to $4600\,\text{K}$ and an oxygen mass in the range from $4.5\,M_\odot$ to $6\,M_\odot$. 
We show some comparison values from the literature in \autoref{tab:mod:oxygen:litcompare}, where one can see that in the context of Type Ib supernovae \supernova{2019odp} is in the very high range of estimated oxygen masses. %
Values seen in Type IIb supernovae are typically also much smaller, but %
some extreme cases of Type Ic supernovae, in particular \supernova{2011bm} \citep{2012ApJ...749L..28V}, which was deduced to be an explosion of a very massive star, have been reported to show even higher oxygen masses.
Most of the oxygen masses in the literature (\autoref{tab:mod:oxygen:litcompare}) were obtained using the method from \cite{1986ApJ...310L..35U} and assuming a temperature in the range of $3000-4500\,$K as well as local thermodynamic equilibrium (LTE) conditions.
If we do the same for \supernova{2019odp} (for T = 3888 K), we get an minimum oxygen mass of $\sim5~M_\odot$.
This would be an uncomfortably large
fraction of the estimated ejecta mass.

Therefore, we expanded on the method used in \cite{2014MNRAS.439.3694J} to take into account deviation from LTE as well.
We first describe the analytic models that we use in Sect. \ref{sec:oxygen:model}, and in Sect. \ref{sec:oxygen:fitting} we describe the fitting method employed to actually apply the improved model.
In Sect. \ref{sec:oxygen:results}, we describe the results of this methodology.

\subsubsection{Constraining the oxygen mass from nebular spectra: Model description}
\label{sec:oxygen:model}

The most important oxygen emission lines for this are the $[\ion{O}{I}]\,\lambda\lambda6300,6364$ 
doublet lines as well as the $[\ion{O}{I}]\, \lambda 5577$ 
line.
We used the total line luminosity, $L_\lambda (N_u, \beta_\lambda)$, where $\beta_\lambda$ is the Sobolev escape probability and $N_u$ is the total number of ions in the excited state, $u$, assuming the Sobolev approximation and uniform density defined in %
\citet[][their equation 34]{2017hsn..book..795J}.

A difference in our method is that we treated the two $[\ion{O}{I}]\,\lambda\lambda6300,6364$ doublet lines separately.
Since both share the same source state and only differ in the split ground state (which is in LTE), they only differ by the radiative decay rate, $A$, as well as in the escape probability, $\beta$.
We assume that all oxygen line emission comes from the same environmental conditions, and thus we can state the $[\ion{O}{I}]\, \lambda 6364$ optical depth in terms of the $[\ion{O}{I}]\, \lambda 6300$ optical depth \citep{2011AcA....61..179E,1992ApJ...387..309L}:%
\begin{equation}
  \tau_{6364} = \frac{\tau_{6300}}{3}
  \label{eqn:oxygen:mod:tau}
.\end{equation}
Since we fitted both lines of the doublet simultaneously, we can use the amplitude ratio between the two lines to constrain the optical depth of both lines.
This replaces the fixed assumption of $\beta_{6300,6364} \approx 0.5$ %
in \cite{2014MNRAS.439.3694J}.
We specified all optical depths in relation to the $[\ion{O}{I}]\, \lambda 6300$ optical depth: $\tau \equiv \tau_{6300}$.

The Sobolev escape probability, $\beta$, in terms of the optical depth, $\tau$, was defined as follows \citep{2017hsn..book..795J,1992ApJ...387..309L}:
\begin{equation}
  \beta_\lambda = \frac{1 - \exp^{-\tau_\lambda}}{\tau_\lambda}
  \label{eqn:sobolev:beta}
.\end{equation}

We assumed the first excited state, $u_1$, to be in LTE and used the Saha equation (as it is defined shortly before equation 42 in 
\citealt{2017hsn..book..795J}) to approximate the state number, $N_{u_1}$.
In the case of state $u_2$, we allowed this population number to fall below the LTE estimate ($N_2^\text{LTE} \rightarrow d_2 N_2^\text{LTE}$, where $d_2$ is the LTE departure coefficient).
Temperatures in the emission regions in the nebular phase are typically assumed to be below $5000$\,K in the high-density regime \citep{2004A&A...426..963E, 1998A&A...337..207S}, but we adopted a more conservative prior range of $1000-8000$\,K, since it is difficult to define any particular cut-off point for the temperature.

We approximated the partition function with the statistical weight, $g_g$, of the ground state
and stated the equation in terms of the oxygen mass, $M_\text{OI}$, using $N = M_\text{OI}\, \mu^{-1} m_p^{-1}$ %
and %
the $\ion{O}{I}\,\lambda 6300$ and $\ion{O}{I}\,\lambda 6364$ line luminosity as

\begin{equation}
  L_\lambda = \left( \frac{A_\lambda ~h ~ c}{\mu ~ m_p ~ \lambda} \frac{g_u}{g_g} \right) \frac{\beta_\lambda M_\text{OI} }{\exp\left(T_\lambda / T\right)},
  \label{eqn:oxygen:mod:lum}
\end{equation}
where $T_\lambda = \frac{E_u}{k_B}$, $\mu=16$ is the mean atomic weight of oxygen, $m_p$ is the mass of the proton, $A_\lambda$ is the radiative decay rate of the transition, $g_u$ is the statistical weight of the upper state, and $g_g$ is the statistical weight of the lower state (values for physical constants can be found in \autoref{tab:mod:neb:oxygen:constants}).

If we assume the first excited state, $u_1$, to be in LTE, we can state the optical depth of $[\ion{O}{I}]\, \lambda 5577$ relative to the $[\ion{O}{I}]\, \lambda 6300$ optical depth:
\begin{equation}
  \frac{\tau_{5577}}{\tau} = \frac{g_2}{g_1} d_2 \frac{A_{5577}}{A_{6300}} \frac{5577^3}{6300^3} \exp{\left(\frac{E_1}{k_B T}\right)} \frac{1- \exp{\frac{-\Delta E_{2\rightarrow1}}{k_B T}}  }{1-\exp{\frac{-E_1}{k_B T}}},
  \label{eqn:oxygen:mod:tau5577}
\end{equation}
where $d_2$ is the NLTE deviation fraction for the second excited state, $u_2$, and $\Delta E_{2 \rightarrow 1} = E_2 - E_1$.
In \cite{2014MNRAS.439.3694J}, the analog for this is the assumed ratio of $\beta_{5577}/\beta_{6300,6364} \approx 1.5$.
The LTE departure coefficient denotes by how much the number density of the $u_2$ state falls below the LTE estimate.
\cite{2014MNRAS.439.3694J} find a range for $d_2$ of 0.8 to 0.3 over the time range of 250 to 450 days post-explosion from their modelling efforts for Type IIP supernovae.
We %
adopted a conservative
range of  $0.1-1.0$ 
as our
prior for $d_2$. %

Combined with Eq. \ref{eqn:sobolev:beta}, we can calculate $\beta_{5577} (d_2, T, \tau)$ and state the $[\ion{O}{I}]\, \lambda 5577$ line luminosity as follows
\citep[based on][their Eq. 2]{2014MNRAS.439.3694J}:
\begin{eqnarray}
  \frac{L_{5577}}{L_{6300}} &=& d_2\, \frac{g_{u_2}}{g_{u_1}} \cdot \exp^{\frac{-\Delta E_{2 \rightarrow 1}}{k_BT}} \frac{A_{5577} \beta_{5577} (d_2, T, \tau)}{A_{6300} \beta_{6300}} \frac{6300}{5577} \\
  &=& d_2 \cdot 51 \cdot \exp^{-25789.8/T} \frac{\beta_{5577} (d_2, T, \tau)}{\beta_{6300}},
  \label{eqn:oxygen:mod:5577ratio}
\end{eqnarray}
with the physical constants given in \autoref{tab:mod:neb:oxygen:constants}.

Based on the observed line luminosities $L_{5577},L_{6300},L_{6364}$, one can then use equations \ref{eqn:oxygen:mod:5577ratio} and \ref{eqn:oxygen:mod:lum} to constrain $M_\text{OI}$, $T$, $d_2$, and $\tau$.
With fewer input parameters than output parameters, some of the parameters are degenerate, which is why the priors have to be carefully considered.
The fitting method is described 
in Sect. \ref{sec:oxygen:fitting}.

\begin{table*}
  \caption{Physical constants used in the oxygen analysis.}
  \begin{tabular}{|c|c|c|c|}
    \hline
    Description & Symbol & Value & Unit \\
    \hline
    \hline
    $u_2 \equiv 2s^2 2p^4 ({}^1 S)$ Energy Level & $E_2/k_B$ & 48620 & K \\
    $u_1 \equiv 2s^2 2p^4 ({}^1 D)$ Energy Level & $E_1/k_B$ & 22830 & K \\
    Statistical weight of $u_2$ & $g_{u_2}$ & 1 & \\
    Statistical weight of $u_1$ & $g_{u_1}$ & 5 & \\
    Statistical weight of ground state $g$ & $g_g$ & 9 & \\
    Radiative Decay Rate of $[\ion{O}{I}]\, \lambda 5577$ ($u_2 \rightarrow u_1$) & $A_{5577}$ & 1.26 & $\text{s}^{-1}$ \\
    Radiative Decay Rate of $[\ion{O}{I}]\, \lambda 6300$ ($u_1 \rightarrow g$) & $A_{6300}$ & $5.63\cdot10^{-3}$ & $\text{s}^{-1}$ \\
    Radiative Decay Rate of $[\ion{O}{I}]\, \lambda 6364$ ($u_1 \rightarrow g$) & $A_{6364}$ & $1.83\cdot10^{-3}$ & $\text{s}^{-1}$ \\
    \hline
  \end{tabular}
  \tablebib{
    \cite{NIST_ASD, 2014MNRAS.439.3694J, 2017hsn..book..795J}.
  }
  \label{tab:mod:neb:oxygen:constants}
\end{table*}

\subsubsection{Constraining the oxygen mass from nebular spectra: Fitting method}
\label{sec:oxygen:fitting}

\begin{table}
  \caption{Priors for the second stage of the oxygen mass estimation fit.}
  \begin{tabular}{|c|c|c|c|}
    \hline
    Parameter & Symbol & Unit & Prior \\
    \hline
    \hline
    Oxygen Mass & $M_\ion{O}{I}$ & $M_\odot$ & $\mathcal{U}(0, 6)$ \\
    Oxygen Temperature & $T$ & K & $\mathcal{U}(1000,8000)$ \\
    Distance & $D$ & cm & $\mathcal{U}(D_\text{min}, D_\text{max})$ \\
    $[\ion{O}{I}]\, \lambda 6300$ Opt. Depth & $\log \tau$ & & $\mathcal{U}(-5, 1)$ \\
    LTE Departure & $d_2$ & & $\mathcal{U}(0.1, 1)$ \\
    \hline
  \end{tabular}
  \tablefoot{For \supernova{2019odp}, we adopted the distance prior from Sect. \ref{sec:discovery}.}
  \label{tab:mod:neb:oxygen:priors}
\end{table}

While our oxygen estimation method improves upon \cite{2014MNRAS.439.3694J}, it requires the disentangling of the flux from the $[\ion{O}{I}]\,\lambda 6300$ and the $[\ion{O}{I}]\,\lambda 6364$ lines, which overlap due to the large velocities present in supernovae.
This means we have to make assumptions about the line profile function, $\mathcal{P}$.
To make this whole process more tractable, we split the problem into two parts.
First, we estimated the line fluxes, $\{F\}$, for the three relevant oxygen lines using empirical models for the spectral line profiles.
Then, we used the model described in Sect. \ref{sec:oxygen:model} to derive the physical parameters of the model from the line fluxes.

We used the nested sampler \textit{dynesty} to fit the parameter vector, $\Xi$.
Since in the early nebular spectra the $\ion{O}{I}\,\lambda7774$ recombination line is still present, we used this line to estimate the line profile of the forbidden oxygen lines for those epochs. 
We projected a section around the $\ion{O}{I}\,\lambda7774$ line into velocity space, normalised it, and then used it as an empirical line profile function, $\mathcal{P}_\lambda$.
For the last ($+348\,$d) spectrum, the recombination line was no longer visible and we used a parametric Gaussian as the line profile function instead:
\begin{equation}
  \mathcal{P}_\lambda = \mathcal{G}_\lambda (\lambda_c, \sigma)
.\end{equation}

To this central line profile function, $\mathcal{P}$, we added a thick shell function, $\mathcal{S}$, to model any additional emission in the outer regions of the supernova ejecta, since the simple Gaussian does not capture all of the flux of the line complex:
\begin{equation}
  \mathcal{P}^\prime_\lambda (\lambda_c, \Xi) = F_{6300}\,\mathcal{P}_\lambda(\lambda_c, \Xi) + F_\text{shell}\, \mathcal{S}_\lambda(\lambda_c, \Xi),
\end{equation}
where $F$ are the line fluxes.
The line profile function for the thick shell is approximated by an elongated Gaussian function:
\begin{equation}
  \mathcal{S}_\lambda(\lambda_c,\Xi) = \frac{1}{k\,\sigma\,\sqrt{2\pi}} 
  \begin{cases}
    \exp\left( \frac{-\left(\lambda-\lambda_c+\lambda_s\right)^2}{2\sigma^2} \right) & \lambda \le -\lambda_s \\
    \exp\left( \frac{-\left(\lambda-\lambda_c-\lambda_s\right)^2}{2\sigma^2} \right) & \lambda \ge \lambda_s \\
    1 & -\lambda_s < \lambda < \lambda_s
  \end{cases},
\end{equation}
where $\lambda_c$ is the centre position, $\lambda_s$ is the elongation width (corresponding to the inner cut-off velocity of the shell), $\sigma$ is the Gaussian width (corresponding to the width of the shell), and $k$ is the numerically derived normalisation constant.

We segmented the spectra into two spectral regions: the $[\ion{O}{I}]\, \lambda 5577$ region and the $[\ion{O}{I}]\, \lambda\lambda 6300,6364$ region.
These line profile functions were then (re)projected onto the observed wavelength grid around the $[\ion{O}{I}]\, \lambda 5577$ region and the $[\ion{O}{I}]\, \lambda\lambda 6300,6364$ regions.
We allowed for a global wavelength offset, $\Delta\lambda$.
For the $[\ion{O}{I}]\, \lambda\lambda 6300,6364$ region, we have the following model spectral flux function:
\begin{equation}
  M_{\lambda-\Delta\lambda}(\Xi) = \mathcal{P}_\lambda^\prime (6300, \Xi) + \mathcal{R}_6\, \mathcal{P}_\lambda^\prime(6364, \Xi) + C_\lambda,
\end{equation}
with $\mathcal{R}_6 \equiv F_{6364}/F_{6300}$, which is constrained to the range between what is fully optically thin ($\mathcal{R}_6=1/3$) and fully optically thick ($\mathcal{R}_6=1$).
For the $[\ion{O}{I}]\, \lambda 5577$ region, we have the following model spectral flux function:
\begin{equation}
  M_{\lambda-\Delta\lambda}(\Xi) = \mathcal{R}_5\, \mathcal{P}_\lambda^\prime(5577, \Xi) + C_\lambda,
\end{equation}
with $\mathcal{R}_5 \equiv F_{5577}/F_{6300}$.
For each region, we assumed a separate linear continuum:
\begin{equation}
  C_\lambda = \alpha_R + \beta_R\cdot\left(\lambda - \lambda_\text{RC}\right),
\end{equation}
where $\lambda_\text{RC}$ is the centre of the region ($5577~\AA$ and $6330~\AA$, respectively).

We marginalised over all nuisance parameters to yield a three-dimensional posterior distribution, which only contains the three line fluxes, $F_{5577}$, $F_{6300}$, and $F_{6364}$.
Next, we calculated the mean and covariance matrix of this three-dimensional posterior distribution, which was then used as an input for the second stage of the analysis.

For the second stage Bayesian model, the parameter vector, $\theta$, only consists of a few parameters: the oxygen mass, $M_{OI}$, temperature, $T$, optical depth, $\tau$, distance, $D$, and LTE departure coefficient, $d_2$.
We used the nested sampler \textit{dynesty} to estimate the posterior distribution using the priors given in \autoref{tab:mod:neb:oxygen:priors} and the line fluxes measured in the previous stage.
In each iteration of the sampler, we did the following (using the parameter vector, $\theta$, as an input):
\begin{enumerate}
\item Calculate the line luminosities, $L_{6300}$ and $L_{6364}$, using Eq. \ref{eqn:oxygen:mod:lum} and the oxygen mass, $M_{OI}$, temperature, $T$, and optical depth, $\tau$, as inputs from the parameter vector.
\item Calculate the optical depth ratio, $\tau_{5577}/\tau$, using Eq. \ref{eqn:oxygen:mod:tau5577}, which depends on the LTE departure coefficient, $d_2$, and temperature, $T$.
\item Calculate the luminosity, $L_{5577}$, using Eq. \ref{eqn:oxygen:mod:5577ratio}, which depends on $L_{6300}$, $d_2$, $T$, and $\tau$.
\item Convert the line luminosities to line fluxes using the distance, $D$.
\item Calculate the log likelihood using a Gaussian multivariate and the derived mean and covariance matrix from the previous stage.
\end{enumerate}

As the number of output parameters is actually larger than the number of input parameters and the main goal of the study is to constrain the oxygen mass of the supernova, we marginalised over all parameters except for the oxygen mass.
This allowed us to take into consideration any prior information for these parameters, without having to assume specific values.

We applied this methodology to the three early-to-late nebular phase spectra ($+128\,$d, $+138\,$d, and $+348\,$d) for \supernova{2019odp}.
The estimated line luminosities from the spectral line fitting are summarised in \autoref{tab:mod:neb:oxygen:fluxes} and the detailed fitting results (spectral line fits) can be found in \autoref{appendix:oxygen:spectralfits}.
The used models for spectral fitting are also outlined in \autoref{tab:mod:neb:oxygen:fluxes}.
For the $+348~$d spectrum, the $[\ion{O}{I}]\, \lambda 5577$ line was not detected, and thus the line luminosity should be taken as an upper limit.
Using the measured line luminosities, we %
estimated the physical parameters.

\subsubsection{Constraining the oxygen mass from nebular spectra: Results}
\label{sec:oxygen:results}

\begin{figure}
  \includegraphics[width=\linewidth]{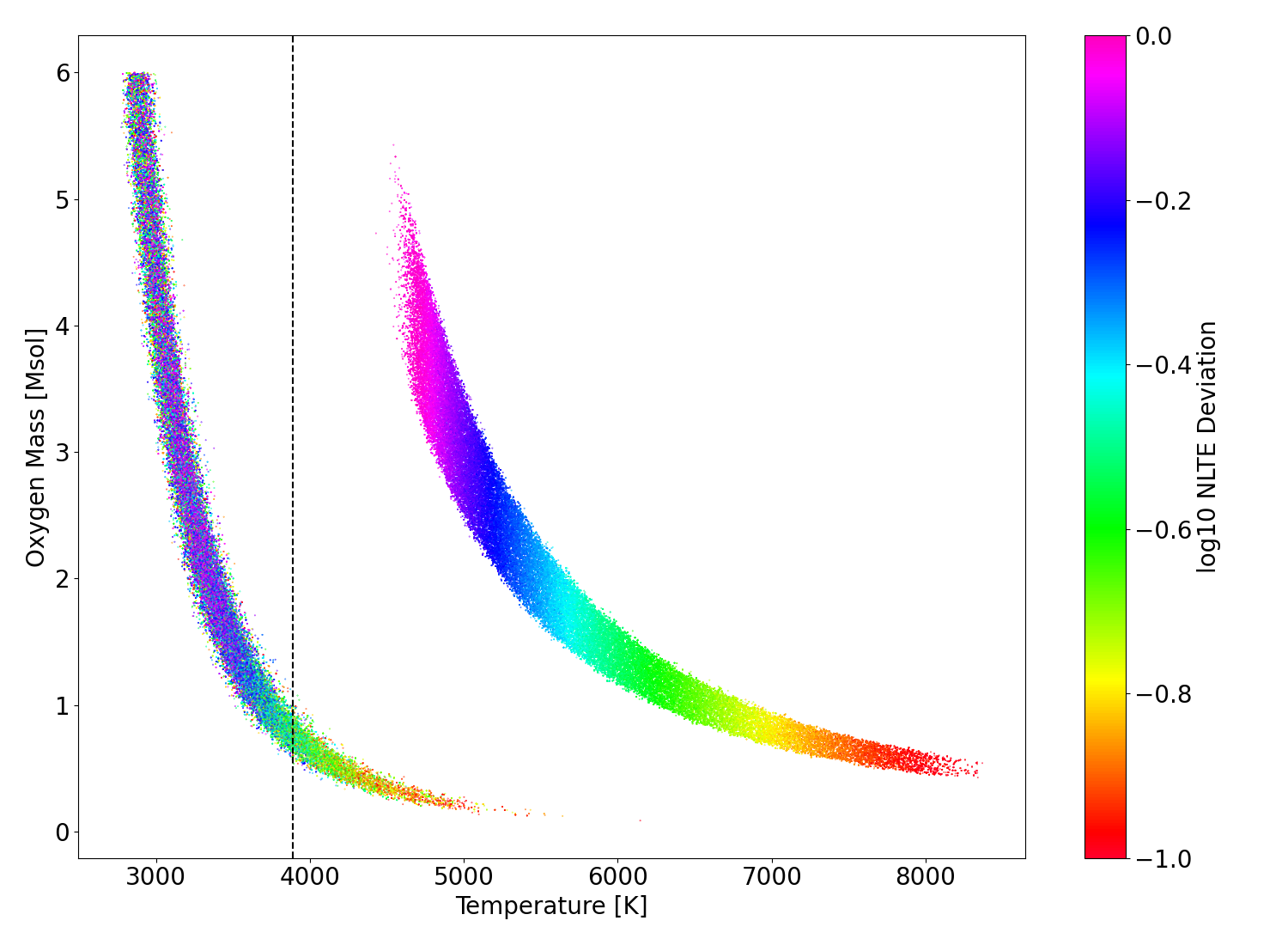}
  
  \caption{Scatter plot of the marginalised posterior distribution showing the oxygen mass as a function of the temperature for the last two observed spectra (the left-most trace corresponds to the very late time spectrum at $+348$\,d). The colour denotes the LTE departure coefficient. The estimated value range for the LTE departure coefficient from the $\ion{O}{I}\,\lambda 7774$ recombination line (Sect. \ref{sec:a:oxygen:nlte}) is marked in the colour bar. One can see that the earlier spectrum provides a lower mass constraint, while the late one does not. }
  \label{fig:mod:neb:oxygen:moit}
\end{figure}

The resulting posterior distributions for the $+138\,$d and $+348\,$d spectra are shown in Fig. \ref{fig:mod:neb:oxygen:moit}.

While the distribution is fairly broad, we can state without any additional constraints:
\begin{eqnarray}
  \min M_\text{OI}(T) &=& 0.47\,M_\odot \label{eqn:oxygen:result:strict}, \\
  \max M_\text{OI}(T) &=& 5\,M_\odot.
  \label{eqn:oxygen:result:strict:upper}
\end{eqnarray}
The lower limit on the oxygen mass corresponds to the high-temperature end of our prior.
The LTE departure coefficient, $d_2$, of $0.1$ corresponds to an electron number density, $n_e$, of around $10^7\,\text{cm}^{-3}$ \citep{2007ApJ...666.1069M}.
While Eq. \ref{eqn:oxygen:result:strict:upper} corresponds to a strict upper limit on the oxygen mass, it is only a strict limit on the light-emitting neutral oxygen mass and not on the total oxygen mass.

To further constrain the oxygen mass, one can try to roughly estimate the electron density (and thus LTE deviation) in the oxygen zone based on the observed $\ion{O}{I}\,\lambda 7774$ recombination line, which we do in Sect. \ref{sec:a:oxygen:nlte}.
The recombination line analysis points to an LTE deviation, $d_2$, in the range of $0.6$ to $0.72$ (or even higher in the case in which the oxygen zone is clumpy), which thus points towards an oxygen mass of
\begin{eqnarray}
  \min_{\text{rec}} M_\text{OI}(T) \approx 2.1\,M_\odot
  \label{eqn:oxygen:result:recomb}
.\end{eqnarray}
However, the analysis for this is less rigorous, 
which is why we have also kept the assumption-free limit.
One caveat to this is that the emission for the recombination line does not necessarily have to come from the same zone in the ejecta as the forbidden line emission (although in the spectra used for this analysis the spectral shape of the forbidden and recombination line match fairly well).

\subsubsection{Constraining the oxygen mass from nebular spectra: Zero-age main sequence mass estimate}
\label{sec:oxygen:discuss}

\begin{figure}
  \includegraphics[width=\linewidth]{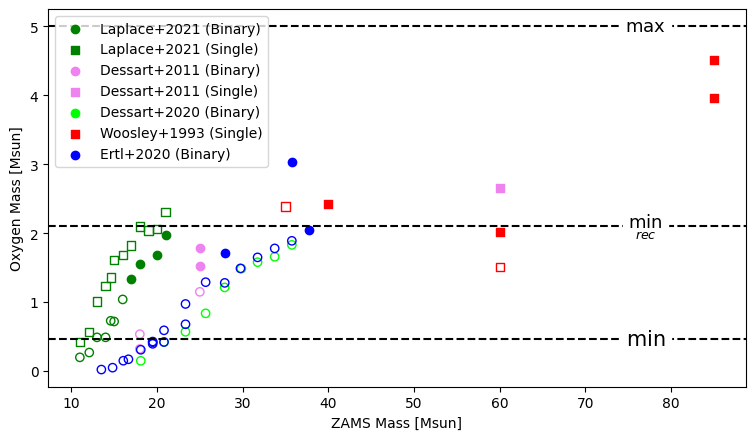}
  \caption{
    Relation between the supernova ejecta oxygen yield and the ZAMS mass of different stellar evolution and supernova explosion models.
    The lower dashed line denotes the constraint-free lower limit on the oxygen mass (Eq. \ref{eqn:oxygen:result:strict}),
    while the upper dashed line denotes the minimum oxygen mass assuming the NLTE conditions derived from the recombination line analysis (Eq. \ref{eqn:oxygen:result:recomb}).
    Binary evolution models are shown using circles and single massive star models are denoted by square markers.
    Models inconsistent with the estimated ejecta mass or not fully stripped of hydrogen are shown using unfilled symbols.
  }
  \label{fig:mod:neb:oxygen:litcompare}
\end{figure}

\begin{table*}[ht]
  \caption{ZAMS mass estimates using different model grids.}
  \begin{tabular}{|c|c|c|c|c|}
    \hline
    & & & \multicolumn{2}{|c|}{Consistent with oxygen mass} \\
    \textbf{Model Grid} & \textbf{Progenitor Channel} & \textbf{Grid Range} & \textbf{Strict Limit} & \textbf{Recombination Limit}  \\
    & & $(M_\odot)$ & $(M_\odot)$ & $(M_\odot)$  \\
    \hline
    \hline
    \cite{2021AA...656A..58L} & Binary  & $11-21$ & $17-21$ & $(>21)$  \\
    \cite{1993ApJ...411..823W} & Single & $35-85$ & $40-85$ & $60-85$  \\
    \cite{2011MNRAS.414.2985D} & Binary & $18,25$ & $25$ & $(>25)$ \\
    \cite{2011MNRAS.414.2985D} & Single & $25,60$ & $60$ & $60$ \\
    \cite{2020AA...642A.106D} & Binary  & $18-36$ & $(>36)$ & $(>36)$  \\
    \cite{2020ApJ...890...51E} & Binary & $13-38$ & $27-38$ & $36-38$ \\
    \hline
  \end{tabular}
  \tablefoot{The upper mass limit may be limited by the range of values in the model grid. Mass 
  limits within parentheses
  are extrapolated limits on the ZAMS mass, where none of the models were consistent with the ejecta mass as well as the oxygen mass. }
  \label{tab:mod:neb:oxygen:zamsgrids}
\end{table*}

In this section, we try to put the constraints we derived on the oxygen mass from the nebular spectroscopy in context with some stellar evolution models from the literature, since the ultimate goal is to derive ZAMS mass estimates for the progenitor star.
We have selected five different studies to highlight the complexity and range of results depending on methodology and assumptions, and we discuss these in conjunction with our derived properties for SN 2019odp.
As was outlined in Sect. \ref{sec:intro}, there are two primary progenitor channels to consider: single massive star and binary stars.
Several model grids for both progenitor channels are summarised in Fig. \ref{fig:mod:neb:oxygen:litcompare}, showing the well-known monotonic evolution of ejected oxygen mass versus ZAMS mass, with
both oxygen mass lower limits and the upper limit derived in Sect. \ref{sec:oxygen:results} marked using dashed lines.
Aside from consistency with the oxygen mass estimates, we also require that the models be roughly consistent with our estimates for the ejecta mass (Sect. \ref{sec:mod:phot:results}) and be mostly stripped of the hydrogen layers in the explosion (Sect. \ref{sec:evo:spec} and Sect. \ref{sec:classification}).
The models that are not consistent with these observables are shown using unfilled circles or squares 
in Fig. \ref{fig:mod:neb:oxygen:litcompare} for completeness.

As can be seen in Fig. \ref{fig:mod:neb:oxygen:litcompare}, there is a considerable spread in the ZAMS mass to oxygen mass relation (whereas the positive correlation is common to all model grids).
This can be attributed to the differing assumptions between the different model grids and the general complexity of stellar evolution and supernova explosion physics.
We thus estimate ZAMS mass ranges for each model grid separately and try to provide an extrapolated lower limit on the ZAMS mass in case the model grid contains no models that are massive enough.
This is summarised in \autoref{tab:mod:neb:oxygen:zamsgrids} for both of the derived lower limits on the oxygen mass for SN 2019odp.
Despite the above-mentioned shortcomings, we can use this discussion to constrain the ZAMS mass
based on our measurements.
In the binary scenario, we are restricted to a range above $17-38\,M_\odot$ for the strictest limit on the oxygen mass. For the 
$  \min_\text{rec} M_\text{OI}(T) = 2.1\,M_\odot $
limit, Fig. \ref{fig:mod:neb:oxygen:litcompare}
shows that only very massive stars are feasible.
For the single massive star scenario, the range permitted by our strictest oxygen mass is larger with $12-60\,M_\odot$, but for single star models
we can also exclude models below a ZAMS mass of $\sim 40\,M_\odot$, since this is currently believed to be the threshold to fully strip the hydrogen in this scenario.
The single mass models from \cite{2011MNRAS.414.2985D} can fulfil our strictest oxygen limit, but again only very massive stars are compatible with our highest limit on the ejected oxygen mass.
All evidence thus points to \supernova{2019odp} being the explosion of a massive star. We emphasise that although a binary scenario is not ruled out, SN 2019odp is a stripped envelope supernova with observables consistent with a massive single progenitor star.

{
  \renewcommand{\arraystretch}{1.4}
  \begin{table}
    \caption{Inferred line luminosities for the nebular phase spectra obtained.}
    \centering

    \begin{tabular}{|c|c|c|c|}
      \hline
       & Luminosity & Luminosity & Luminosity \\
      Phase & $+127.9$d & $+138.2$d & $+348.5$d \\
      & $(10^{38}\,\text{erg}\,\text{s}^{-1})$ & $(10^{38}\,\text{erg}\,\text{s}^{-1})$ & $(10^{38}\,\text{erg}\,\text{s}^{-1})$ \\
      \hline
      \hline
      Lineprofile $\rightarrow$  & 7774 & 7774 & Gaussian \\
      \hline
      $[\ion{O}{I}]\, \lambda 5577$ & $13.5_{-3.9}^{+7.9}$ & $26_{-15.3}^{+21.9}$ & $0.18_{-0.11}^{+0.58}$ \\
      $[\ion{O}{I}]\, \lambda 6300$ & $127_{-14.1}^{+29.5}$ & $129_{-19.8}^{+40.8}$ & $8_{-1.6}^{+4.1}$ \\
      $[\ion{O}{I}]\, \lambda 6364$ & $44_{-5.2}^{+10.9}$ & $44_{-7.5}^{+22.2}$ & $3_{-0.66}^{+2.33}$ \\
      \hline
    \end{tabular}
    
    \tablefoot{The distance uncertainty is folded in. The two line profile models are described in Sect. \ref{sec:oxygen:model}.}
    \label{tab:mod:neb:oxygen:fluxes}
  \end{table}
}

\section{Summary and conclusions}
\label{sec:summary}

We have presented photometric and spectroscopic observations of the \Ib{\supernova{2019odp}}.
Below, we list some of the main conclusions based on our analysis of these observations.
\begin{enumerate}
\item Based on the identification of helium absorption features in spectroscopic observations near the peak, we have reclassified \supernova{2019odp} as a \Ib{supernova} (Sect. \ref{sec:classification}).
  The pre-peak spectra, however, bear great spectral similarity to \IcBL{supernovae} as well as other transitional supernovae (\supernova{2008D}, \supernova{2016coi}, and \supernova{2017ens}).
  This may suggest a common scenario for all these supernovae early on.
\item Using optical photometric observations, we have constructed a (pseudo-)bolometric light curve (Sect. \ref{sec:evo:qbol}) and estimated the peak luminosity to be $4.4_{-0.7}^{+1} \times 10^{42} \ergs$ ($M_\text{qbol} \simeq -17.9$~mag).
  This corresponds to the bright end of the distribution for Type Ib supernovae and the faint end of the distribution for Type Ic-BL supernovae.
\item Using analytic bolometric light curve models, we estimate the \supernova{2019odp} ejecta mass to be $5.4\pm1.5~M_\odot$, with a nickel yield of $0.25\pm0.04~M_\odot$ and a kinetic energy of $6.4\pm1.5 \times 10^{51} \erg$ (Sect. \ref{sec:mod:phot:results}).
  These values are at the extreme end of values for Type Ib supernovae and more in line with values typically seen in Type Ic-BL supernovae.
\item We identify the presence of a pronounced plateau in the early light curve that is $2 - 5$~d in duration (Sect. \ref{sec:evo:phot}) and $2.1_{-0.4}^{+0.4} \times 10^{41} \ergs$ in luminosity (Sect. \ref{sec:evo:qbol}).
  Using analytic shock cooling models, we identify this plateau as the recombination plateau following an undetected shock breakout (Sect. \ref{sec:model:plateau}).
  Based on this, we estimate the progenitor radius to be $\sim 1~R_\odot$.
  Alternatively, the plateau may also be explained by insufficient nickel mixing during the explosion (Sect. \ref{sec:model:plateau}).
\item We expand upon existing methods for estimating oxygen masses from nebular spectra and describe both the improved model (Sect. \ref{sec:oxygen:model}) as well as the full Bayesian fitting methodology (Sect. \ref{sec:oxygen:fitting}).
  Using this new method, we derive a strict range of 
M$_\text{O,strict} = 0.5 - 5~M_\odot$ for \supernova{2019odp} (Sect. \ref{sec:oxygen:results}).
Based on studies on single massive star evolution, this points to a ZAMS mass of at least $40-60~M_\odot$ for binary star evolution to a ZAMS mass of at least $17-38~M_\odot$ (Sect. \ref{sec:oxygen:discuss}).
If one takes into account further information from the analysis of oxygen recombination lines (Sect. \ref{sec:a:oxygen:nlte}), one can further constrain the oxygen mass range, $M_\text{O,rec} = 2.1 - 5~M_\odot$.
In that case, low-mass binary models can be excluded and only binary models more massive than $M_\text{ZAMS} \approx 35~M_\odot$ are consistent.

\item Many observational sample papers \citep{2014MNRAS.437.3848L, 2018A&A...609A.136T,2021A&A...651A..81B} have found relatively low ejecta masses for Type Ibc supernovae, which in combination with stellar evolution models hint at low ZAMS masses, and 
therefore binary progenitors for those objects.
  For \supernova{2019odp}, we have deduced both a high ejecta mass from the diffusion phase light curve and a large mass of ejected oxygen from the nebular spectrum analysis, which is compatible with a single massive scenario.
  Thus, whereas we cannot rule out binary evolution for this supernova, SN 2019odp is one of the few stripped envelope supernovae for which such a scenario does not seem to be required.
  The existence of such objects is of importance for understanding the fate of the most massive stars.
\end{enumerate}

\begin{acknowledgements}

We thank Jakob Nordin and the anonymous referee for comments.
Based on observations obtained with the Samuel Oschin Telescope 48-inch and the 60-inch Telescope at the Palomar Observatory as part of the \textit{Zwicky} Transient Facility project.
ZTF is supported by the National Science Foundation under Grant No. AST-1440341 and a collaboration including Caltech, IPAC, the Weizmann Institute of Science, the Oskar Klein Center at Stockholm University, the University of Maryland, the University of Washington, Deutsches Elektronen-Synchrotron and Humboldt University, Los Alamos National Laboratories, the TANGO Consortium of Taiwan, the University of Wisconsin at Milwaukee, and Lawrence Berkeley National Laboratories.
Operations are conducted by COO, IPAC, and UW.
SED Machine is based upon work supported by the National Science Foundation under Grant No. 1106171.
This work was supported by the GROWTH project \citep{2019PASP..131c8003K} funded by the National Science Foundation under Grant No 1545949.
Part of the funding for GROND (both hardware as well as personnel) was generously granted from the Leibniz-Prize to Prof. G. Hasinger (DFG grant HA 1850/28-1).
Based on observations made with the Nordic Optical Telescope, owned in collaboration by the University of Turku and Aarhus University, and operated jointly by Aarhus University, the University of Turku and the University of Oslo, representing Denmark, Finland and Norway, the University of Iceland and Stockholm University at the Observatorio del Roque de los Muchachos, La Palma, Spain,
of the Instituto de Astrofisica de Canarias.
Some of the data presented here were obtained in part with ALFOSC, which is provided by the Instituto de Astrofisica de Andalucia (IAA) under a joint agreement with the University of Copenhagen and NOT.
Some of the data presented herein were obtained at the \textit{W. M. Keck Observatory}, which is operated as a scientific partnership among the California Institute of Technology, the University of California and the National Aeronautics and Space Administration.
The Observatory was made possible by the generous financial support of the \textit{W. M. Keck Foundation}.
M.~W.~Coughlin acknowledges support from the National Science Foundation with grant numbers PHY-2010970 and OAC-2117997.
P.~Rosnet acknowledges the support received from the Agence Nationale de la Recherche of the French government through the program ANR-21-CE31-0016-03.
This research has made use of the NASA/IPAC Infrared Science Archive, which is operated by the Jet Propulsion Laboratory, California Institute of Technology, under contract with the National Aeronautics and Space Administration.
This research has made use of NASA's Astrophysics Data System.
The acknowledgements were compiled using the Astronomy Acknowledgement Generator.
This research has made use of the NASA/IPAC Extragalactic Database (NED) which is operated by the Jet Propulsion Laboratory, California Institute of Technology, under contract with the National Aeronautics and Space Administration.
This research has made use of the VizieR catalogue access tool, CDS, Strasbourg, France.
This research made use of Astroquery \citep{2019AJ....157...98G}.
This research made use of Astropy, a community-developed core Python package for Astronomy \citep{2018AJ....156..123A, 2013A&A...558A..33A}.
PyRAF is a product of the Space Telescope Science Institute, which is operated by AURA for NASA.
This research made use of APLpy, an open-source plotting package for Python hosted at \url{http://aplpy.github.com}.
This research made use of SciPy \citep{Virtanen_2020}.
This research made use of NumPy \citep{harris2020array}.
IRAF is distributed by the National Optical Astronomy Observatory, which is operated by the Association of Universities for Research in Astronomy (AURA) under cooperative agreement with the National Science Foundation \citep{1993ASPC...52..173T}.
This research made use of matplotlib, a Python library for publication quality graphics \citep{Hunter:2007}.
This work made use of the IPython package \citep{PER-GRA:2007}.
We acknowledge the use of public data from the \textit{Swift} data archive. 

\end{acknowledgements}

\section*{Data availability}

The final reduced and flux-calibrated spectra are available on \texttt{WISeREP}\footnote{\url{https://wiserep.weizmann.ac.il/}} \citep{2012PASP..124..668Y}.
The code and (most) data reduction inputs are available at \url{https://gitlab.com/welterde/ccsn-sn2019odp} (with a snapshot archived on Zenodo at \url{https://zenodo.org/record/7568627}).
The data reduction products and data reduction diagnostics are available on Zenodo at \url{https://zenodo.org/record/7554926}.

\bibliography{combined}

\begin{thebibliography}{142}
\expandafter\ifx\csname natexlab\endcsname\relax\def\natexlab#1{#1}\fi

\bibitem[{{Aguado} {et~al.}(2019){Aguado}, {Ahumada}, {Almeida}, {Anderson},
  {Andrews}, {Anguiano}, {Aquino Ort{\'\i}z}, {Arag{\'o}n-Salamanca},
  {Argudo-Fern{\'a}ndez}, {Aubert}, {Avila-Reese}, {Badenes}, {Barboza
  Rembold}, {Barger}, {Barrera-Ballesteros}, {Bates}, {Bautista}, {Beaton},
  {Beers}, {Belfiore}, {Bernardi}, {Bershady}, {Beutler}, {Bird}, {Bizyaev},
  {Blanc}, {Blanton}, {Blomqvist}, {Bolton}, {Boquien}, {Borissova}, {Bovy},
  {Brandt}, {Brinkmann}, {Brownstein}, {Bundy}, {Burgasser}, {Byler}, {Cano
  Diaz}, {Cappellari}, {Carrera}, {Cervantes Sodi}, {Chen}, {Cherinka}, {Choi},
  {Chung}, {Coffey}, {Comerford}, {Comparat}, {Covey}, {da Silva Ilha}, {da
  Costa}, {Dai}, {Damke}, {Darling}, {Davies}, {Dawson}, {de Sainte Agathe},
  {Deconto Machado}, {Del Moro}, {De Lee}, {Diamond-Stanic}, {Dom{\'\i}nguez
  S{\'a}nchez}, {Donor}, {Drory}, {du Mas des Bourboux}, {Duckworth}, {Dwelly},
  {Ebelke}, {Emsellem}, {Escoffier}, {Fern{\'a}ndez-Trincado}, {Feuillet},
  {Fischer}, {Fleming}, {Fraser-McKelvie}, {Freischlad}, {Frinchaboy}, {Fu},
  {Galbany}, {Garcia-Dias}, {Garc{\'\i}a-Hern{\'a}ndez}, {Garma Oehmichen},
  {Geimba Maia}, {Gil-Mar{\'\i}n}, {Grabowski}, {Gu}, {Guo}, {Ha},
  {Harrington}, {Hasselquist}, {Hayes}, {Hearty}, {Hernandez Toledo}, {Hicks},
  {Hogg}, {Holley-Bockelmann}, {Holtzman}, {Hsieh}, {Hunt}, {Hwang},
  {Ibarra-Medel}, {Jimenez Angel}, {Johnson}, {Jones}, {J{\"o}nsson},
  {Kinemuchi}, {Kollmeier}, {Krawczyk}, {Kreckel}, {Kruk}, {Lacerna}, {Lan},
  {Lane}, {Law}, {Lee}, {Li}, {Lian}, {Lin}, {Lin}, {Lintott}, {Long},
  {Longa-Pe{\~n}a}, {Mackereth}, {de la Macorra}, {Majewski}, {Malanushenko},
  {Manchado}, {Maraston}, {Mariappan}, {Marinelli}, {Marques-Chaves},
  {Masseron}, {Masters}, {McDermid}, {Medina Pe{\~n}a}, {Meneses-Goytia},
  {Merloni}, {Merrifield}, {Meszaros}, {Minniti}, {Minsley}, {Muna}, {Myers},
  {Nair}, {Correa do Nascimento}, {Newman}, {Nitschelm}, {Olmstead}, {Oravetz},
  {Oravetz}, {Ortega Minakata}, {Pace}, {Padilla}, {Palicio}, {Pan}, {Pan},
  {Parikh}, {Parker}, {Peirani}, {Penny}, {Percival}, {Perez-Fournon},
  {Peterken}, {Pinsonneault}, {Prakash}, {Raddick}, {Raichoor}, {Riffel},
  {Riffel}, {Rix}, {Robin}, {Roman-Lopes}, {Rose}, {Ross}, {Rossi}, {Rowlands},
  {Rubin}, {S{\'a}nchez}, {S{\'a}nchez-Gallego}, {Sayres}, {Schaefer},
  {Schiavon}, {Schimoia}, {Schlafly}, {Schlegel}, {Schneider}, {Schultheis},
  {Seo}, {Shamsi}, {Shao}, {Shen}, {Shetty}, {Simonian}, {Smethurst}, {Sobeck},
  {Souter}, {Spindler}, {Stark}, {Stassun}, {Steinmetz}, {Storchi-Bergmann},
  {Stringfellow}, {Su{\'a}rez}, {Sun}, {Taghizadeh-Popp}, {Talbot}, {Tayar},
  {Thakar}, {Thomas}, {Tissera}, {Tojeiro}, {Troup}, {Unda-Sanzana},
  {Valenzuela}, {Vargas-Maga{\~n}a}, {V{\'a}zquez-Mata}, {Wake}, {Weaver},
  {Weijmans}, {Westfall}, {Wild}, {Wilson}, {Woods}, {Yan}, {Yang}, {Zamora},
  {Zasowski}, {Zhang}, {Zheng}, {Zheng}, {Zhu}, {Zinn}, \&
  {Zou}}]{2019ApJS..240...23A}
{Aguado}, D.~S., {Ahumada}, R., {Almeida}, A., {et~al.} 2019, \apjs, 240, 23

\bibitem[{{Arcavi}(2022)}]{2022ApJ...937...75A}
{Arcavi}, I. 2022, \apj, 937, 75

\bibitem[{{Arcavi} {et~al.}(2011){Arcavi}, {Gal-Yam}, {Yaron}, {Sternberg},
  {Rabinak}, {Waxman}, {Kasliwal}, {Quimby}, {Ofek}, {Horesh}, {Kulkarni},
  {Filippenko}, {Silverman}, {Cenko}, {Li}, {Bloom}, {Sullivan}, {Nugent},
  {Poznanski}, {Gorbikov}, {Fulton}, {Howell}, {Bersier}, {Riou},
  {Lamotte-Bailey}, {Griga}, {Cohen}, {Hachinger}, {Polishook}, {Xu},
  {Ben-Ami}, {Manulis}, {Walker}, {Maguire}, {Pan}, {Matheson}, {Mazzali},
  {Pian}, {Fox}, {Gehrels}, {Law}, {James}, {Marchant}, {Smith}, {Mottram},
  {Barnsley}, {Kandrashoff}, \& {Clubb}}]{2011ApJ...742L..18A}
{Arcavi}, I., {Gal-Yam}, A., {Yaron}, O., {et~al.} 2011, \apjl, 742, L18

\bibitem[{{Arcavi} {et~al.}(2017){Arcavi}, {Hosseinzadeh}, {Brown}, {Smartt},
  {Valenti}, {Tartaglia}, {Piro}, {Sanchez}, {Nicholls}, {Monard}, {Howell},
  {McCully}, {Sand}, {Tonry}, {Denneau}, {Stalder}, {Heinze}, {Rest}, {Smith},
  \& {Bishop}}]{2017ApJ...837L...2A}
{Arcavi}, I., {Hosseinzadeh}, G., {Brown}, P.~J., {et~al.} 2017, \apjl, 837, L2

\bibitem[{{Armstrong} {et~al.}(2021){Armstrong}, {Tucker}, {Rest},
  {Ridden-Harper}, {Zenati}, {Piro}, {Hinton}, {Lidman}, {Margheim}, {Narayan},
  {Shaya}, {Garnavich}, {Kasen}, {Villar}, {Zenteno}, {Arcavi}, {Drout},
  {Foley}, {Wheeler}, {Anais}, {Campillay}, {Coulter}, {Dimitriadis}, {Jones},
  {Kilpatrick}, {Mu{\~n}oz-Elgueta}, {Rojas-Bravo}, {Vargas-Gonz{\'a}lez},
  {Bulger}, {Chambers}, {Huber}, {Lowe}, {Magnier}, {Shappee}, {Smartt},
  {Smith}, {Barclay}, {Barentsen}, {Dotson}, {Gully-Santiago}, {Hedges},
  {Howell}, {Cody}, {Auchettl}, {B{\'o}di}, {Bogn{\'a}r}, {Brimacombe},
  {Brown}, {Cseh}, {Galbany}, {Hiramatsu}, {Holoien}, {Howell}, {Jha},
  {K{\"o}nyves-T{\'o}th}, {Kriskovics}, {McCully}, {Milne}, {Mu{\~n}oz}, {Pan},
  {P{\'a}l}, {Sai}, {S{\'a}rneczky}, {Smith}, {S{\'o}dor}, {Szab{\'o}},
  {Szak{\'a}ts}, {Valenti}, {Vink{\'o}}, {Wang}, {Zhang}, \&
  {Zsidi}}]{2021MNRAS.507.3125A}
{Armstrong}, P., {Tucker}, B.~E., {Rest}, A., {et~al.} 2021, \mnras, 507, 3125

\bibitem[{{Arnett}(1980)}]{1980ApJ...237..541A}
{Arnett}, W.~D. 1980, \apj, 237, 541

\bibitem[{{Arnett}(1982)}]{1982ApJ...253..785A}
{Arnett}, W.~D. 1982, \apj, 253, 785

\bibitem[{{Astropy Collaboration} {et~al.}(2018){Astropy Collaboration},
  {Price-Whelan}, {Sip{\H o}cz}, {G{\"u}nther}, {Lim}, {Crawford}, {Conseil},
  {Shupe}, {Craig}, {Dencheva}, {Ginsburg}, {VanderPlas}, {Bradley},
  {P{\'e}rez-Su{\'a}rez}, {de Val-Borro}, {Aldcroft}, {Cruz}, {Robitaille},
  {Tollerud}, {Ardelean}, {Babej}, {Bach}, {Bachetti}, {Bakanov}, {Bamford},
  {Barentsen}, {Barmby}, {Baumbach}, {Berry}, {Biscani}, {Boquien}, {Bostroem},
  {Bouma}, {Brammer}, {Bray}, {Breytenbach}, {Buddelmeijer}, {Burke},
  {Calderone}, {Cano Rodr{\'{\i}}guez}, {Cara}, {Cardoso}, {Cheedella},
  {Copin}, {Corrales}, {Crichton}, {D'Avella}, {Deil}, {Depagne}, {Dietrich},
  {Donath}, {Droettboom}, {Earl}, {Erben}, {Fabbro}, {Ferreira}, {Finethy},
  {Fox}, {Garrison}, {Gibbons}, {Goldstein}, {Gommers}, {Greco}, {Greenfield},
  {Groener}, {Grollier}, {Hagen}, {Hirst}, {Homeier}, {Horton}, {Hosseinzadeh},
  {Hu}, {Hunkeler}, {Ivezi{\'c}}, {Jain}, {Jenness}, {Kanarek}, {Kendrew},
  {Kern}, {Kerzendorf}, {Khvalko}, {King}, {Kirkby}, {Kulkarni}, {Kumar},
  {Lee}, {Lenz}, {Littlefair}, {Ma}, {Macleod}, {Mastropietro}, {McCully},
  {Montagnac}, {Morris}, {Mueller}, {Mumford}, {Muna}, {Murphy}, {Nelson},
  {Nguyen}, {Ninan}, {N{\"o}the}, {Ogaz}, {Oh}, {Parejko}, {Parley}, {Pascual},
  {Patil}, {Patil}, {Plunkett}, {Prochaska}, {Rastogi}, {Reddy Janga},
  {Sabater}, {Sakurikar}, {Seifert}, {Sherbert}, {Sherwood-Taylor}, {Shih},
  {Sick}, {Silbiger}, {Singanamalla}, {Singer}, {Sladen}, {Sooley},
  {Sornarajah}, {Streicher}, {Teuben}, {Thomas}, {Tremblay}, {Turner},
  {Terr{\'o}n}, {van Kerkwijk}, {de la Vega}, {Watkins}, {Weaver}, {Whitmore},
  {Woillez}, {Zabalza}, \& {Astropy Contributors}}]{2018AJ....156..123A}
{Astropy Collaboration}, {Price-Whelan}, A.~M., {Sip{\H o}cz}, B.~M., {et~al.}
  2018, \aj, 156, 123

\bibitem[{{Astropy Collaboration} {et~al.}(2013){Astropy Collaboration},
  {Robitaille}, {Tollerud}, {Greenfield}, {Droettboom}, {Bray}, {Aldcroft},
  {Davis}, {Ginsburg}, {Price-Whelan}, {Kerzendorf}, {Conley}, {Crighton},
  {Barbary}, {Muna}, {Ferguson}, {Grollier}, {Parikh}, {Nair}, {Unther},
  {Deil}, {Woillez}, {Conseil}, {Kramer}, {Turner}, {Singer}, {Fox}, {Weaver},
  {Zabalza}, {Edwards}, {Azalee Bostroem}, {Burke}, {Casey}, {Crawford},
  {Dencheva}, {Ely}, {Jenness}, {Labrie}, {Lim}, {Pierfederici}, {Pontzen},
  {Ptak}, {Refsdal}, {Servillat}, \& {Streicher}}]{2013A&A...558A..33A}
{Astropy Collaboration}, {Robitaille}, T.~P., {Tollerud}, E.~J., {et~al.} 2013,
  \aap, 558, A33

\bibitem[{{Barbarino} {et~al.}(2021){Barbarino}, {Sollerman}, {Taddia},
  {Fremling}, {Karamehmetoglu}, {Arcavi}, {Gal-Yam}, {Laher}, {Schulze},
  {Wozniak}, \& {Yan}}]{2021A&A...651A..81B}
{Barbarino}, C., {Sollerman}, J., {Taddia}, F., {et~al.} 2021, \aap, 651, A81

\bibitem[{Barbary(2016)}]{barbary_kyle_2016_804967}
Barbary, K. 2016, extinction v0.3.0

\bibitem[{{Becker}(2015)}]{2015ascl.soft04004B}
{Becker}, A. 2015, {HOTPANTS: High Order Transform of PSF ANd Template
  Subtraction}

\bibitem[{{Bellm} {et~al.}(2019{\natexlab{a}}){Bellm}, {Kulkarni}, {Barlow},
  {Feindt}, {Graham}, {Goobar}, {Kupfer}, {Ngeow}, {Nugent}, {Ofek}, {Prince},
  {Riddle}, {Walters}, \& {Ye}}]{2019PASP..131f8003B}
{Bellm}, E.~C., {Kulkarni}, S.~R., {Barlow}, T., {et~al.} 2019{\natexlab{a}},
  \pasp, 131, 068003

\bibitem[{{Bellm} {et~al.}(2019{\natexlab{b}}){Bellm}, {Kulkarni}, {Graham},
  {Dekany}, {Smith}, {Riddle}, {Masci}, {Helou}, {Prince}, {Adams},
  {Barbarino}, {Barlow}, {Bauer}, {Beck}, {Belicki}, {Biswas}, {Blagorodnova},
  {Bodewits}, {Bolin}, {Brinnel}, {Brooke}, {Bue}, {Bulla}, {Burruss}, {Cenko},
  {Chang}, {Connolly}, {Coughlin}, {Cromer}, {Cunningham}, {De}, {Delacroix},
  {Desai}, {Duev}, {Eadie}, {Farnham}, {Feeney}, {Feindt}, {Flynn},
  {Franckowiak}, {Frederick}, {Fremling}, {Gal-Yam}, {Gezari}, {Giomi},
  {Goldstein}, {Golkhou}, {Goobar}, {Groom}, {Hacopians}, {Hale}, {Henning},
  {Ho}, {Hover}, {Howell}, {Hung}, {Huppenkothen}, {Imel}, {Ip}, {Ivezi{\'c}},
  {Jackson}, {Jones}, {Juric}, {Kasliwal}, {Kaspi}, {Kaye}, {Kelley},
  {Kowalski}, {Kramer}, {Kupfer}, {Landry}, {Laher}, {Lee}, {Lin}, {Lin},
  {Lunnan}, {Giomi}, {Mahabal}, {Mao}, {Miller}, {Monkewitz}, {Murphy},
  {Ngeow}, {Nordin}, {Nugent}, {Ofek}, {Patterson}, {Penprase}, {Porter},
  {Rauch}, {Rebbapragada}, {Reiley}, {Rigault}, {Rodriguez}, {van Roestel},
  {Rusholme}, {van Santen}, {Schulze}, {Shupe}, {Singer}, {Soumagnac}, {Stein},
  {Surace}, {Sollerman}, {Szkody}, {Taddia}, {Terek}, {Van Sistine}, {van
  Velzen}, {Vestrand}, {Walters}, {Ward}, {Ye}, {Yu}, {Yan}, \&
  {Zolkower}}]{2019PASP..131a8002B}
{Bellm}, E.~C., {Kulkarni}, S.~R., {Graham}, M.~J., {et~al.}
  2019{\natexlab{b}}, \pasp, 131, 018002

\bibitem[{{Bengyat} \& {Gal-Yam}(2022)}]{2022ApJ...930...31B}
{Bengyat}, O. \& {Gal-Yam}, A. 2022, \apj, 930, 31

\bibitem[{{Bersten} {et~al.}(2014){Bersten}, {Benvenuto}, {Folatelli},
  {Nomoto}, {Kuncarayakti}, {Srivastav}, {Anupama}, {Quimby}, \&
  {Sahu}}]{2014AJ....148...68B}
{Bersten}, M.~C., {Benvenuto}, O.~G., {Folatelli}, G., {et~al.} 2014, \aj, 148,
  68

\bibitem[{{Bersten} {et~al.}(2018){Bersten}, {Folatelli}, {Garc{\'\i}a}, {van
  Dyk}, {Benvenuto}, {Orellana}, {Buso}, {S{\'a}nchez}, {Tanaka}, {Maeda},
  {Filippenko}, {Zheng}, {Brink}, {Cenko}, {de Jaeger}, {Kumar}, {Moriya},
  {Nomoto}, {Perley}, {Shivvers}, \& {Smith}}]{2018Natur.554..497B}
{Bersten}, M.~C., {Folatelli}, G., {Garc{\'\i}a}, F., {et~al.} 2018, \nat, 554,
  497

\bibitem[{{Bertin} {et~al.}(2002){Bertin}, {Mellier}, {Radovich}, {Missonnier},
  {Didelon}, \& {Morin}}]{2002ASPC..281..228B}
{Bertin}, E., {Mellier}, Y., {Radovich}, M., {et~al.} 2002, in Astronomical
  Society of the Pacific Conference Series, Vol. 281, Astronomical Data
  Analysis Software and Systems XI, ed. D.~A. {Bohlender}, D.~{Durand}, \&
  T.~H. {Handley}, 228

\bibitem[{{Bianco} {et~al.}(2014){Bianco}, {Modjaz}, {Hicken}, {Friedman},
  {Kirshner}, {Bloom}, {Challis}, {Marion}, {Wood-Vasey}, \&
  {Rest}}]{2014ApJS..213...19B}
{Bianco}, F.~B., {Modjaz}, M., {Hicken}, M., {et~al.} 2014, \apjs, 213, 19

\bibitem[{{Blagorodnova} {et~al.}(2018){Blagorodnova}, {Neill}, {Walters},
  {Kulkarni}, {Fremling}, {Ben-Ami}, {Dekany}, {Fucik}, {Konidaris}, {Nash},
  {Ngeow}, {Ofek}, {O' Sullivan}, {Quimby}, {Ritter}, \&
  {Vyhmeister}}]{2018PASP..130c5003B}
{Blagorodnova}, N., {Neill}, J.~D., {Walters}, R., {et~al.} 2018, \pasp, 130,
  035003

\bibitem[{{Blanton} \& {Roweis}(2007)}]{2007AJ....133..734B}
{Blanton}, M.~R. \& {Roweis}, S. 2007, \aj, 133, 734

\bibitem[{{Blondin} \& {Tonry}(2007)}]{2007ApJ...666.1024B}
{Blondin}, S. \& {Tonry}, J.~L. 2007, \apj, 666, 1024

\bibitem[{Bradley {et~al.}(2021)Bradley, Sipőcz, Robitaille, Tollerud,
  Vinícius, Deil, Barbary, Wilson, Busko, Donath, Günther, Cara, krachyon,
  Conseil, Bostroem, Droettboom, Bray, Lim, Bratholm, Barentsen, Craig, Rathi,
  Pascual, Perren, Georgiev, de~Val-Borro, Kerzendorf, Bach, Quint, \&
  Souchereau}]{larry_bradley_2021_5525286}
Bradley, L., Sipőcz, B., Robitaille, T., {et~al.} 2021, astropy/photutils:
  1.2.0

\bibitem[{{Breeveld} {et~al.}(2011){Breeveld}, {Landsman}, {Holland}, {Roming},
  {Kuin}, \& {Page}}]{2011AIPC.1358..373B}
{Breeveld}, A.~A., {Landsman}, W., {Holland}, S.~T., {et~al.} 2011, in American
  Institute of Physics Conference Series, Vol. 1358, Gamma Ray Bursts 2010, ed.
  J.~E. {McEnery}, J.~L. {Racusin}, \& N.~{Gehrels}, 373--376

\bibitem[{{Brennan} {et~al.}(2019){Brennan}, {Fraser}, {Ihanec}, {Gromadzki},
  \& {Yaron}}]{2019TNSCR1595....1B}
{Brennan}, S., {Fraser}, M., {Ihanec}, N., {Gromadzki}, M., \& {Yaron}, O.
  2019, Transient Name Server Classification Report, 2019-1595, 1

\bibitem[{{Brown} {et~al.}(2014){Brown}, {Breeveld}, {Holland}, {Kuin}, \&
  {Pritchard}}]{2014Ap&SS.354...89B}
{Brown}, P.~J., {Breeveld}, A.~A., {Holland}, S., {Kuin}, P., \& {Pritchard},
  T. 2014, \apss, 354, 89

\bibitem[{{Bruch} {et~al.}(2021){Bruch}, {Gal-Yam}, {Schulze}, {Yaron}, {Yang},
  {Soumagnac}, {Rigault}, {Strotjohann}, {Ofek}, {Sollerman}, {Masci},
  {Barbarino}, {Ho}, {Fremling}, {Perley}, {Nordin}, {Cenko}, {Adams},
  {Adreoni}, {Bellm}, {Blagorodnova}, {Bulla}, {Burdge}, {De}, {Dhawan},
  {Drake}, {Duev}, {Dugas}, {Graham}, {Graham}, {Irani}, {Jencson},
  {Karamehmetoglu}, {Kasliwal}, {Kim}, {Kulkarni}, {Kupfer}, {Liang},
  {Mahabal}, {Miller}, {Prince}, {Riddle}, {Sharma}, {Smith}, {Taddia},
  {Taggart}, {Walters}, \& {Yan}}]{2021ApJ...912...46B}
{Bruch}, R.~J., {Gal-Yam}, A., {Schulze}, S., {et~al.} 2021, \apj, 912, 46

\bibitem[{{Cardelli} {et~al.}(1989){Cardelli}, {Clayton}, \&
  {Mathis}}]{1989ApJ...345..245C}
{Cardelli}, J.~A., {Clayton}, G.~C., \& {Mathis}, J.~S. 1989, \apj, 345, 245

\bibitem[{{Chambers} {et~al.}(2016){Chambers}, {Magnier}, {Metcalfe},
  {Flewelling}, {Huber}, {Waters}, {Denneau}, {Draper}, {Farrow}, {Finkbeiner},
  {Holmberg}, {Koppenhoefer}, {Price}, {Rest}, {Saglia}, {Schlafly}, {Smartt},
  {Sweeney}, {Wainscoat}, {Burgett}, {Chastel}, {Grav}, {Heasley}, {Hodapp},
  {Jedicke}, {Kaiser}, {Kudritzki}, {Luppino}, {Lupton}, {Monet}, {Morgan},
  {Onaka}, {Shiao}, {Stubbs}, {Tonry}, {White}, {Ba{\~n}ados}, {Bell},
  {Bender}, {Bernard}, {Boegner}, {Boffi}, {Botticella}, {Calamida},
  {Casertano}, {Chen}, {Chen}, {Cole}, {Deacon}, {Frenk}, {Fitzsimmons},
  {Gezari}, {Gibbs}, {Goessl}, {Goggia}, {Gourgue}, {Goldman}, {Grant},
  {Grebel}, {Hambly}, {Hasinger}, {Heavens}, {Heckman}, {Henderson}, {Henning},
  {Holman}, {Hopp}, {Ip}, {Isani}, {Jackson}, {Keyes}, {Koekemoer}, {Kotak},
  {Le}, {Liska}, {Long}, {Lucey}, {Liu}, {Martin}, {Masci}, {McLean}, {Mindel},
  {Misra}, {Morganson}, {Murphy}, {Obaika}, {Narayan}, {Nieto-Santisteban},
  {Norberg}, {Peacock}, {Pier}, {Postman}, {Primak}, {Rae}, {Rai}, {Riess},
  {Riffeser}, {Rix}, {R{\"o}ser}, {Russel}, {Rutz}, {Schilbach}, {Schultz},
  {Scolnic}, {Strolger}, {Szalay}, {Seitz}, {Small}, {Smith}, {Soderblom},
  {Taylor}, {Thomson}, {Taylor}, {Thakar}, {Thiel}, {Thilker}, {Unger},
  {Urata}, {Valenti}, {Wagner}, {Walder}, {Walter}, {Watters}, {Werner},
  {Wood-Vasey}, \& {Wyse}}]{2016arXiv161205560C}
{Chambers}, K.~C., {Magnier}, E.~A., {Metcalfe}, N., {et~al.} 2016, arXiv
  e-prints, arXiv:1612.05560

\bibitem[{{Chatzopoulos} {et~al.}(2012){Chatzopoulos}, {Wheeler}, \&
  {Vinko}}]{2012ApJ...746..121C}
{Chatzopoulos}, E., {Wheeler}, J.~C., \& {Vinko}, J. 2012, \apj, 746, 121

\bibitem[{{Chen} {et~al.}(2018){Chen}, {Inserra}, {Fraser}, {Moriya}, {Schady},
  {Schweyer}, {Filippenko}, {Perley}, {Ruiter}, {Seitenzahl}, {Sollerman},
  {Taddia}, {Anderson}, {Foley}, {Jerkstrand}, {Ngeow}, {Pan}, {Pastorello},
  {Points}, {Smartt}, {Smith}, {Taubenberger}, {Wiseman}, {Young}, {Benetti},
  {Berton}, {Bufano}, {Clark}, {Della Valle}, {Galbany}, {Gal-Yam},
  {Gromadzki}, {Guti{\'e}rrez}, {Heinze}, {Kankare}, {Kilpatrick},
  {Kuncarayakti}, {Leloudas}, {Lin}, {Maguire}, {Mazzali}, {McBrien},
  {Prentice}, {Rau}, {Rest}, {Siebert}, {Stalder}, {Tonry}, \&
  {Yu}}]{2018ApJ...867L..31C}
{Chen}, T.~W., {Inserra}, C., {Fraser}, M., {et~al.} 2018, \apjl, 867, L31

\bibitem[{{Chevalier} \& {Fransson}(2008)}]{2008ApJ...683L.135C}
{Chevalier}, R.~A. \& {Fransson}, C. 2008, \apjl, 683, L135

\bibitem[{{Clocchiatti} {et~al.}(2011){Clocchiatti}, {Suntzeff}, {Covarrubias},
  \& {Candia}}]{2011AJ....141..163C}
{Clocchiatti}, A., {Suntzeff}, N.~B., {Covarrubias}, R., \& {Candia}, P. 2011,
  \aj, 141, 163

\bibitem[{{Contardo} {et~al.}(2000){Contardo}, {Leibundgut}, \&
  {Vacca}}]{2000A&A...359..876C}
{Contardo}, G., {Leibundgut}, B., \& {Vacca}, W.~D. 2000, \aap, 359, 876

\bibitem[{{Das} {et~al.}(2023){Das}, {Kasliwal}, {Fremling}, {Yang}, {Schulze},
  {Sollerman}, {Sit}, {De}, {Tzanidakis}, {Perley}, {Anand}, {Andreoni},
  {Barbarino}, {Brudge}, {Drake}, {Gal-Yam}, {Laher}, {Karambelkar},
  {Kulkarni}, {Masci}, {Medford}, {Polin}, {Reedy}, {Riddle}, {Sharma},
  {Smith}, {Yan}, {Yang}, \& {Yao}}]{2023ApJ...959...12D}
{Das}, K.~K., {Kasliwal}, M.~M., {Fremling}, C., {et~al.} 2023, \apj, 959, 12

\bibitem[{{Das} {et~al.}(2024){Das}, {Kasliwal}, {Sollerman}, {Fremling},
  {Irani}, {Leung}, {Yang}, {Wu}, {Fuller}, {Anand}, {Andreoni}, {Barbarino},
  {Brink}, {De}, {Dugas}, {Groom}, {Helou}, {Hinds}, {Ho}, {Karambelkar},
  {Kulkarni}, {Perley}, {Purdum}, {Regnault}, {Schulze}, {Sharma}, {Sit},
  {Sravan}, {Srinivasaragavan}, {Stein}, {Taggart}, {Tartaglia}, {Tzanidakis},
  {Wold}, {Yan}, {Yao}, \& {Zolkower}}]{2023arXiv230604698D}
{Das}, K.~K., {Kasliwal}, M.~M., {Sollerman}, J., {et~al.} 2024, \apj, 972, 91

\bibitem[{{Dekany} {et~al.}(2020){Dekany}, {Smith}, {Riddle}, {Feeney},
  {Porter}, {Hale}, {Zolkower}, {Belicki}, {Kaye}, {Henning}, {Walters},
  {Cromer}, {Delacroix}, {Rodriguez}, {Reiley}, {Mao}, {Hover}, {Murphy},
  {Burruss}, {Baker}, {Kowalski}, {Reif}, {Mueller}, {Bellm}, {Graham}, \&
  {Kulkarni}}]{2020PASP..132c8001D}
{Dekany}, R., {Smith}, R.~M., {Riddle}, R., {et~al.} 2020, \pasp, 132, 038001

\bibitem[{{Dessart} {et~al.}(2011){Dessart}, {Hillier}, {Livne}, {Yoon},
  {Woosley}, {Waldman}, \& {Langer}}]{2011MNRAS.414.2985D}
{Dessart}, L., {Hillier}, D.~J., {Livne}, E., {et~al.} 2011, \mnras, 414, 2985

\bibitem[{{Dessart} {et~al.}(2016){Dessart}, {Hillier}, {Woosley}, {Livne},
  {Waldman}, {Yoon}, \& {Langer}}]{2016MNRAS.458.1618D}
{Dessart}, L., {Hillier}, D.~J., {Woosley}, S., {et~al.} 2016, \mnras, 458,
  1618

\bibitem[{{Dessart} {et~al.}(2020){Dessart}, {Yoon}, {Aguilera-Dena}, \&
  {Langer}}]{2020AA...642A.106D}
{Dessart}, L., {Yoon}, S.-C., {Aguilera-Dena}, D.~R., \& {Langer}, N. 2020,
  \aap, 642, A106

\bibitem[{{Elmhamdi}(2011)}]{2011AcA....61..179E}
{Elmhamdi}, A. 2011, \actaa, 61, 179

\bibitem[{{Elmhamdi} {et~al.}(2004){Elmhamdi}, {Danziger}, {Cappellaro}, {Della
  Valle}, {Gouiffes}, {Phillips}, \& {Turatto}}]{2004A&A...426..963E}
{Elmhamdi}, A., {Danziger}, I.~J., {Cappellaro}, E., {et~al.} 2004, \aap, 426,
  963

\bibitem[{{Ergon} {et~al.}(2015){Ergon}, {Jerkstrand}, {Sollerman},
  {Elias-Rosa}, {Fransson}, {Fraser}, {Pastorello}, {Kotak}, {Taubenberger},
  {Tomasella}, {Valenti}, {Benetti}, {Helou}, {Kasliwal}, {Maund}, {Smartt}, \&
  {Spyromilio}}]{2015A&A...580A.142E}
{Ergon}, M., {Jerkstrand}, A., {Sollerman}, J., {et~al.} 2015, \aap, 580, A142

\bibitem[{{Ertl} {et~al.}(2020){Ertl}, {Woosley}, {Sukhbold}, \&
  {Janka}}]{2020ApJ...890...51E}
{Ertl}, T., {Woosley}, S.~E., {Sukhbold}, T., \& {Janka}, H.~T. 2020, \apj,
  890, 51

\bibitem[{{Fang} {et~al.}(2019){Fang}, {Maeda}, {Kuncarayakti}, {Sun}, \&
  {Gal-Yam}}]{2019NatAs...3..434F}
{Fang}, Q., {Maeda}, K., {Kuncarayakti}, H., {Sun}, F., \& {Gal-Yam}, A. 2019,
  Nature Astronomy, 3, 434

\bibitem[{{Filippenko}(1997)}]{1997ARA&A..35..309F}
{Filippenko}, A.~V. 1997, \araa, 35, 309

\bibitem[{{Foley} {et~al.}(2003){Foley}, {Papenkova}, {Swift}, {Filippenko},
  {Li}, {Mazzali}, {Chornock}, {Leonard}, \& {Van Dyk}}]{2003PASP..115.1220F}
{Foley}, R.~J., {Papenkova}, M.~S., {Swift}, B.~J., {et~al.} 2003, \pasp, 115,
  1220

\bibitem[{{Fransson} \& {Chevalier}(1989)}]{1989ApJ...343..323F}
{Fransson}, C. \& {Chevalier}, R.~A. 1989, \apj, 343, 323

\bibitem[{{Fremling} {et~al.}(2016){Fremling}, {Sollerman}, {Taddia}, {Ergon},
  {Fraser}, {Karamehmetoglu}, {Valenti}, {Jerkstrand}, {Arcavi}, {Bufano},
  {Elias Rosa}, {Filippenko}, {Fox}, {Gal-Yam}, {Howell}, {Kotak}, {Mazzali},
  {Milisavljevic}, {Nugent}, {Nyholm}, {Pian}, \&
  {Smartt}}]{2016A&A...593A..68F}
{Fremling}, C., {Sollerman}, J., {Taddia}, F., {et~al.} 2016, \aap, 593, A68

\bibitem[{{Gehrels} {et~al.}(2004){Gehrels}, {Chincarini}, {Giommi}, {Mason},
  {Nousek}, {Wells}, {White}, {Barthelmy}, {Burrows}, {Cominsky}, {Hurley},
  {Marshall}, {M{\'e}sz{\'a}ros}, {Roming}, {Angelini}, {Barbier}, {Belloni},
  {Campana}, {Caraveo}, {Chester}, {Citterio}, {Cline}, {Cropper}, {Cummings},
  {Dean}, {Feigelson}, {Fenimore}, {Frail}, {Fruchter}, {Garmire}, {Gendreau},
  {Ghisellini}, {Greiner}, {Hill}, {Hunsberger}, {Krimm}, {Kulkarni}, {Kumar},
  {Lebrun}, {Lloyd-Ronning}, {Markwardt}, {Mattson}, {Mushotzky}, {Norris},
  {Osborne}, {Paczynski}, {Palmer}, {Park}, {Parsons}, {Paul}, {Rees},
  {Reynolds}, {Rhoads}, {Sasseen}, {Schaefer}, {Short}, {Smale}, {Smith},
  {Stella}, {Tagliaferri}, {Takahashi}, {Tashiro}, {Townsley}, {Tueller},
  {Turner}, {Vietri}, {Voges}, {Ward}, {Willingale}, {Zerbi}, \&
  {Zhang}}]{2004ApJ...611.1005G}
{Gehrels}, N., {Chincarini}, G., {Giommi}, P., {et~al.} 2004, \apj, 611, 1005

\bibitem[{{Ginsburg} {et~al.}(2019){Ginsburg}, {Sip{\H o}cz}, {Brasseur},
  {Cowperthwaite}, {Craig}, {Deil}, {Guillochon}, {Guzman}, {Liedtke}, {Lian
  Lim}, {Lockhart}, {Mommert}, {Morris}, {Norman}, {Parikh}, {Persson},
  {Robitaille}, {Segovia}, {Singer}, {Tollerud}, {de Val-Borro}, {Valtchanov},
  {Woillez}, {Astroquery Collaboration}, \& {a subset of the astropy
  Collaboration}}]{2019AJ....157...98G}
{Ginsburg}, A., {Sip{\H o}cz}, B.~M., {Brasseur}, C.~E., {et~al.} 2019, \aj,
  157, 98

\bibitem[{{Graham} {et~al.}(2019){Graham}, {Kulkarni}, {Bellm}, {Adams},
  {Barbarino}, {Blagorodnova}, {Bodewits}, {Bolin}, {Brady}, {Cenko}, {Chang},
  {Coughlin}, {De}, {Eadie}, {Farnham}, {Feindt}, {Franckowiak}, {Fremling},
  {Gezari}, {Ghosh}, {Goldstein}, {Golkhou}, {Goobar}, {Ho}, {Huppenkothen},
  {Ivezi{\'c}}, {Jones}, {Juric}, {Kaplan}, {Kasliwal}, {Kelley}, {Kupfer},
  {Lee}, {Lin}, {Lunnan}, {Mahabal}, {Miller}, {Ngeow}, {Nugent}, {Ofek},
  {Prince}, {Rauch}, {van Roestel}, {Schulze}, {Singer}, {Sollerman}, {Taddia},
  {Yan}, {Ye}, {Yu}, {Barlow}, {Bauer}, {Beck}, {Belicki}, {Biswas}, {Brinnel},
  {Brooke}, {Bue}, {Bulla}, {Burruss}, {Connolly}, {Cromer}, {Cunningham},
  {Dekany}, {Delacroix}, {Desai}, {Duev}, {Feeney}, {Flynn}, {Frederick},
  {Gal-Yam}, {Giomi}, {Groom}, {Hacopians}, {Hale}, {Helou}, {Henning},
  {Hover}, {Hillenbrand}, {Howell}, {Hung}, {Imel}, {Ip}, {Jackson}, {Kaspi},
  {Kaye}, {Kowalski}, {Kramer}, {Kuhn}, {Landry}, {Laher}, {Mao}, {Masci},
  {Monkewitz}, {Murphy}, {Nordin}, {Patterson}, {Penprase}, {Porter},
  {Rebbapragada}, {Reiley}, {Riddle}, {Rigault}, {Rodriguez}, {Rusholme}, {van
  Santen}, {Shupe}, {Smith}, {Soumagnac}, {Stein}, {Surace}, {Szkody}, {Terek},
  {Van Sistine}, {van Velzen}, {Vestrand}, {Walters}, {Ward}, {Zhang}, \&
  {Zolkower}}]{2019PASP..131g8001G}
{Graham}, M.~J., {Kulkarni}, S.~R., {Bellm}, E.~C., {et~al.} 2019, \pasp, 131,
  078001

\bibitem[{{Greiner} {et~al.}(2008){Greiner}, {Bornemann}, {Clemens}, {Deuter},
  {Hasinger}, {Honsberg}, {Huber}, {Huber}, {Krauss}, {Kr{\"u}hler},
  {K{\"u}pc{\"u} Yolda{\textcommabelow s}}, {Mayer-Hasselwand er}, {Mican},
  {Primak}, {Schrey}, {Steiner}, {Szokoly}, {Th{\"o}ne}, {Yolda{\textcommabelow
  s}}, {Klose}, {Laux}, \& {Winkler}}]{2008PASP..120..405G}
{Greiner}, J., {Bornemann}, W., {Clemens}, C., {et~al.} 2008, \pasp, 120, 405

\bibitem[{Görtler {et~al.}(2019)Görtler, Kehlbeck, \& Deussen}]{gortler2019a}
Görtler, J., Kehlbeck, R., \& Deussen, O. 2019, Distill,
  https://distill.pub/2019/visual-exploration-gaussian-processes

\bibitem[{Harris {et~al.}(2020)Harris, Millman, van~der Walt, Gommers,
  Virtanen, Cournapeau, Wieser, Taylor, Berg, Smith, Kern, Picus, Hoyer, van
  Kerkwijk, Brett, Haldane, del R{'{\i}}o, Wiebe, Peterson,
  G{'{e}}rard-Marchant, Sheppard, Reddy, Weckesser, Abbasi, Gohlke, \&
  Oliphant}]{harris2020array}
Harris, C.~R., Millman, K.~J., van~der Walt, S.~J., {et~al.} 2020, Nature, 585,
  357

\bibitem[{{Heger} \& {Woosley}(2010)}]{2010ApJ...724..341H}
{Heger}, A. \& {Woosley}, S.~E. 2010, \apj, 724, 341

\bibitem[{{Higson} {et~al.}(2019){Higson}, {Handley}, {Hobson}, \&
  {Lasenby}}]{2019S&C....29..891H}
{Higson}, E., {Handley}, W., {Hobson}, M., \& {Lasenby}, A. 2019, Statistics
  and Computing, 29, 891

\bibitem[{{Hirschi} {et~al.}(2005){Hirschi}, {Meynet}, \&
  {Maeder}}]{2005A&A...433.1013H}
{Hirschi}, R., {Meynet}, G., \& {Maeder}, A. 2005, \aap, 433, 1013

\bibitem[{{Houck} \& {Fransson}(1996)}]{1996ApJ...456..811H}
{Houck}, J.~C. \& {Fransson}, C. 1996, \apj, 456, 811

\bibitem[{Hunter(2007)}]{Hunter:2007}
Hunter, J.~D. 2007, Computing In Science \& Engineering, 9, 90

\bibitem[{{Jerkstrand}(2017)}]{2017hsn..book..795J}
{Jerkstrand}, A. 2017, in Handbook of Supernovae, ed. A.~W. {Alsabti} \&
  P.~{Murdin}, 795

\bibitem[{{Jerkstrand} {et~al.}(2015){Jerkstrand}, {Ergon}, {Smartt},
  {Fransson}, {Sollerman}, {Taubenberger}, {Bersten}, \&
  {Spyromilio}}]{2015A&A...573A..12J}
{Jerkstrand}, A., {Ergon}, M., {Smartt}, S.~J., {et~al.} 2015, \aap, 573, A12

\bibitem[{{Jerkstrand} {et~al.}(2014){Jerkstrand}, {Smartt}, {Fraser},
  {Fransson}, {Sollerman}, {Taddia}, \& {Kotak}}]{2014MNRAS.439.3694J}
{Jerkstrand}, A., {Smartt}, S.~J., {Fraser}, M., {et~al.} 2014, \mnras, 439,
  3694

\bibitem[{{Jerkstrand} {et~al.}(2017){Jerkstrand}, {Smartt}, {Inserra},
  {Nicholl}, {Chen}, {Kr{\"u}hler}, {Sollerman}, {Taubenberger}, {Gal-Yam},
  {Kankare}, {Maguire}, {Fraser}, {Valenti}, {Sullivan}, {Cartier}, \&
  {Young}}]{2017ApJ...835...13J}
{Jerkstrand}, A., {Smartt}, S.~J., {Inserra}, C., {et~al.} 2017, \apj, 835, 13

\bibitem[{{Karamehmetoglu} {et~al.}(2023){Karamehmetoglu}, {Sollerman},
  {Taddia}, {Barbarino}, {Feindt}, {Fremling}, {Gal-Yam}, {Kasliwal},
  {Petrushevska}, {Schulze}, {Stritzinger}, \&
  {Zapartas}}]{2023A&A...678A..87K}
{Karamehmetoglu}, E., {Sollerman}, J., {Taddia}, F., {et~al.} 2023, \aap, 678,
  A87

\bibitem[{{Kasliwal} {et~al.}(2019){Kasliwal}, {Cannella}, {Bagdasaryan},
  {Hung}, {Feindt}, {Singer}, {Coughlin}, {Fremling}, {Walters}, {Duev},
  {Itoh}, \& {Quimby}}]{2019PASP..131c8003K}
{Kasliwal}, M.~M., {Cannella}, C., {Bagdasaryan}, A., {et~al.} 2019, \pasp,
  131, 038003

\bibitem[{{Khatami} \& {Kasen}(2019)}]{2019ApJ...878...56K}
{Khatami}, D.~K. \& {Kasen}, D.~N. 2019, \apj, 878, 56

\bibitem[{{Kilpatrick} {et~al.}(2021){Kilpatrick}, {Drout}, {Auchettl},
  {Dimitriadis}, {Foley}, {Jones}, {DeMarchi}, {French}, {Gall}, {Hjorth},
  {Jacobson-Gal{\'a}n}, {Margutti}, {Piro}, {Ramirez-Ruiz}, {Rest}, \&
  {Rojas-Bravo}}]{2021MNRAS.504.2073K}
{Kilpatrick}, C.~D., {Drout}, M.~R., {Auchettl}, K., {et~al.} 2021, \mnras,
  504, 2073

\bibitem[{Kramida {et~al.}(2021)Kramida, {Yu.~Ralchenko}, Reader, \& {and NIST
  ASD Team}}]{NIST_ASD}
Kramida, A., {Yu.~Ralchenko}, Reader, J., \& {and NIST ASD Team}. 2021, {NIST
  Atomic Spectra Database (ver. 5.9), [Online]. Available:
  {\tt{https://physics.nist.gov/asd}} [2022, March 17]. National Institute of
  Standards and Technology, Gaithersburg, MD.}

\bibitem[{{Kr{\"u}hler} {et~al.}(2008){Kr{\"u}hler}, {K{\"u}pc{\"u}
  Yolda{\c{s}}}, {Greiner}, {Clemens}, {McBreen}, {Primak}, {Savaglio},
  {Yolda{\c{s}}}, {Szokoly}, \& {Klose}}]{2008ApJ...685..376K}
{Kr{\"u}hler}, T., {K{\"u}pc{\"u} Yolda{\c{s}}}, A., {Greiner}, J., {et~al.}
  2008, \apj, 685, 376

\bibitem[{{Kumar} {et~al.}(2022){Kumar}, {Singh}, {Sahu}, \&
  {Anupama}}]{2022ApJ...927...61K}
{Kumar}, B., {Singh}, A., {Sahu}, D.~K., \& {Anupama}, G.~C. 2022, \apj, 927,
  61

\bibitem[{{Kuncarayakti} {et~al.}(2015){Kuncarayakti}, {Maeda}, {Bersten},
  {Folatelli}, {Morrell}, {Hsiao}, {Gonz{\'a}lez-Gait{\'a}n}, {Anderson},
  {Hamuy}, {de Jaeger}, {Guti{\'e}rrez}, \& {Kawabata}}]{2015A&A...579A..95K}
{Kuncarayakti}, H., {Maeda}, K., {Bersten}, M.~C., {et~al.} 2015, \aap, 579,
  A95

\bibitem[{{Laplace} {et~al.}(2021){Laplace}, {Justham}, {Renzo}, {G{\"o}tberg},
  {Farmer}, {Vartanyan}, \& {de Mink}}]{2021AA...656A..58L}
{Laplace}, E., {Justham}, S., {Renzo}, M., {et~al.} 2021, \aap, 656, A58

\bibitem[{{Li} \& {McCray}(1992)}]{1992ApJ...387..309L}
{Li}, H. \& {McCray}, R. 1992, \apj, 387, 309

\bibitem[{{Limongi} \& {Chieffi}(2003)}]{2003ApJ...592..404L}
{Limongi}, M. \& {Chieffi}, A. 2003, \apj, 592, 404

\bibitem[{{Liu} {et~al.}(2018){Liu}, {Zhang}, {Wang}, \&
  {Dai}}]{2018ApJ...868L..24L}
{Liu}, L.-D., {Zhang}, B., {Wang}, L.-J., \& {Dai}, Z.-G. 2018, \apjl, 868, L24

\bibitem[{{Lucy}(1991)}]{1991ApJ...383..308L}
{Lucy}, L.~B. 1991, \apj, 383, 308

\bibitem[{{Lunnan} {et~al.}(2016){Lunnan}, {Chornock}, {Berger},
  {Milisavljevic}, {Jones}, {Rest}, {Fong}, {Fransson}, {Margutti}, {Drout},
  {Blanchard}, {Challis}, {Cowperthwaite}, {Foley}, {Kirshner}, {Morrell},
  {Riess}, {Roth}, {Scolnic}, {Smartt}, {Smith}, {Villar}, {Chambers},
  {Draper}, {Huber}, {Kaiser}, {Kudritzki}, {Magnier}, {Metcalfe}, \&
  {Waters}}]{2016ApJ...831..144L}
{Lunnan}, R., {Chornock}, R., {Berger}, E., {et~al.} 2016, \apj, 831, 144

\bibitem[{{Lyman} {et~al.}(2014){Lyman}, {Bersier}, \&
  {James}}]{2014MNRAS.437.3848L}
{Lyman}, J.~D., {Bersier}, D., \& {James}, P.~A. 2014, \mnras, 437, 3848

\bibitem[{{Lyman} {et~al.}(2016){Lyman}, {Bersier}, {James}, {Mazzali},
  {Eldridge}, {Fraser}, \& {Pian}}]{2016MNRAS.457..328L}
{Lyman}, J.~D., {Bersier}, D., {James}, P.~A., {et~al.} 2016, \mnras, 457, 328

\bibitem[{{Maeda} {et~al.}(2007){Maeda}, {Tanaka}, {Nomoto}, {Tominaga},
  {Kawabata}, {Mazzali}, {Umeda}, {Suzuki}, \& {Hattori}}]{2007ApJ...666.1069M}
{Maeda}, K., {Tanaka}, M., {Nomoto}, K., {et~al.} 2007, \apj, 666, 1069

\bibitem[{{Masci} {et~al.}(2019){Masci}, {Laher}, {Rusholme}, {Shupe}, {Groom},
  {Surace}, {Jackson}, {Monkewitz}, {Beck}, {Flynn}, {Terek}, {Landry},
  {Hacopians}, {Desai}, {Howell}, {Brooke}, {Imel}, {Wachter}, {Ye}, {Lin},
  {Cenko}, {Cunningham}, {Rebbapragada}, {Bue}, {Miller}, {Mahabal}, {Bellm},
  {Patterson}, {Juri{\'c}}, {Golkhou}, {Ofek}, {Walters}, {Graham}, {Kasliwal},
  {Dekany}, {Kupfer}, {Burdge}, {Cannella}, {Barlow}, {Van Sistine}, {Giomi},
  {Fremling}, {Blagorodnova}, {Levitan}, {Riddle}, {Smith}, {Helou}, {Prince},
  \& {Kulkarni}}]{2019PASP..131a8003M}
{Masci}, F.~J., {Laher}, R.~R., {Rusholme}, B., {et~al.} 2019, \pasp, 131,
  018003

\bibitem[{{Mazzali} {et~al.}(2002){Mazzali}, {Deng}, {Maeda}, {Nomoto},
  {Umeda}, {Hatano}, {Iwamoto}, {Yoshii}, {Kobayashi}, {Minezaki}, {Doi},
  {Enya}, {Tomita}, {Smartt}, {Kinugasa}, {Kawakita}, {Ayani}, {Kawabata},
  {Yamaoka}, {Qiu}, {Motohara}, {Gerardy}, {Fesen}, {Kawabata}, {Iye},
  {Kashikawa}, {Kosugi}, {Ohyama}, {Takada-Hidai}, {Zhao}, {Chornock},
  {Filippenko}, {Benetti}, \& {Turatto}}]{2002ApJ...572L..61M}
{Mazzali}, P.~A., {Deng}, J., {Maeda}, K., {et~al.} 2002, \apjl, 572, L61

\bibitem[{{Mazzali} {et~al.}(2001){Mazzali}, {Nomoto}, {Patat}, \&
  {Maeda}}]{2001ApJ...559.1047M}
{Mazzali}, P.~A., {Nomoto}, K., {Patat}, F., \& {Maeda}, K. 2001, \apj, 559,
  1047

\bibitem[{{Medler} {et~al.}(2022){Medler}, {Mazzali}, {Teffs}, {Ashall},
  {Anderson}, {Arcavi}, {Benetti}, {Bostroem}, {Burke}, {Cai},
  {Charalampopoulos}, {Elias-Rosa}, {Ergon}, {Galbany}, {Gromadzki},
  {Hiramatsu}, {Howell}, {Inserra}, {Lundqvist}, {McCully}, {M{\"u}ller-Bravo},
  {Newsome}, {Nicholl}, {Padilla Gonzalez}, {Paraskeva}, {Pastorello},
  {Pellegrino}, {Pessi}, {Reguitti}, {Reynolds}, {Roy}, {Terreran},
  {Tomasella}, \& {Young}}]{2022MNRAS.513.5540M}
{Medler}, K., {Mazzali}, P.~A., {Teffs}, J., {et~al.} 2022, \mnras, 513, 5540

\bibitem[{{Modjaz} {et~al.}(2014){Modjaz}, {Blondin}, {Kirshner}, {Matheson},
  {Berlind}, {Bianco}, {Calkins}, {Challis}, {Garnavich}, {Hicken}, {Jha},
  {Liu}, \& {Marion}}]{2014AJ....147...99M}
{Modjaz}, M., {Blondin}, S., {Kirshner}, R.~P., {et~al.} 2014, \aj, 147, 99

\bibitem[{{Modjaz} {et~al.}(2019){Modjaz}, {Guti{\'e}rrez}, \&
  {Arcavi}}]{2019NatAs...3..717M}
{Modjaz}, M., {Guti{\'e}rrez}, C.~P., \& {Arcavi}, I. 2019, Nature Astronomy,
  3, 717

\bibitem[{{Modjaz} {et~al.}(2009){Modjaz}, {Li}, {Butler}, {Chornock},
  {Perley}, {Blondin}, {Bloom}, {Filippenko}, {Kirshner}, {Kocevski},
  {Poznanski}, {Hicken}, {Foley}, {Stringfellow}, {Berlind}, {Barrado y
  Navascues}, {Blake}, {Bouy}, {Brown}, {Challis}, {Chen}, {de Vries},
  {Dufour}, {Falco}, {Friedman}, {Ganeshalingam}, {Garnavich}, {Holden},
  {Illingworth}, {Lee}, {Liebert}, {Marion}, {Olivier}, {Prochaska},
  {Silverman}, {Smith}, {Starr}, {Steele}, {Stockton}, {Williams}, \&
  {Wood-Vasey}}]{2009ApJ...702..226M}
{Modjaz}, M., {Li}, W., {Butler}, N., {et~al.} 2009, \apj, 702, 226

\bibitem[{{Modjaz} {et~al.}(2016){Modjaz}, {Liu}, {Bianco}, \&
  {Graur}}]{2016ApJ...832..108M}
{Modjaz}, M., {Liu}, Y.~Q., {Bianco}, F.~B., \& {Graur}, O. 2016, \apj, 832,
  108

\bibitem[{{Mould} {et~al.}(2000){Mould}, {Huchra}, {Freedman}, {Kennicutt},
  {Ferrarese}, {Ford}, {Gibson}, {Graham}, {Hughes}, {Illingworth}, {Kelson},
  {Macri}, {Madore}, {Sakai}, {Sebo}, {Silbermann}, \&
  {Stetson}}]{2000ApJ...529..786M}
{Mould}, J.~R., {Huchra}, J.~P., {Freedman}, W.~L., {et~al.} 2000, \apj, 529,
  786

\bibitem[{{Nadyozhin}(1994)}]{1994ApJS...92..527N}
{Nadyozhin}, D.~K. 1994, \apjs, 92, 527

\bibitem[{{Nakar} \& {Sari}(2010)}]{2010ApJ...725..904N}
{Nakar}, E. \& {Sari}, R. 2010, \apj, 725, 904

\bibitem[{{National Optical Astronomy
  Observatories}(1999)}]{1999ascl.soft11002N}
{National Optical Astronomy Observatories}. 1999, {IRAF: Image Reduction and
  Analysis Facility}, Astrophysics Source Code Library, record ascl:9911.002

\bibitem[{{Nordin} {et~al.}(2019){Nordin}, {Brinnel}, {Giomi}, {Santen},
  {Gal-Yam}, {Yaron}, \& {Schulze}}]{2019TNSTR1585....1N}
{Nordin}, J., {Brinnel}, V., {Giomi}, M., {et~al.} 2019, Transient Name Server
  Discovery Report, 2019-1585, 1

\bibitem[{{Oke} {et~al.}(1995){Oke}, {Cohen}, {Carr}, {Cromer}, {Dingizian},
  {Harris}, {Labrecque}, {Lucinio}, {Schaal}, {Epps}, \&
  {Miller}}]{1995PASP..107..375O}
{Oke}, J.~B., {Cohen}, J.~G., {Carr}, M., {et~al.} 1995, \pasp, 107, 375

\bibitem[{{Oke} \& {Gunn}(1982)}]{1982PASP...94..586O}
{Oke}, J.~B. \& {Gunn}, J.~E. 1982, \pasp, 94, 586

\bibitem[{{Patat} {et~al.}(2001){Patat}, {Cappellaro}, {Danziger}, {Mazzali},
  {Sollerman}, {Augusteijn}, {Brewer}, {Doublier}, {Gonzalez}, {Hainaut},
  {Lidman}, {Leibundgut}, {Nomoto}, {Nakamura}, {Spyromilio}, {Rizzi},
  {Turatto}, {Walsh}, {Galama}, {van Paradijs}, {Kouveliotou}, {Vreeswijk},
  {Frontera}, {Masetti}, {Palazzi}, \& {Pian}}]{2001ApJ...555..900P}
{Patat}, F., {Cappellaro}, E., {Danziger}, J., {et~al.} 2001, \apj, 555, 900

\bibitem[{P\'erez \& Granger(2007)}]{PER-GRA:2007}
P\'erez, F. \& Granger, B.~E. 2007, Computing in Science and Engineering, 9, 21

\bibitem[{{Perley}(2019)}]{2019PASP..131h4503P}
{Perley}, D.~A. 2019, \pasp, 131, 084503

\bibitem[{{Piro} {et~al.}(2021){Piro}, {Haynie}, \&
  {Yao}}]{2021ApJ...909..209P}
{Piro}, A.~L., {Haynie}, A., \& {Yao}, Y. 2021, \apj, 909, 209

\bibitem[{{Piro} \& {Nakar}(2013)}]{2013ApJ...769...67P}
{Piro}, A.~L. \& {Nakar}, E. 2013, \apj, 769, 67

\bibitem[{{Prentice} {et~al.}(2018){Prentice}, {Ashall}, {Mazzali}, {Zhang},
  {James}, {Wang}, {Vink{\'o}}, {Percival}, {Short}, {Piascik}, {Huang}, {Mo},
  {Rui}, {Wang}, {Xiang}, {Xin}, {Yi}, {Yu}, {Zhai}, {Zhang}, {Hosseinzadeh},
  {Howell}, {McCully}, {Valenti}, {Cseh}, {Hanyecz}, {Kriskovics}, {P{\'a}l},
  {S{\'a}rneczky}, {S{\'o}dor}, {Szak{\'a}ts}, {Sz{\'e}kely},
  {Varga-Vereb{\'e}lyi}, {Vida}, {Bradac}, {Reichart}, {Sand}, \&
  {Tartaglia}}]{2018MNRAS.478.4162P}
{Prentice}, S.~J., {Ashall}, C., {Mazzali}, P.~A., {et~al.} 2018, \mnras, 478,
  4162

\bibitem[{{Prentice} {et~al.}(2022){Prentice}, {Maguire}, {Siebenaler}, \&
  {Jerkstrand}}]{2022MNRAS.514.5686P}
{Prentice}, S.~J., {Maguire}, K., {Siebenaler}, L., \& {Jerkstrand}, A. 2022,
  \mnras, 514, 5686

\bibitem[{{Prentice} \& {Mazzali}(2017)}]{2017MNRAS.469.2672P}
{Prentice}, S.~J. \& {Mazzali}, P.~A. 2017, \mnras, 469, 2672

\bibitem[{{Puls} {et~al.}(2008){Puls}, {Vink}, \&
  {Najarro}}]{2008A&ARv..16..209P}
{Puls}, J., {Vink}, J.~S., \& {Najarro}, F. 2008, \aapr, 16, 209

\bibitem[{{Rabinak} \& {Waxman}(2011)}]{2011ApJ...728...63R}
{Rabinak}, I. \& {Waxman}, E. 2011, \apj, 728, 63

\bibitem[{{Rigault} {et~al.}(2019){Rigault}, {Neill}, {Blagorodnova}, {Dugas},
  {Feeney}, {Walters}, {Brinnel}, {Copin}, {Fremling}, {Nordin}, \&
  {Sollerman}}]{2019AA...627A.115R}
{Rigault}, M., {Neill}, J.~D., {Blagorodnova}, N., {et~al.} 2019, \aap, 627,
  A115

\bibitem[{{Roming} {et~al.}(2005){Roming}, {Kennedy}, {Mason}, {Nousek}, {Ahr},
  {Bingham}, {Broos}, {Carter}, {Hancock}, {Huckle}, {Hunsberger}, {Kawakami},
  {Killough}, {Koch}, {McLelland}, {Smith}, {Smith}, {Soto}, {Boyd},
  {Breeveld}, {Holland}, {Ivanushkina}, {Pryzby}, {Still}, \&
  {Stock}}]{2005SSRv..120...95R}
{Roming}, P. W.~A., {Kennedy}, T.~E., {Mason}, K.~O., {et~al.} 2005, \ssr, 120,
  95

\bibitem[{{Sahu} {et~al.}(2011){Sahu}, {Gurugubelli}, {Anupama}, \&
  {Nomoto}}]{2011MNRAS.413.2583S}
{Sahu}, D.~K., {Gurugubelli}, U.~K., {Anupama}, G.~C., \& {Nomoto}, K. 2011,
  \mnras, 413, 2583

\bibitem[{{Schlafly} \& {Finkbeiner}(2011)}]{2011ApJ...737..103S}
{Schlafly}, E.~F. \& {Finkbeiner}, D.~P. 2011, \apj, 737, 103

\bibitem[{{Schneider} {et~al.}(1990){Schneider}, {Thuan}, {Magri}, \&
  {Wadiak}}]{1990ApJS...72..245S}
{Schneider}, S.~E., {Thuan}, T.~X., {Magri}, C., \& {Wadiak}, J.~E. 1990,
  \apjs, 72, 245

\bibitem[{{Science Software Branch at STScI}(2012)}]{2012ascl.soft07011S}
{Science Software Branch at STScI}. 2012, {PyRAF: Python alternative for IRAF},
  Astrophysics Source Code Library, record ascl:1207.011

\bibitem[{{Shivvers} {et~al.}(2019){Shivvers}, {Filippenko}, {Silverman},
  {Zheng}, {Foley}, {Chornock}, {Barth}, {Cenko}, {Clubb}, {Fox},
  {Ganeshalingam}, {Graham}, {Kelly}, {Kleiser}, {Leonard}, {Li}, {Matheson},
  {Mauerhan}, {Modjaz}, {Serduke}, {Shields}, {Steele}, {Swift}, {Wong}, \&
  {Yuk}}]{2019MNRAS.482.1545S}
{Shivvers}, I., {Filippenko}, A.~V., {Silverman}, J.~M., {et~al.} 2019, \mnras,
  482, 1545

\bibitem[{{Skilling}(2004)}]{2004AIPC..735..395S}
{Skilling}, J. 2004, in American Institute of Physics Conference Series, Vol.
  735, Bayesian Inference and Maximum Entropy Methods in Science and
  Engineering: 24th International Workshop on Bayesian Inference and Maximum
  Entropy Methods in Science and Engineering, ed. R.~{Fischer}, R.~{Preuss}, \&
  U.~V. {Toussaint}, 395--405

\bibitem[{Skilling(2006)}]{10.1214/06-BA127}
Skilling, J. 2006, Bayesian Analysis, 1, 833

\bibitem[{{Skrutskie} {et~al.}(2006){Skrutskie}, {Cutri}, {Stiening},
  {Weinberg}, {Schneider}, {Carpenter}, {Beichman}, {Capps}, {Chester},
  {Elias}, {Huchra}, {Liebert}, {Lonsdale}, {Monet}, {Price}, {Seitzer},
  {Jarrett}, {Kirkpatrick}, {Gizis}, {Howard}, {Evans}, {Fowler}, {Fullmer},
  {Hurt}, {Light}, {Kopan}, {Marsh}, {McCallon}, {Tam}, {Van Dyk}, \&
  {Wheelock}}]{2006AJ....131.1163S}
{Skrutskie}, M.~F., {Cutri}, R.~M., {Stiening}, R., {et~al.} 2006, \aj, 131,
  1163

\bibitem[{{Smartt} {et~al.}(2015){Smartt}, {Valenti}, {Fraser}, {Inserra},
  {Young}, {Sullivan}, {Pastorello}, {Benetti}, {Gal-Yam}, {Knapic},
  {Molinaro}, {Smareglia}, {Smith}, {Taubenberger}, {Yaron}, {Anderson},
  {Ashall}, {Balland}, {Baltay}, {Barbarino}, {Bauer}, {Baumont}, {Bersier},
  {Blagorodnova}, {Bongard}, {Botticella}, {Bufano}, {Bulla}, {Cappellaro},
  {Campbell}, {Cellier-Holzem}, {Chen}, {Childress}, {Clocchiatti},
  {Contreras}, {Dall'Ora}, {Danziger}, {de Jaeger}, {De Cia}, {Della Valle},
  {Dennefeld}, {Elias-Rosa}, {Elman}, {Feindt}, {Fleury}, {Gall},
  {Gonzalez-Gaitan}, {Galbany}, {Morales Garoffolo}, {Greggio}, {Guillou},
  {Hachinger}, {Hadjiyska}, {Hage}, {Hillebrandt}, {Hodgkin}, {Hsiao}, {James},
  {Jerkstrand}, {Kangas}, {Kankare}, {Kotak}, {Kromer}, {Kuncarayakti},
  {Leloudas}, {Lundqvist}, {Lyman}, {Hook}, {Maguire}, {Manulis}, {Margheim},
  {Mattila}, {Maund}, {Mazzali}, {McCrum}, {McKinnon}, {Moreno-Raya},
  {Nicholl}, {Nugent}, {Pain}, {Pignata}, {Phillips}, {Polshaw}, {Pumo},
  {Rabinowitz}, {Reilly}, {Romero-Ca{\~n}izales}, {Scalzo}, {Schmidt},
  {Schulze}, {Sim}, {Sollerman}, {Taddia}, {Tartaglia}, {Terreran},
  {Tomasella}, {Turatto}, {Walker}, {Walton}, {Wyrzykowski}, {Yuan}, \&
  {Zampieri}}]{2015A&A...579A..40S}
{Smartt}, S.~J., {Valenti}, S., {Fraser}, M., {et~al.} 2015, \aap, 579, A40

\bibitem[{{Soderberg} {et~al.}(2008){Soderberg}, {Berger}, {Page}, {Schady},
  {Parrent}, {Pooley}, {Wang}, {Ofek}, {Cucchiara}, {Rau}, {Waxman}, {Simon},
  {Bock}, {Milne}, {Page}, {Barentine}, {Barthelmy}, {Beardmore}, {Bietenholz},
  {Brown}, {Burrows}, {Burrows}, {Byrngelson}, {Cenko}, {Chandra}, {Cummings},
  {Fox}, {Gal-Yam}, {Gehrels}, {Immler}, {Kasliwal}, {Kong}, {Krimm},
  {Kulkarni}, {Maccarone}, {M{\'e}sz{\'a}ros}, {Nakar}, {O'Brien}, {Overzier},
  {de Pasquale}, {Racusin}, {Rea}, \& {York}}]{2008Natur.453..469S}
{Soderberg}, A.~M., {Berger}, E., {Page}, K.~L., {et~al.} 2008, \nat, 453, 469

\bibitem[{{Soffitta} {et~al.}(1998){Soffitta}, {Feroci}, {Piro}, {in 't Zand},
  {Heise}, {di Ciolo}, {Muller}, {Palazzi}, \&
  {Frontera}}]{1998IAUC.6884....1S}
{Soffitta}, P., {Feroci}, M., {Piro}, L., {et~al.} 1998, \iaucirc, 6884, 1

\bibitem[{{Sollerman} {et~al.}(1998){Sollerman}, {Leibundgut}, \&
  {Spyromilio}}]{1998A&A...337..207S}
{Sollerman}, J., {Leibundgut}, B., \& {Spyromilio}, J. 1998, \aap, 337, 207

\bibitem[{{Speagle}(2020)}]{2020MNRAS.493.3132S}
{Speagle}, J.~S. 2020, \mnras, 493, 3132

\bibitem[{{Stritzinger} {et~al.}(2002){Stritzinger}, {Hamuy}, {Suntzeff},
  {Smith}, {Phillips}, {Maza}, {Strolger}, {Antezana}, {Gonz{\'a}lez},
  {Wischnjewsky}, {Candia}, {Espinoza}, {Gonz{\'a}lez}, {Stubbs}, {Becker},
  {Rubenstein}, \& {Galaz}}]{2002AJ....124.2100S}
{Stritzinger}, M., {Hamuy}, M., {Suntzeff}, N.~B., {et~al.} 2002, \aj, 124,
  2100

\bibitem[{{Stritzinger} {et~al.}(2009){Stritzinger}, {Mazzali}, {Phillips},
  {Immler}, {Soderberg}, {Sollerman}, {Boldt}, {Braithwaite}, {Brown}, {Burns},
  {Contreras}, {Covarrubias}, {Folatelli}, {Freedman}, {Gonz{\'a}lez}, {Hamuy},
  {Krzeminski}, {Madore}, {Milne}, {Morrell}, {Persson}, {Roth}, {Smith}, \&
  {Suntzeff}}]{2009ApJ...696..713S}
{Stritzinger}, M., {Mazzali}, P., {Phillips}, M.~M., {et~al.} 2009, \apj, 696,
  713

\bibitem[{{Stritzinger} {et~al.}(2018){Stritzinger}, {Taddia}, {Burns},
  {Phillips}, {Bersten}, {Contreras}, {Folatelli}, {Holmbo}, {Hsiao},
  {Hoeflich}, {Leloudas}, {Morrell}, {Sollerman}, \&
  {Suntzeff}}]{2018A&A...609A.135S}
{Stritzinger}, M.~D., {Taddia}, F., {Burns}, C.~R., {et~al.} 2018, \aap, 609,
  A135

\bibitem[{{Taddia} {et~al.}(2019){Taddia}, {Sollerman}, {Fremling},
  {Barbarino}, {Karamehmetoglu}, {Arcavi}, {Cenko}, {Filippenko}, {Gal-Yam},
  {Hiramatsu}, {Hosseinzadeh}, {Howell}, {Kulkarni}, {Laher}, {Lunnan},
  {Masci}, {Nugent}, {Nyholm}, {Perley}, {Quimby}, \&
  {Silverman}}]{2019A&A...621A..71T}
{Taddia}, F., {Sollerman}, J., {Fremling}, C., {et~al.} 2019, \aap, 621, A71

\bibitem[{{Taddia} {et~al.}(2015){Taddia}, {Sollerman}, {Leloudas},
  {Stritzinger}, {Valenti}, {Galbany}, {Kessler}, {Schneider}, \&
  {Wheeler}}]{2015A&A...574A..60T}
{Taddia}, F., {Sollerman}, J., {Leloudas}, G., {et~al.} 2015, \aap, 574, A60

\bibitem[{{Taddia} {et~al.}(2018){Taddia}, {Stritzinger}, {Bersten}, {Baron},
  {Burns}, {Contreras}, {Holmbo}, {Hsiao}, {Morrell}, {Phillips}, {Sollerman},
  \& {Suntzeff}}]{2018A&A...609A.136T}
{Taddia}, F., {Stritzinger}, M.~D., {Bersten}, M., {et~al.} 2018, \aap, 609,
  A136

\bibitem[{{Takada-Hidai} {et~al.}(2002){Takada-Hidai}, {Aoki}, \&
  {Zhao}}]{2002PASJ...54..899T}
{Takada-Hidai}, M., {Aoki}, W., \& {Zhao}, G. 2002, \pasj, 54, 899

\bibitem[{{Tartaglia} {et~al.}(2017){Tartaglia}, {Fraser}, {Sand}, {Valenti},
  {Smartt}, {McCully}, {Anderson}, {Arcavi}, {Elias-Rosa}, {Galbany},
  {Gal-Yam}, {Haislip}, {Hosseinzadeh}, {Howell}, {Inserra}, {Jha}, {Kankare},
  {Lundqvist}, {Maguire}, {Mattila}, {Reichart}, {Smith}, {Smith},
  {Stritzinger}, {Sullivan}, {Taddia}, \& {Tomasella}}]{2017ApJ...836L..12T}
{Tartaglia}, L., {Fraser}, M., {Sand}, D.~J., {et~al.} 2017, \apjl, 836, L12

\bibitem[{{Tauris} {et~al.}(2015){Tauris}, {Langer}, \&
  {Podsiadlowski}}]{2015MNRAS.451.2123T}
{Tauris}, T.~M., {Langer}, N., \& {Podsiadlowski}, P. 2015, \mnras, 451, 2123

\bibitem[{{Tody}(1993)}]{1993ASPC...52..173T}
{Tody}, D. 1993, in Astronomical Society of the Pacific Conference Series,
  Vol.~52, Astronomical Data Analysis Software and Systems II, ed. R.~J.
  {Hanisch}, R.~J.~V. {Brissenden}, \& J.~{Barnes}, 173

\bibitem[{{Uomoto}(1986)}]{1986ApJ...310L..35U}
{Uomoto}, A. 1986, \apjl, 310, L35

\bibitem[{{Valenti} {et~al.}(2012){Valenti}, {Taubenberger}, {Pastorello},
  {Aramyan}, {Botticella}, {Fraser}, {Benetti}, {Smartt}, {Cappellaro},
  {Elias-Rosa}, {Ergon}, {Magill}, {Magnier}, {Kotak}, {Price}, {Sollerman},
  {Tomasella}, {Turatto}, \& {Wright}}]{2012ApJ...749L..28V}
{Valenti}, S., {Taubenberger}, S., {Pastorello}, A., {et~al.} 2012, \apjl, 749,
  L28

\bibitem[{{Virtanen} {et~al.}(2020){Virtanen}, {Gommers}, {Oliphant},
  {Haberland}, {Reddy}, {Cournapeau}, {Burovski}, {Peterson}, {Weckesser},
  {Bright}, {van der Walt}, {Brett}, {Wilson}, {Jarrod Millman}, {Mayorov},
  {Nelson}, {Jones}, {Kern}, {Larson}, {Carey}, {Polat}, {Feng}, {Moore}, {Vand
  erPlas}, {Laxalde}, {Perktold}, {Cimrman}, {Henriksen}, {Quintero}, {Harris},
  {Archibald}, {Ribeiro}, {Pedregosa}, {van Mulbregt}, \&
  {Contributors}}]{Virtanen_2020}
{Virtanen}, P., {Gommers}, R., {Oliphant}, T.~E., {et~al.} 2020, Nature
  Methods, 17, 261

\bibitem[{{Wheeler} {et~al.}(2015){Wheeler}, {Johnson}, \&
  {Clocchiatti}}]{2015MNRAS.450.1295W}
{Wheeler}, J.~C., {Johnson}, V., \& {Clocchiatti}, A. 2015, \mnras, 450, 1295

\bibitem[{{Williamson} {et~al.}(2019){Williamson}, {Modjaz}, \&
  {Bianco}}]{2019ApJ...880L..22W}
{Williamson}, M., {Modjaz}, M., \& {Bianco}, F.~B. 2019, \apjl, 880, L22

\bibitem[{{Woosley} {et~al.}(1993){Woosley}, {Langer}, \&
  {Weaver}}]{1993ApJ...411..823W}
{Woosley}, S.~E., {Langer}, N., \& {Weaver}, T.~A. 1993, \apj, 411, 823

\bibitem[{{Wygoda} {et~al.}(2019){Wygoda}, {Elbaz}, \&
  {Katz}}]{2019MNRAS.484.3941W}
{Wygoda}, N., {Elbaz}, Y., \& {Katz}, B. 2019, \mnras, 484, 3941

\bibitem[{{Yamanaka} {et~al.}(2017){Yamanaka}, {Nakaoka}, {Tanaka}, {Maeda},
  {Honda}, {Hanayama}, {Morokuma}, {Imai}, {Kinugasa}, {Murata}, {Nishimori},
  {Hashimoto}, {Gima}, {Hosoya}, {Ito}, {Karita}, {Kawabata}, {Morihana},
  {Morikawa}, {Murakami}, {Nagayama}, {Ono}, {Onozato}, {Sarugaku}, {Sato},
  {Suzuki}, {Takahashi}, {Takayama}, {Yaguchi}, {Akitaya}, {Asakura},
  {Kawabata}, {Kuroda}, {Nogami}, {Oasa}, {Omodaka}, {Saito}, {Sekiguchi},
  {Tominaga}, {Uemura}, \& {Watanabe}}]{2017ApJ...837....1Y}
{Yamanaka}, M., {Nakaoka}, T., {Tanaka}, M., {et~al.} 2017, \apj, 837, 1

\bibitem[{{Yaron} \& {Gal-Yam}(2012)}]{2012PASP..124..668Y}
{Yaron}, O. \& {Gal-Yam}, A. 2012, \pasp, 124, 668

\bibitem[{{Yoon} {et~al.}(2019){Yoon}, {Chun}, {Tolstov}, {Blinnikov}, \&
  {Dessart}}]{2019ApJ...872..174Y}
{Yoon}, S.-C., {Chun}, W., {Tolstov}, A., {Blinnikov}, S., \& {Dessart}, L.
  2019, \apj, 872, 174

\bibitem[{{Yoon} {et~al.}(2012){Yoon}, {Gr{\"a}fener}, {Vink}, {Kozyreva}, \&
  {Izzard}}]{2012A&A...544L..11Y}
{Yoon}, S.~C., {Gr{\"a}fener}, G., {Vink}, J.~S., {Kozyreva}, A., \& {Izzard},
  R.~G. 2012, \aap, 544, L11

\end{thebibliography}

\begin{appendix}

\section{Arnett model description}
\label{sec:model:main:arnett}

The most used analytic model for estimating ejecta properties for supernovae from the %
light curve peak is the \citet[][hereafter A82]{1982ApJ...253..785A} model.
The model is based on the following assumptions:
\begin{enumerate}
\item Homologous expansion of the ejecta:
  \begin{equation}
    R(t) \simeq R(t=0) + v_\text{sc}\,t
  \end{equation}
\item The ejecta is a radiation-dominated gas
\item The diffusion approximation applies
\item The opacity $\kappa_\text{opt}$ is constant throughout the ejecta and also constant in time
\item Spherical symmetry
\item Volume emission of radioactive decay energy is proportional to the radiation energy per unit volume. In this case the radioactive decay energy refers to the ${}^{56}$Ni and ${}^{56}$Co decays.
\end{enumerate}

Apart from these assumptions further approximations can be made, such as the commonly used approximation that the initial radius $R(t=0)$ is small and thus $R(t) \simeq v_\text{sc}\,t$.

Using the formalism from \citet[][their eq. 11\footnote{We use $Q_\text{dep}$ instead of $L_\text{heat}$ and $\tau_m$ instead of $\tau_d$.}]{2019ApJ...878...56K} we can state the luminosity as follows:
\begin{equation}
  L(t) = \frac{2}{\tau_m^2} \exp^{-\left(t / \tau_m\right)^2} \int_0^t t^\prime Q_\text{dep}(t) \exp^{\left({t^\prime}/\tau_m\right)^2} dt^\prime,
\end{equation}
where $\tau_m$ is the diffusion timescale and $Q_\text{dep}(t)$ is the time-dependent heating term.

The diffusion timescale relates to the ejecta mass and scale-velocity $v_\text{sc}$ in the following way 
\citep[][eqs. 18 and 22]{1982ApJ...253..785A}:

\begin{equation}
  \tau_m^2 = \frac{2\kappa_\text{opt} M_\text{ej}}{\beta\,c\,v_\text{sc}},
\end{equation}
where $M_\text{ej}$ is the ejecta mass, $\beta$ is a density-profile integration constant and $\kappa_\text{opt}$ is the opacity.

Since the scale velocity is not a quantity that can be measured directly we need to find some proxy for it.
We can use the `representative' mean expansion rate $v_m$ from \cite{2016MNRAS.458.1618D} (which corresponds to $<v^2>$ in \cite{1980ApJ...237..541A}):
\begin{equation}
  v_m^2 = \frac{2 E_K}{M_\text{ej}}, %
\end{equation}
where $E_K$ is the total kinetic energy of the explosion.

We use \citet[their Eq. 31]{1980ApJ...237..541A} to relate the mean expansion rate to the scale velocity:
\begin{equation}
  v_m^2 = \zeta^2 v_\text{sc}^2,
\end{equation}
where $\zeta$ is a density-profile dependent integration constant\footnote{
\cite{1980ApJ...237..541A} calculate $\zeta^2 \equiv I_K/I_M$ for uniform and different exponential density distributions.
\cite{2018ApJ...868L..24L} provides analytic expressions for broken power-law distributions.
}.

We use the following relation between \ion{He}{I}\,$\lambda\,5876$ line velocity $v_\text{abs}$ (see Sect. \ref{sec:ana:evo:velocity}) and the mean expansion rate $v_m$ from \citet[their sec.~5.3]{2016MNRAS.458.1618D}:
\begin{equation}
  \frac{v_\text{abs}(\ion{He}{I}\,\lambda5876)}{1000~\kms} = 2.64 + 0.765 \frac{v_m}{1000~\kms},
\end{equation}
we estimate the expansion rate of \supernova{2019odp} at light curve peak to be $v_m = 10912 \pm 1435$ \kms.

Thus the ejecta mass and kinetic energy can be given in terms of the mean expansion rate $v_\text{m}$:
\begin{eqnarray}
  M_\text{ej} &=& \frac{\tau_m^2 \,\beta c v_\text{m}}{2 \zeta \kappa_\text{opt}} \\
  E_K &=&  \frac{\tau_m^2 \,\beta c v_\text{m}^3}{4 \zeta \kappa_\text{opt}}
  \label{eqn:a:arnett:ekin}
\end{eqnarray}

Assuming a power-law density profile (with a limited range of the power-law index around uniform density) we approximate $\beta / \zeta$ as a constant value ($17.8$ for this paper\footnote{This corresponds to the often used $\beta=13.8$ and $\zeta^2=\frac{3}{5}$ for uniform density.}), and
adopt a mean opacity $\kappa_\text{opt} = 0.07$ $\text{cm}^2 \text{g}^{-1}$.

{
  \renewcommand{\arraystretch}{1.2}
  \begin{table}[h!]
  \centering
  \caption{Nuclear decay constants used in the radioactive decay chain heating function.}
  \begin{tabular}{|c|c|c|}
    \hline
    Description & Symbol & Value \\
    \hline
    \hline
    ${}^{56}$Ni Lifetime & $\tau_\text{Ni}$ & $8.77$~d\\
    ${}^{56}$Co Lifetime & $\tau_\text{Co}$ & $111.45$~d\\
    ${}^{56}$Ni Decay Lum. & $q_\text{Ni}$ & $6.45\times 10^{43}$~$\text{erg}~M_\odot^{-1}~\text{s}^{-1}$ \\
    ${}^{56}$Co Gamma Decay Lum. & $q_{\text{Co},\gamma}$ & $1.38\times 10^{43}$~$\text{erg}~M_\odot^{-1}~\text{s}^{-1}$ \\
    ${}^{56}$Co Positron Decay Lum. & $q_{\text{Co},e^+}$ & $4.64\times 10^{41}$~$\text{erg}~M_\odot^{-1}~\text{s}^{-1}$ \\
    \hline
    
  \end{tabular}
  \tablebib{\cite{2019MNRAS.484.3941W}}
  \label{tab:nucdata}
\end{table}

}

We use the formalism for the heating function from \cite{2019MNRAS.484.3941W} for the radioactive decay heating of ${}^{56}$Ni and ${}^{56}$Co (their Eqs. 11 and 12):

\begin{equation}
  Q_\gamma(t) = M_\text{Ni} \left( q_\text{Ni} \exp^{-t/\tau_\text{Ni}} + q_{\text{Co},\gamma} \exp^{-t/\tau_\text{Co}} \right) %
\end{equation}
\begin{equation}
  Q_{e^+}(t) = M_\text{Ni} ~ q_{\text{Co},e^+} \left( \exp^{-t/\tau_\text{Co}} - \exp^{-t/\tau_\text{Ni}} \right),
\end{equation}
where $\epsilon_\text{Ni}$ is the specific heating rate for ${}^{56}$Ni decay, $\epsilon_{\text{Co},\gamma}$ is the gamma-ray specific heating rate for ${}^{56}$Co decay, $\epsilon_{\text{Co},e^+}$ is the positron specific heating rate for ${}^{56}$Co decay, $\tau_\text{Ni,Co}$ are the decay timescales \citep{1994ApJS...92..527N}.
The used physical constants for the heating function(s) are summarised in \autoref{tab:nucdata}.
The total heating function is given by:
\begin{equation}
  Q_\text{dep}(t) = Q_\gamma(t) + Q_{e^+}(t)
\end{equation}

\subsection{Arnett model: Large initial radius variant}
\label{sec:model:main:arnett:large}

While the A82 model as previously described assumes the initial radius $R_0$ to be negligible ($R_0 \ll 1000~R_\odot$) one can relax this assumption.

For this we introduce an additional timescale:
\begin{equation}
  \tau_r = \frac{v_\text{sc} \tau_m^2}{2 R_0}
\end{equation}

and then following the approach from \cite{2012ApJ...746..121C} (but using the more generalised notation from before) we get the following function for the luminosity:
\begin{equation}
  L(t) = \frac{2}{\tau_m^2} \exp^{-t^2 / \tau_m^2 - t / \tau_r} \int_0^t (t^\prime+\frac{R_0}{v_\text{sc}}) Q_\text{dep}(t) \exp^{{t^\prime}^2/\tau_m^2 + t^\prime/\tau_r} dt^\prime
\end{equation}

\subsection{Photometric fitting framework}
\label{sec:mod:phot:fitting}

Based on the observed photometric evolution of the supernovae we want to use a Bayesian framework to estimate the parameters of the models discussed in Sect. \ref{sec:model:main:arnett}.
We use the combined and interpolated photometric dataset described in Sect. \ref{sec:obs:interpolated} as input.
We use the Python nested sampling framework \textit{dynesty} to both estimate the posterior as well as the evidence of the model under consideration.
The methodology propagates all the correlated uncertainties (extinction, distance, Lyman bolometric correction factor) into the derived posterior.

We pregenerate 12000 realisations of the bolometric light curve based on the following algorithm\footnote{This pregeneration step is purely for performance reasons, since the Gaussian process step is quite expensive. We chose a large enough number of realisations for us to be able to replace the call to the Gaussian Process sampler function inside the likelihood function with a random choice of one such realisation without affecting the resulting posterior distribution}:
\begin{enumerate}
\item We sample $N_t=50$ uniform times ($\mathcal{U}(t_\text{min}, t_\text{max})$) from the temporal validity range of the underlying interpolated light curves (to ensure we are only interpolating them and not extrapolating).
\item We create a pseudo-bolometric light curve realisation by sampling the Gaussian process of the combined light curves (see Sect. \ref{sec:obs:interpolated}) and then applying the bolometric method to convert the apparent magnitudes into pseudo-bolometric magnitudes (see Sect. \ref{sec:evo:qbol}).
\item Next we sample from each of the uncertainty distributions of the correlated uncertainties mentioned before and apply the offset to the whole realisation of the bolometric light curve (since we assume them to be the same for each point in the light curve).
\end{enumerate}
In the actual likelihood function we then randomly select one of those light curve realisations $I_X$ in each iteration and calculate the model magnitudes $M$ based on the given parameter vector $\theta$ for the iteration.
The likelihood is then given as
\begin{equation}
  \log \mathcal{L} (\theta) = -2.5 \sum_{i=1}^{N=50} \left( I_{X,i} - M_i (\theta) \right)^2
\end{equation}

Using this method we can assure that all correlated uncertainties and the uncertainty from the Gaussian-process interpolation step are correctly propagated.

The used priors for the model are shown in \autoref{tab:mod:phot:arnett:r0small:prior}.
Since the model does not contain any treatment for gamma-ray leakage we restrict the fitting range to roughly the diffusion timescale.
The resulting parameter estimates for \supernova{2019odp} and the comparison objects for all models are shown in \autoref{tab:mod:phot:results} (we also include literature values where available as a consistency check).

\begin{table}[hb]
  \caption{Priors for the small $R_0$ Arnett model fit.}
  \begin{tabular}{|c|c|c|c|}
    \hline
    Parameter & Symbol & Unit & Prior \\
    \hline
    \hline
    Nickel Mass & \Mni & $M_\odot$ & $\mathcal{U}(0.01,1.4)$ \\
    Diffusion Timescale & $\tau_m$ & d & $\mathcal{U}(5,55)$ \\
    Explosion Epoch & $t_\text{expl}$ & d & $\mathcal{U}(t_\text{min},t_\text{detect})$ \\
    \hline
  \end{tabular}
  \tablefoot{The root-mean-square velocity values are different per object: For \supernova{2019odp} the value can be found in Sect. \ref{sec:ana:evo:velocity} and for the comparison objects they are taken from \autoref{tab:ana:comparison:params}. For \supernova{2019odp} we use the explosion epoch prior given in Sect. \ref{sec:discovery} and for the comparison objects the one listed in \autoref{tab:ana:comparison:params}. }
  \label{tab:mod:phot:arnett:r0small:prior}
\end{table}

\section{Photometric models}
\label{appendix:photmodel}

\begin{table}[h]
  \caption{Used photometric models per supernova and band.}
  \begin{tabular}{|c|c|c|}
    \hline
    Transient & Band & Model \\
    \hline
    \hline
    \supernova{2019odp} & ugri & plateau-contardo \\
    \supernova{2019odp} & zJHK & linear \\
    \hline
    iPTF13bvn & Ugriz & plateau-contardo \\
    \supernova{2008D} & UBVri & prebump-contardo \\
    \supernova{1998bw} & UBV $\text{R}_C\,\text{I}_C$ & plateau-conardo \\
    \supernova{2002ap} & UBVRI & plateau-contardo \\
    \hline
  \end{tabular}
  \label{tab:a:photmodel:assignment}
\end{table}

In this section we describe the used analytic photometry light curve models and priors for the models.
We use different models depending on the light curve coverage and the shape of the early excess (if there was one detected) with the decision being specific to the photometric band.
The assignment of photometric model to a given transient and band is listed in \autoref{tab:a:photmodel:assignment}.
All analytic functions use the phase relative to the prior peak estimate: $\Delta t \equiv t - t_\text{peak}$.

Our (complex) light curve models are based on the light curve model by \cite{2000A&A...359..876C} (hereafter C20):
\begin{equation}
  m_\text{C20}(\Delta t) = \delta(\Delta t) \left( \alpha + \beta \Delta t + A_\text{DF} \exp{\frac{-(\Delta t - t_{0,\text{DF}})}{2 \sigma_\text{DF}^2}} \right),
\end{equation}
where $\Delta t$ is the phase relative to the main peak prior time, $\delta(t)$ is the explosion scaling function, $\alpha$ is the linear intercept, $\beta$ is the linear slope, $A_\text{DF}$ is the amplitude of the main diffusion peak Gaussian, $t_{0,\text{DF}}$ is the phase offset of the main diffusion peak Gaussian and $\sigma_\text{DF}$ is the width of the Gaussian.
The rise scaling function is given as
\begin{equation}
  \delta(\Delta t)^{-1} = 1 - \exp{\frac{-(\Delta t - t_{0,\text{rise}})}{\tau_\text{rise}}},
\end{equation}
where $t_{0,\text{rise}}$ is the phase offset of the rise scaling function, $\tau_\text{rise}$ is the rise timescale.

We derive two modified light curve models from the C20 depending on the shape of the early excess:
\begin{enumerate}
\item The `plateau-contardo' model adds a smoothing function $g(t)$ to interpolate between the plateau magnitude $m_\text{plat}$ and the $m_\text{C20}$ function:
  \begin{eqnarray}
    g(\Delta t) =\left(
    \arctan\left( \frac{\Delta t - t_{0,\text{plat}}}{\tau_\text{smooth}} \right) \frac{1}{\pi} + 0.5
    \right)^2 \\
    m_\text{PLC}(\Delta t) = m_\text{plat} + (m_\text{C20}(\Delta t)-m_\text{plat}) g(\Delta t)
  \end{eqnarray}
  This introduces three additional parameters to the C20 model that we allow to vary in a reasonable range: plateau magnitude $m_\text{plat}$, plateau end time $t_{0,\text{plat}}$ and the smoothing timescale $\tau_\text{smooth}$.
\item The `prebump-contardo' model adds a secondary Gaussian peak at a peak relative to the main peak:
  \begin{equation}
    m_\text{PBC}(\Delta t) = p_\text{C20} + \delta(t) A_\text{PB} \exp{ \frac{-(\Delta t - t_{0,\text{PB}})}{2 \sigma_\text{PB}^2} }
  \end{equation}
  This introduces three additional parameters to the model: pre-bump amplitude $A_\text{PB}$, pre-bump width $\sigma_\text{PB}$ and centre time of the pre-bump $t_{0,\text{PB}}$.
\end{enumerate}

If the data only covers a small time range before/after peak we use a linear model instead:
\begin{equation}
  m_\text{lin}(\Delta t) = \alpha + \frac{\beta}{1000} \Delta t
\end{equation}

\FloatBarrier

\section{Oxygen mass modelling}
\label{sec:a:oxygen}

\subsection{NLTE deviation factor}
\label{sec:a:oxygen:nlte}

The LTE departure coefficient is defined as follows:
\begin{equation}
  d_2 = \frac{n_2}{n_2^\text{LTE}}
\end{equation}

\begin{figure}
  \includegraphics[width=\linewidth]{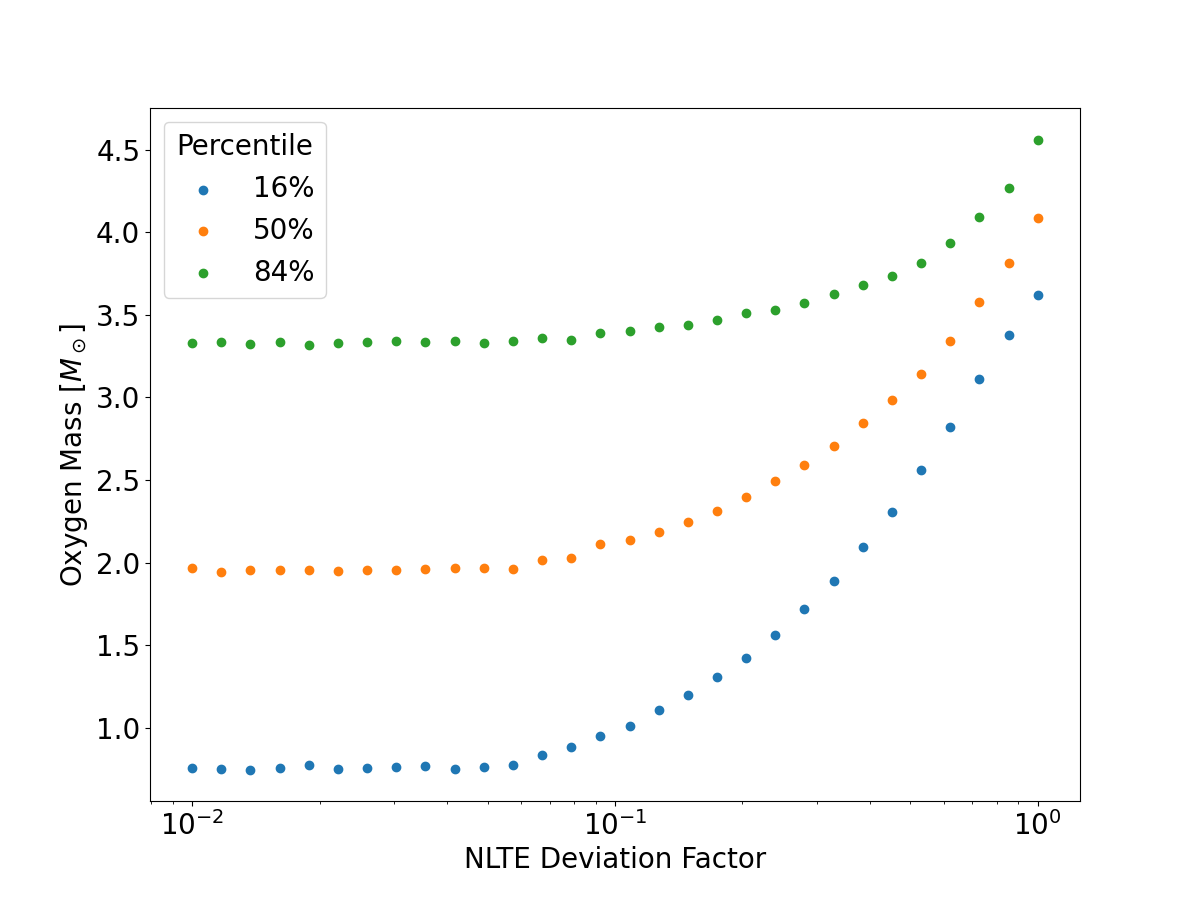}
  \caption{Sensitivity of the oxygen mass percentiles as a function of the minimum LTE departure coefficient $d_2$.}
  \label{fig:a:oxygen:nlte:sens}
\end{figure}

We estimate the sensitivity to deviations from LTE conditions by running the fitting procedure for a spectrum multiple times in a sequence where we vary the allowed maximum LTE departure coefficient $d_2$ from 0.01 to 1.0 (where 1.0 means LTE conditions).
The resulting change due to the change of the LTE departure can be seen in Fig. \ref{fig:a:oxygen:nlte:sens}.

We get a rough estimate for the LTE departure coefficient $d_2$ by estimating the electron density $n_e$ from the $\ion{O}{I}\,\lambda 7774$ recombination line and using the following relation:
\begin{equation}
  d_2 \approx \left( 1 + 1.44 \left( \frac{T}{1000\,\text{K}} \right)^{-0.034} \left( \frac{n_e}{10^8\,\text{cm}^{-3}} \right)^{-1} \right)^{-1}
  \label{eqn:a:oxygen:nlte:ne}
\end{equation}
which we got by dividing Eq. 2 from \cite{1996ApJ...456..811H} by Eq. 2 from \cite{2015A&A...573A..12J}.
To approximate the electron density we use the oxygen recombination lines that are visible in the earlier spectra.
We can use the following relation from \cite{2015A&A...573A..12J} to relate the line luminosity to the electron density (their Eq. 3):
\begin{equation}
  L_\text{rec} = \frac{4\pi}{3} \left( V_\text{core} t \right)^3 \Psi \alpha_\text{eff} f_O n_e^2 h \nu,
\end{equation}
where $V_\text{core}$ is the line width, $t$ is the time since explosion, $\Psi$ is the fraction of electrons provided by oxygen initiations and $f_O$ is the oxygen zone filling factor.

We assume $\Psi$ to be of order unity (from \citealt{2015A&A...573A..12J}).
We estimate $V_\text{core}$ from the measured Gaussian line width.
Since $\alpha_\text{eff}$ depends on the temperature, we use the values from \citet[their sect. C.1]{2015A&A...573A..12J} for three different temperatures.
Using $\alpha(T=2500\,\text{K})=2.8 \times 10 ^{-13}\,\text{cm}^3\,\text{s}^{-1}$, $\alpha(T=5000\,\text{K})=1.6 \times 10^{-13}\,\text{cm}^3\,\text{s}^{-1}$ and $\alpha(T=7500\,\text{K})=1.1 \times 10^{-13}\,\text{cm}^3\,\text{s}^{-1}$ and a crudely estimated line luminosity of $L \simeq 27 \cdot 10^{38}$ erg\,$\text{s}^{-1}$ as well as the line width estimated velocity of 2464 \kms  (both from the $+158$\,d Keck spectrum) we estimate $n_e \sqrt{f_O}$ to be
$2.2 \cdot 10^{8}$\,$\text{cm}^{-3}$ (2500~K), $2.9 \cdot 10^{8}$\,$\text{cm}^{-3}$ (5000~K) and $3.5 \cdot 10^{8}$\,$\text{cm}^{-3}$ (7500~K).
This is in line to the values seen in \cite{2015A&A...573A..12J}.
Using Eq. \ref{eqn:a:oxygen:nlte:ne} we get a range of $0.6$ to $0.72$ for LTE departure coefficient $d_2$ for an assumed zone filling factor of one.
Higher filling factors would yield even higher values.
However the recombination lines can be emitted from higher density regions than the nebular lines we are using in the main analysis and are thus not really suitable for quantitative analysis and we thus do not use this estimated  $d_2$ value/range in our main analysis.

\subsection{Spectral fitting results}
\label{appendix:oxygen:spectralfits}

\begin{figure}[h]
  \includegraphics[width=\linewidth]{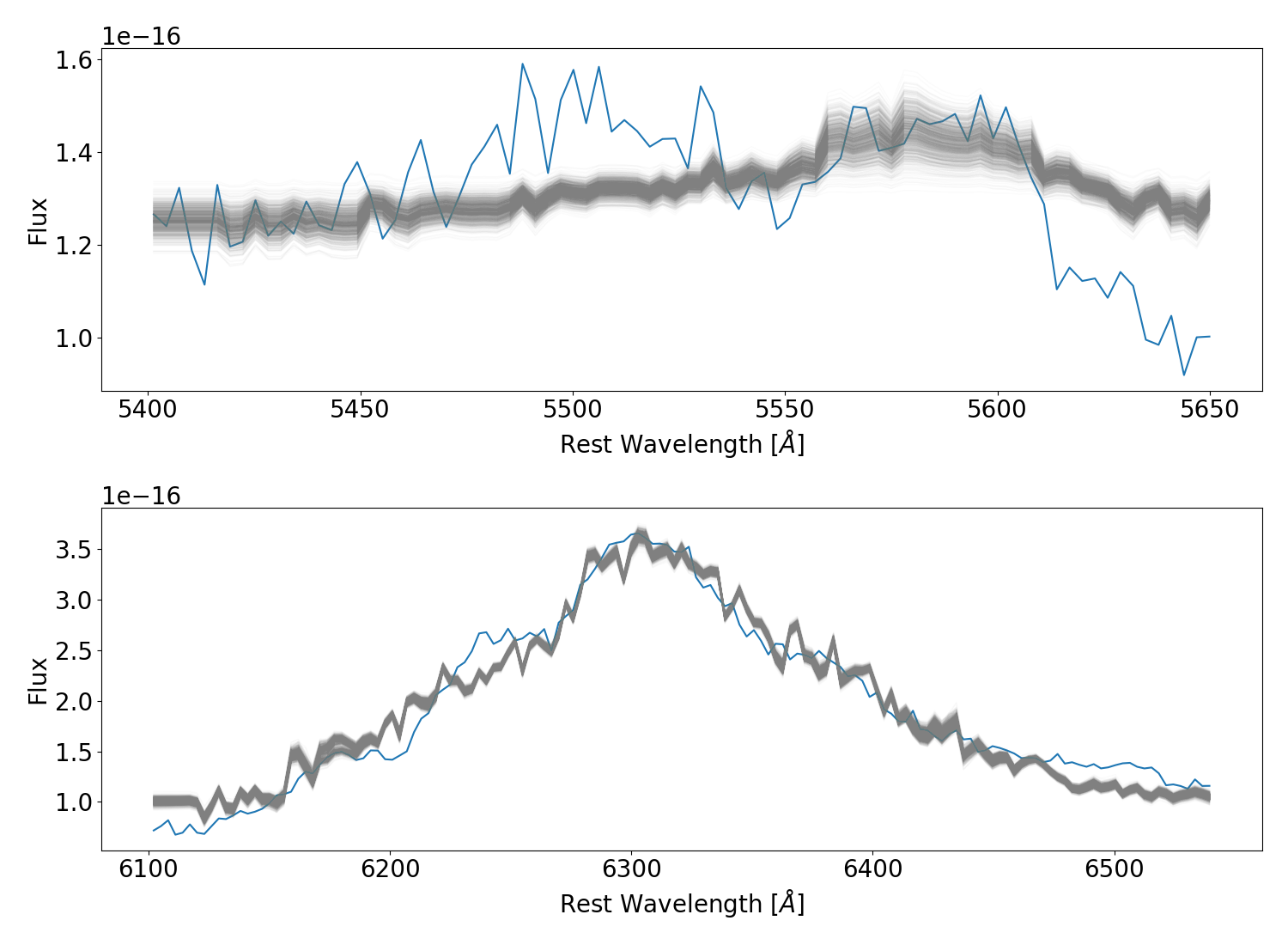}
  \caption{NOT/ALFOSC spectrum at 128 days post-peak with model spectra drawn from the posterior distribution of the model overlaid in grey.}
  \label{fig:a:oxygen:spec:diag:notlate}
\end{figure}

\begin{figure}[h]
  \includegraphics[width=\linewidth]{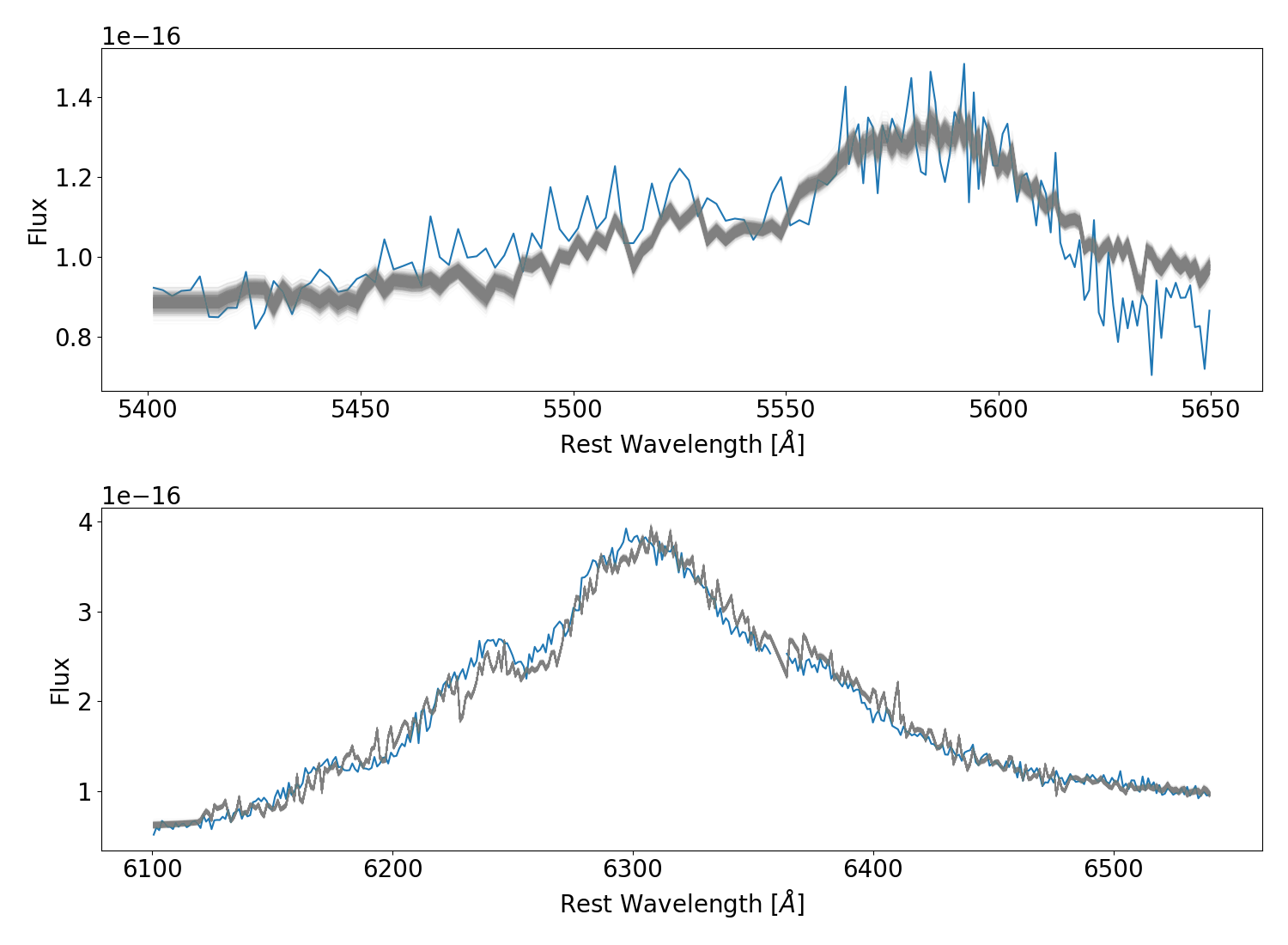}
  \caption{Early Keck/LRIS spectrum at 138 days post-peak with model spectra drawn from the posterior distribution of the model overlaid in grey.}
  \label{fig:a:oxygen:spec:diag:keckearly}
\end{figure}

\begin{figure}[h]
  \includegraphics[width=\linewidth]{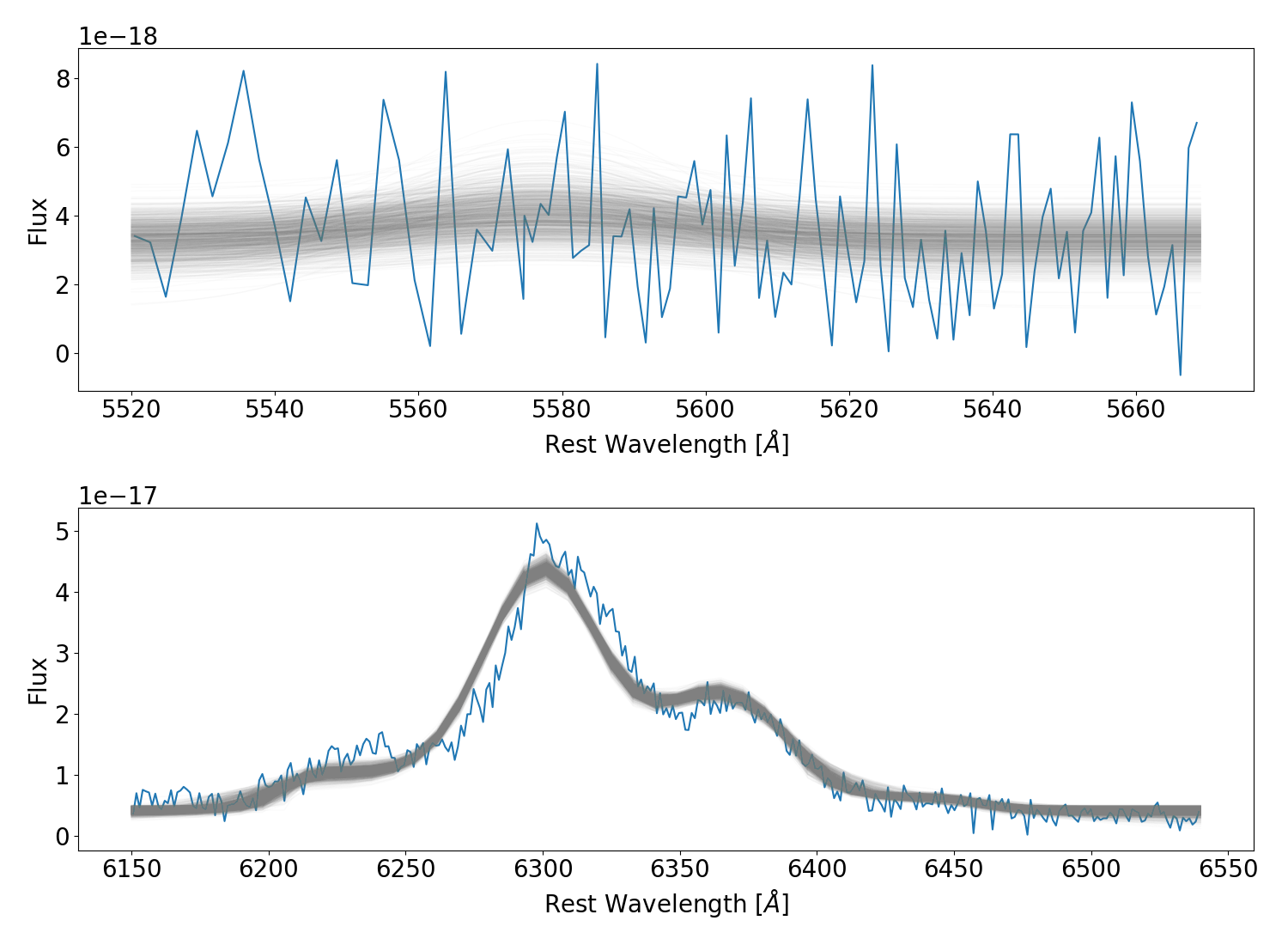}
  \caption{Late Keck/LRIS spectrum at 358 days post-peak with model spectra drawn from the posterior distribution of the model overlaid in grey.}
  \label{fig:a:oxygen:spec:diag:kecklate}
\end{figure}

In this section we show the best fit results from the spectral line profile models.
In Figs. \ref{fig:a:oxygen:spec:diag:notlate}, \ref{fig:a:oxygen:spec:diag:keckearly}, and Fig. \ref{fig:a:oxygen:spec:diag:kecklate} we show the relevant spectral fitting regions with model spectra drawn from the posterior distribution overplotted.
The corresponding corner plots visualising the posterior distributions (as well as the trace files) can be found on Zenodo at \url{https://zenodo.org/record/7554926}.  %

\FloatBarrier

\FloatBarrier
\newpage

\section{Photometric blackbody fitting}
\label{appendix:bbfit}

We use the interpolated and pre-processed light curve constructed by the method described in Sect. \ref{sec:obs:interpolated}.
We construct the time grid to perform the fitting by selecting all observations between the first time all selected filter bands had at least one detection and the peak and then selecting all days post-peak that had at least one observation.
This ensures we are only interpolating the light curve and never have to rely on extrapolating it (which is highly uncertain at the very early epochs).

For each time on the time grid we estimate the extinction-corrected magnitude in all filter bands.
Next we fit a three-parameter Bayesian model to these observations: temperature $\log\,T$, radius $\log\, R$ and the distance $D$.
We follow the suggestion by \cite{2022ApJ...937...75A} and use a log-uniform prior for the temperature (although we use nested sampling instead of MCMC).
The distance is a nuisance parameter, which is constrained by the redshift uncertainty.
In the likelihood function we generate a blackbody SED using the temperature and radius.
We then perform synthetic photometry using the corresponding filter curves to compare to the light curve.
We perform the posterior calculation using the nested sampling code \textit{dynesty}.

\subsection{Validation}

\begin{figure}
  \includegraphics[width=\linewidth]{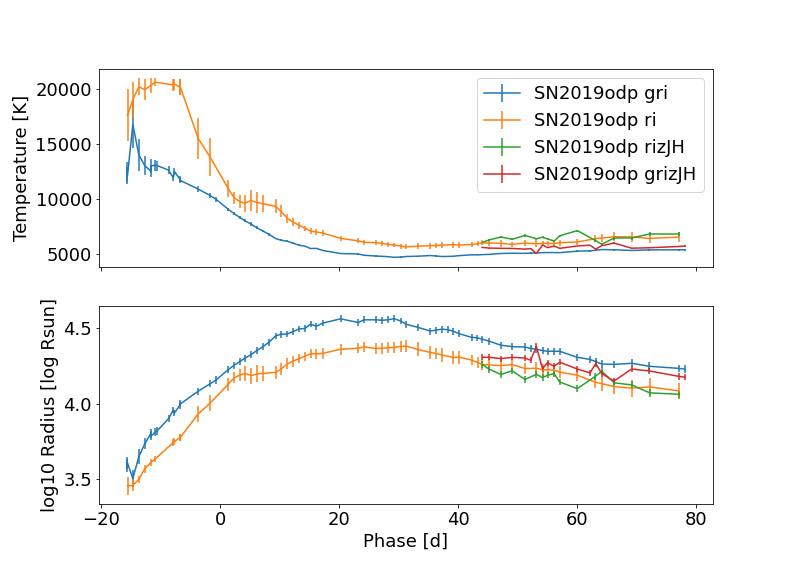}
  \caption{Comparison between three filter sets.}
  \label{fig:bbfit:validation:comparison}
\end{figure}

We compare the resulting temperature and radius between three different filter combinations: $gri$, $ri$ and $rizJH$.
The comparison is shown in Fig. \ref{fig:bbfit:validation:comparison}.
We only have a full spectral coverage from optical to NIR between 40 and 80 days.

\FloatBarrier

\section{Observation logs}

\begin{table*}
  \centering
  \caption{Full forced photometry dataset obtained for SN\,2019odp.}
\begin{tabular}{|c|c|c|c|c|c|c|c|c|c|c|}
\hline
UT & MJD & $\Delta t_\text{expl}$ & $\Delta t_g$ & Filter & Telescope/Instrument & $m$ & $\Delta m$ & $m_\text{lim}$ & $F$ & $\Delta F$ \\
& (d) & (d) & (d) & & & (mag) & (mag) & (mag) & ($\mu$Jy) & ($\mu$Jy) \\ 
\hline
\hline
2019-08-17 09:01 & 58712.4 & -2.1 & -21.6 & r & P48/ZTF & nan & nan & 20.11 & -1.0 & 7.4 \\
2019-08-17 11:26 & 58712.5 & -2.0 & -21.5 & g & P48/ZTF & nan & nan & 19.89 & -1.5 & 9.3 \\
2019-08-18 10:28 & 58713.4 & -1.1 & -20.6 & i & P48/ZTF & 20.91 & 0.76 & 19.83 & 15.7 & 10.6 \\
2019-08-21 09:19 & 58716.4 & 1.9 & -17.6 & r & P48/ZTF & 18.72 & 0.05 & 21.26 & 117.7 & 3.1 \\
2019-08-21 09:23 & 58716.4 & 1.9 & -17.6 & r & P48/ZTF & 18.76 & 0.05 & 21.21 & 114.2 & 3.2 \\
2019-08-21 09:24 & 58716.4 & 1.9 & -17.6 & r & P48/ZTF & 18.67 & 0.05 & 21.23 & 123.4 & 3.1 \\
2019-08-22 07:52 & 58717.3 & 2.8 & -16.7 & r & P48/ZTF & 18.70 & 0.05 & 21.12 & 120.6 & 3.1 \\
2019-08-22 09:28 & 58717.4 & 2.9 & -16.6 & g & P48/ZTF & 18.69 & 0.07 & 20.95 & 121.8 & 4.2 \\
2019-08-22 09:52 & 58717.4 & 2.9 & -16.6 & g & P48/ZTF & 18.69 & 0.07 & 21.04 & 121.7 & 3.8 \\
2019-08-22 09:53 & 58717.4 & 2.9 & -16.6 & g & P48/ZTF & 18.69 & 0.07 & 21.13 & 120.9 & 3.6 \\
2019-08-22 09:53 & 58717.4 & 2.9 & -16.6 & g & P48/ZTF & 18.60 & 0.06 & 21.16 & 131.7 & 3.3 \\
2019-08-23 04:07 & 58718.2 & 3.7 & -15.8 & UVW1 & \textit{Swift}/UVOT & 21.17 & 0.35 & 21.24 & 12.4 & 4.0 \\
2019-08-23 04:09 & 58718.2 & 3.7 & -15.8 & U & \textit{Swift}/UVOT & 20.49 & 0.34 & 20.62 & 23.1 & 7.1 \\
2019-08-23 04:10 & 58718.2 & 3.7 & -15.8 & B & \textit{Swift}/UVOT & 18.91 & 0.18 & 19.91 & 99.1 & 16.7 \\
2019-08-23 04:13 & 58718.2 & 3.7 & -15.8 & UVW2 & \textit{Swift}/UVOT & 22.81 & 0.59 & 22.05 & 2.7 & 1.5 \\
2019-08-23 04:17 & 58718.2 & 3.7 & -15.8 & V & \textit{Swift}/UVOT & 18.59 & 0.28 & 18.98 & 133.0 & 34.0 \\
2019-08-23 04:28 & 58718.2 & 3.7 & -15.8 & UVM2 & \textit{Swift}/UVOT & 22.41 & 0.29 & 22.79 & 3.9 & 1.0 \\
2019-08-23 08:55 & 58718.4 & 3.9 & -15.6 & r & P48/ZTF & 18.63 & 0.07 & 20.62 & 128.1 & 4.6 \\
2019-08-23 09:19 & 58718.4 & 3.9 & -15.6 & i & P48/ZTF & 18.93 & 0.08 & 20.72 & 97.3 & 4.9 \\
2019-08-23 09:35 & 58718.4 & 3.9 & -15.6 & g & P48/ZTF & 18.77 & 0.07 & 21.01 & 112.8 & 3.8 \\
2019-08-24 09:10 & 58719.4 & 4.9 & -14.6 & r & P48/ZTF & 18.35 & 0.04 & 21.34 & 165.3 & 2.7 \\

\hline
\end{tabular}
\tablefoot{
ZTF photometry is based on image subtraction forced photometry and contains pre-explosion epochs.
Both magnitudes as well as fluxes are given.
No foreground/host extinction correction nor any other secondary correction steps are applied.
The full photometry table are available on Zenodo at \url{https://zenodo.org/record/7554926} (which also contains a second file with all corrections applied).}
\label{tab:obslog:phot}
\end{table*}

\begin{table*}
  \caption{Observation log of the spectroscopic observations.}
\begin{tabular}{|c|c|c|c|c|c|c|c|}
\hline
UT & MJD & $\Delta t_\text{expl}$ & $\Delta t_g$ & Telescope/Instrument & Setup & Airmass & Exp. Time \\
& (d) & (d) & (d) & & & & (s) \\ 
\hline
\hline
2019-08-22 03:52 & 58717.2 & 2.7 & -16.8 & P60/SEDM & IFU & 3.0 & 2250 \\
2019-08-23 05:23 & 58718.2 & 3.7 & -15.8 & ESO-NTT/EFOSC & 1.0-slit/Gr13 & 1.4 & 900 \\
2019-08-23 10:55 & 58718.5 & 4.0 & -15.5 & P60/SEDM & IFU & 1.2 & 2250 \\
2019-08-25 04:15 & 58720.2 & 5.7 & -13.8 & P60/SEDM & IFU & 2.2 & 2250 \\
2019-08-27 08:44 & 58722.4 & 7.9 & -11.6 & P200/DBSP & 600/4000 & 1.1 & 300 \\
2019-08-28 04:06 & 58723.2 & 8.7 & -10.8 & P60/SEDM & IFU & 2.2 & 2250 \\
2019-08-30 23:35 & 58726.0 & 11.5 & -8.0 & NOT/ALFOSC & Gr4 & 1.2 & 1200 \\
2019-09-01 04:56 & 58727.2 & 12.7 & -6.8 & P60/SEDM & IFU & 1.5 & 1800 \\
2019-09-10 07:19 & 58736.3 & 21.8 & 2.3 & P60/SEDM & IFU & 1.1 & 1800 \\
2019-09-17 07:44 & 58743.3 & 28.8 & 9.3 & P60/SEDM & IFU & 1.1 & 1800 \\
2019-09-23 02:48 & 58749.1 & 34.6 & 15.1 & P60/SEDM & IFU & 1.9 & 1800 \\
2019-09-28 05:56 & 58754.2 & 39.7 & 20.2 & P60/SEDM & IFU & 1.1 & 1800 \\
2019-10-03 22:20 & 58759.9 & 45.4 & 25.9 & NOT/ALFOSC & Gr4 & 1.1 & 1800 \\
2019-10-07 05:35 & 58763.2 & 48.7 & 29.2 & P60/SEDM & IFU & 1.1 & 1800 \\
2019-10-13 07:06 & 58769.3 & 54.8 & 35.3 & P60/SEDM & IFU & 1.1 & 1800 \\
2019-10-13 07:47 & 58769.3 & 54.8 & 35.3 & P60/SEDM & IFU & 1.3 & 1800 \\
2019-10-20 05:46 & 58776.2 & 61.7 & 42.2 & P60/SEDM & IFU & 1.1 & 2250 \\
2019-10-22 00:46 & 58778.0 & 63.5 & 44.0 & NOT/ALFOSC & Gr4 & 1.3 & 2200 \\
2019-10-27 03:34 & 58783.1 & 68.6 & 49.1 & P60/SEDM & IFU & 1.1 & 2250 \\
2019-11-04 02:02 & 58791.1 & 76.6 & 57.1 & P60/SEDM & IFU & 1.2 & 2250 \\
2019-11-11 03:33 & 58798.1 & 83.6 & 64.1 & P60/SEDM & IFU & 1.1 & 2250 \\
2019-11-22 22:20 & 58809.9 & 95.4 & 75.9 & NOT/ALFOSC & Gr4 & 1.2 & 2200 \\
2019-11-24 06:12 & 58811.3 & 96.8 & 77.3 & P60/SEDM & IFU & 1.6 & 2250 \\
2019-12-18 02:26 & 58835.1 & 120.6 & 101.1 & P60/SEDM & IFU & 1.1 & 2250 \\
2019-12-21 03:58 & 58838.2 & 123.7 & 104.2 & P60/SEDM & IFU & 1.4 & 2250 \\
2020-01-04 03:21 & 58852.1 & 137.6 & 118.1 & P60/SEDM & IFU & 1.5 & 2250 \\
2020-01-13 20:27 & 58861.9 & 147.4 & 127.9 & NOT/ALFOSC  & Gr4 & 1.6 & 2700 \\
2020-01-24 05:30 & 58872.2 & 157.7 & 138.2 & Keck-I/LRIS & 1.0-slit/400/3400/8500 & 2.0 & 300 \\
2020-08-21 11:58 & 59082.5 & 368.0 & 348.5 & Keck-I/LRIS & 1.0-slit/400/3400/8500 & 1.0 & 1363 \\

\hline
\end{tabular}

\label{tab:obslog:spec}
\end{table*}

\begin{table*}[h!]
  \centering
  \caption{Facilities used for photometric follow-up observations as well as their respective native filter systems and calibration sources.}
  \begin{tabular}{|c|c|c|c|c|}
    \hline
    Telescope/Instrument & Bands & Filter System & Reference System & Calibration Source \\
    \hline
    \hline
    P48/ZTF & $gri$ & ZTF & AB & PS1/Internal \\
    \textit{Swift}/UVOT & UBV\,M2\,W1\,W2 & Custom & Vega & Internal \\
    MPG 2.2m/GROND & $g^\prime r^\prime i^\prime z^\prime$ & Sloan/Custom & AB & SDSS DR12 \\
    MPG 2.2m/GROND & $JH K_s$ & Johnson/Custom & Vega & 2MASS All-Sky DR \\
    P60/SEDM-RC & $ugri$ & Sloan/Astrodon & AB & SDSS/PS1 \\
    NOT/ALFOSC & $gri$ & Sloan & AB & PS1 \\
    \hline
  \end{tabular}
  \label{tab:a:photsystem}
\end{table*}

A partial listing of the forced photometry dataset is given in \autoref{tab:obslog:phot} with the full dataset available on Zenodo at \url{https://zenodo.org/record/7554926}.
The used photometric filters, reference systems and calibration sources are described in \autoref{tab:a:photsystem}.
The observation log of the spectroscopic observations is given in \autoref{tab:obslog:spec}.

\newpage

\FloatBarrier

\end{appendix}

\end{document}